\documentclass[a4paper,fleqn,usenatbib]{mnras}

\usepackage{mathptmx}

\usepackage[T1]{fontenc}
\usepackage{ae,aecompl}

\usepackage{graphicx}	
\usepackage{amsmath}	
\usepackage{amssymb}	
\usepackage{comment}
\title[]{Space reconstruction of the morphology and kinematics of axisymmetric radio sources}

\author[P.N. Diep et al.]{
P.N. Diep\thanks{E-mail: pndiep@mail.vnsc.org.vn}, N.T. Phuong, D.T. Hoai, P.T. Nhung, N.T. Thao, P. Tuan-Anh and
\newauthor{P. Darriulat}
\\
Department of Astrophysics, Vietnam National Satellite Center (VNSC), Vietnam Academy of Science and Technology (VAST),\\ 
18 Hoang Quoc Viet, Ha Noi, Viet Nam
}

\date{Accepted XXX. Received YYY; in original form ZZZ}

\pubyear{2016}

\begin{document}
\label{firstpage}
\pagerange{\pageref{firstpage}--\pageref{lastpage}}
\maketitle

\begin{abstract}
The unprecedented quality of the observations available from the Atacama 
Large Millimetre/sub-millimetre Array (ALMA) calls for analysis methods 
making the best of them. Reconstructing in space the morphology and kinematics
 of radio sources is an underdetermined problem that requires imposing 
additional constraints for its solution. The hypothesis of rotational 
invariance, which is a good approximation to, 
or at least a good reference for the description of the gas envelopes of 
many evolved stars and protostars, is particularly efficient in this role. 
In the first part of the article, a systematic use of simulated observations 
allows for identifying the main problems and for constructing quantities 
aimed at solving them. In particular the evaluation of the
orientation of the star axis in space and the differentiation between 
expansion along the star axis and rotation about it are given 
special attention. The use of polar rather than Cartesian sky coordinates 
to display the results of the analysis is shown to often better match the
morphology and kinematics of actual stars. The radial dependence of the 
gas density and temperature and the possible presence of velocity gradients 
are briefly considered. In the second part, these results are applied to a few stars taken as examples
with the aim of evaluating their usefulness when applied to concrete cases. 
A third part takes stock of what precedes and formulates some guidelines 
for modelling the radio emission of axisymmetric radio sources, 
limited however to the mathematics and geometry of the problem, 
physics considerations being generally ignored.
\end{abstract}



\section{Introduction}
From the end of their lifetime on the Main Sequence to their death 
as White Dwarfs, most stars evolve from a perfectly spherically symmetric 
morphology to the often very irregular shapes of Planetary Nebulae 
dissolving into the Interstellar Medium (ISM). The transition, with Red Giant 
and Asymptotic Giant Branch (AGB) as main intermediate states, is the object 
of intense study. The first departure from spherical symmetry is often 
observed to be the appearance of a bipolar outflow, axisymmetry -- meaning
invariance by rotation (rather than symmetry) about an axis -- replacing 
spherical symmetry. Many observations, whether of the dust from infrared, 
mid-infrared and far-infrared or of the gas from atomic radio or molecular 
millimetre/sub-millimetre emission, in particular from carbon monoxide, 
are interpreted in terms of such symmetry. Namely they aim at giving a model 
of the circumstellar envelope (CSE) displaying invariance with respect 
to rotation about a well-defined star axis, averaging over deviations from it, 
assumed to be small. The physics mechanisms at play in such symmetry breaking 
are not well understood and many observations aim at their clarification. 
The presence of a companion accreting gas from the dying star is very often 
invoked, but the roles of rotation and of magnetic field -- how the 
angular momentum and magnetic flux of the original star are 
distributed between the degenerate core and the CSE -- are still unclear.

The study of protostars displays striking similarities with the study of 
evolved stars. The CSE is now in-falling rather than expanding but axisymmetry 
is again found to be a good working approximation in many cases and the 
mechanisms that govern the transition from gravitational in-fall to formation 
of a rotating disk, accretion by the protostar and emission of a bipolar 
outflow are still partly unclear.
  
In the present article, we focus on the millimetre/sub-millimetre observation 
of the expanding CSE of AGB stars, typically from carbon monoxide emission, 
but many developments apply as well to the in-falling CSEs of protostars.

The data are in the form of flux densities $f$, measured in Jansky/beam in 
bins (pixels) of sky coordinates ($y$ to the east and $z$ to the north) 
and of frequency, the latter being expressed in terms of the Doppler velocity 
$V_x$ measured along the line of sight taken as $x$ axis. Making a model of 
the CSE from such observations can be conceptually separated in two steps, 
even if, in practice, they are generally mixed: the reconstruction in 
space of the gas properties, a purely mathematical problem, and the 
conception of a physical model accounting for the distribution in space of 
the gas density, temperature and velocity.

The former, the mathematical problem, is what is addressed here. 
It aims at measuring, at each point of space coordinates $(x,y,z)$, 
the temperature $T$, density $d$ and velocity vector $(V_x,V_y,V_z)$ 
of the gas. This means, for each $(y,z)$ pixel, to calculate the values taken 
by $T$, $d$ and the three velocity components as a function of $x$. 
While the assumption of rotational invariance about the star axis brings 
a helpful constraint, the problem is obviously unsolvable without making 
additional hypotheses (six unknowns in each pixel, $x$, $T$, $d$ and 
the components of the velocity vector, for only two input data: 
the measured flux density $f(y,z,V_x)$ and the Doppler velocity $V_x$). 
From the beginning, we simplify the density and temperature side 
of the problem by introducing an effective emissivity $\rho$
(in earlier publications we called it effective density, which is improper), 
defined as
\begin{equation} \label{eq1}
\rho (x,y,z)=f(y,z,V_x)dV_x/dx
\end{equation}
implying 
\begin{equation} \label{eq2}
\int f(y,z,V_x)dV_x=\int \rho(x,y,z)dx=F(y,z)
\end{equation}

The effective emissivity is therefore a function of the gas density, 
the abundance of the emitting gas, the population of the emitting 
quantum state and the probabilities of emission and absorption of 
the detected radiation. It is equal to the volume emissivity in the 
optically thin limit. In general, we measure all distances as 
angular distances, including along the line of sight: an angular distance 
of 1$''$ means a distance of 1 AU times the distance of the star from the 
Earth measured in parsec, the effective emissivity being measured in 
Jy\,km\,s\,$^{-1}$\,arcsec$^{-3}$. Also from the beginning, we often restrict 
our considerations to centrally symmetric star morphologies and kinematics, 
implying $f(y,z,V_x)=f(-y,-z,-V_x)$.

The problem of reconstructing CSE properties in space from radio observations 
is not new and has been addressed for decades by radio astronomers. 
What is new, however, is the availability of observations of unprecedented 
spatial and spectral resolution in the millimetre and sub-millimetre ranges, 
with the recent operation of the ALMA interferometer. It opens a new era in 
millimetre and sub-millimetre radio astronomy and requires analysis methods 
that make the best of the excellent quality of the data that it provides, 
namely allowing for more quantitative conclusions to be reliably drawn from 
observations than was previously possible. Indeed, the sensitivity and 
spatial resolution that could be reached before ALMA were such that one 
had often to be satisfied with qualitative statements obtained from the 
visual inspection of channel maps, spectral maps and position-velocity (P-V) 
diagrams. In particular, the reliability and uniqueness of the proposed 
models could rarely be quantitatively discussed.

The general problem of reconstructing (one says ``deprojecting'') 
a 3-D image from its 2-D projection is a topic of broad interest and 
the subject of abundant literature \citep[see for example ][]{Ihrke2008}. 
It has been addressed with particular attention in the domain of 
optical astronomy, often with emphasis on providing tools that can be used, 
interactively if possible, by a broad community of people having little 
expertise in programming, not only for research applications but also for
educational purposes, planetarium simulations, etc... An
archetype of such 3-D modelling tool is the SHAPE software of
\cite{Steffen2011} which makes optimal use of state of the art
knowledge in computer visualisation and obtains excellent results
in terms of accuracy and computer time.

As the problem is obviously underdetermined (it boils down
to solving integral equation 2) additional hypotheses need to be
made. Axisymmetry is the most common 
\citep[][\citeauthor{Magnor2004} \citeyear{Magnor2004} \& \citeyear{Magnor2005}]{Leahy1991,Palmer1994}. 
In this case, deprojection is
relatively straightforward once the orientation of the symmetry axis in 
space is known. But, while the projection of the symmetry axis on the sky 
plane is directly revealed, its inclination with respect to the line of sight 
is more hidden and its evaluation requires algorithms such as chi square 
minimization and/or iterative approximations and may result in significant 
uncertainties. An interesting approach has been explored by \cite{Lintu2007} 
in the case of the circumstellar dust of Planetary Nebulae illuminated by 
the central star. They do not assume any symmetry but make use of the 
properties of the scattered light to obtain a plausible approximation 
of the dust distribution. 

Most of the literature, however, concentrates on the morphology deprojection without considering the kinematic deprojection. Morphology deprojection is from 2-D on the sky plane to 3-D in space. Kinematic deprojection implies, in addition, going from the 1-D Doppler velocity to the 3-D space velocity, or, equivalently, from the 3-D datacube (2-D on the sky plane + 1-D for the Doppler velocity)  to the 6-D phase space (3-D for space coordinates and 3-D for velocity coordinates). Kinematic deprojection is therefore even more underdetermined than morphology deprojection. A notable exception is from \cite{Sabbadin1984} and \cite{Sabbadin2000} 
who evaluate the velocity field under the assumption of radial expansion. The SHAPE software also allows for including kinematic considerations but the accent is on morphology more than on kinematics: it inherits from optical observations, where the measurement of velocities is difficult, requiring methods such as long-slit spectroscopy, while in radio astronomy the full velocity spectrum is directly available in each pixel. Yet, the flexibility of the code makes it possible to produce simulations including arbitrary velocity fields, such as was done recently by \cite{Decin2015} using ALMA data.

In comparison with the existing literature, the present paper displays both 
similarities and differences. The similarities are very general and include the type of problems one is facing, the considerations that they inspire and the analysis of how they should be approached.

The differences result in part from the restriction of our work to the case of 
evolved stars and protostars observed in millimetre and sub-millimetre 
astronomy. In this domain, the information carried by kinematics is usually 
richer than that carried by morphology. This is due both to the power of the
instrument in providing simply accurate velocity spectra and in the physics of 
evolved stars, with a morphology departing rather late from spherical while 
important winds can be already revealed in an early phase of their evolution. 
Much of the existing literature on optical deprojection addresses the case of 
Planetary Nebulae, with morphologies of often extreme complexity. This is
not the case for evolved stars; even in the case of Mira, on which we briefly 
comment, one can describe the sources in terms of simpler objects that can be 
studied separately. Moreover, while lumpiness is frequently observed in AGB 
CSEs, it is less of a problem than in the case of Planetary Nebulae and 
axisymmetry usually remains a good approximation (we address this issue in
Section~\ref{sec3}).

Much of the existing literature on optical deprojection devotes particular 
attention to producing user-friendly algorithms with emphasis on computer 
time and interactivity. We do not have such ambition. Our motivation is simply 
to address issues of data analysis in general, and deprojection in particular, 
in the context of radio astronomy of evolved stars, where much of the current 
literature is limited to displaying channel and spectral maps. We also mean to 
show that much information can be obtained by constructing simple variables 
and simple histograms that do not require complex analysis tools. More 
generally, the paper is written in the spirit of encouraging more quantitative
analyses than commonly found in analyses of radio observations of evolved 
stars using instruments that do not have as good sensitivity and resolution 
as ALMA is now offering.

The paper is organized as follows: in Section~\ref{sec2}, we explore the more important 
features of the problem of deprojection using simulated observations; 
in Section~\ref{sec3}, we apply the information obtained in the first part to a few 
actual stars; Section~\ref{sec4} summarizes the two preceding sections and makes some 
statements aimed at helping and guiding future analyses of 
millimetre/sub-millimetre observations of the CSEs of axisymmetric 
radio sources.
 
For an easier reading, we have collected in an appendix many developments, 
which, although important for an in-depth understanding, are not essential 
for grasping the main message.

\section{An exploration of the main features using simulated observations}
\label{sec2}
\subsection{Transformation relations between space and star coordinates}
\begin{figure}
\begin{center}
	\includegraphics[width=.9\columnwidth,trim={0cm 0cm 0cm 0cm},clip]{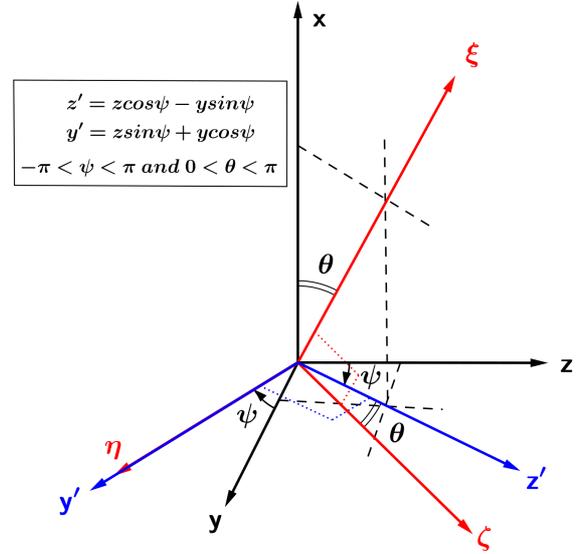}
    \caption{Coordinate transformations.}
    \label{fig1}
\end{center}
\end{figure}

We define (Figure~\ref{fig1}) two orthonormal coordinate systems, $(x,y,z)$ 
and $(\xi,\eta,\zeta)$; $(x,y,z)$ with $x$ along the line of sight 
(positive away from the observer), $y$ pointing east and $z$ pointing north; 
$(\xi,\eta,\zeta)$ with $\xi$ along the star axis obtained from $(x,y,z)$ by 
a first rotation of angle $\psi$ about $x$ and a second rotation of angle 
$\theta$ about $y'$, the transformed of $y$ in the first rotation. 
Transformation relations between the two systems are spelled out in 
Appendix~\ref{seca1}. We also use polar coordinates $(R,\varphi)$ in the 
plane of the sky:
\begin{equation} \label{eq3}
\begin{split} 
&y=R\cos\varphi \\
&z=R\sin\varphi 
\end{split}
\end{equation}
and we define 
\begin{equation} \label{eq4}
r=\sqrt{x^2+y^2+z^2}
\end{equation}

\subsection{Effective emissivity, wind velocity and their symmetries}
\label{sec2.2}
The flux densities, $f(y,z,V_x)$ or $f(R,\varphi,V_x)$, available to analysis 
are three-dimensional objects, which makes their visualization difficult. 
Traditionally, one uses channel-maps ($z$ vs $y$ distributions in each $V_x$ 
bin or group of neighbour bins) or spectral maps ($V_x$ distributions in each 
pixel or group of neighbour pixels). However, when the number of velocity 
bins and of pixels is large, it becomes increasingly difficult to obtain useful 
information by direct inspection of an increasingly large number of maps. 
It is therefore convenient to reduce the problem to two dimensions by 
introducing moments of order $k$ obtained by integrating the measured flux 
densities over one of the variables after having multiplied them by that 
variable raised to power $k$. In practice, only a few of these moments are 
commonly used, usually under other names. Integrating over $V_x$ maps the flux
\begin{equation} \label{eq5}
 F(R,\varphi)=\int_{-\infty}^{+\infty}f(R,\varphi,V_x)dV_x
\end{equation}
As $\rho(x,y,z)$ decreases usually rapidly with $r$, typically as $r^{-2}$, 
mapping the quantity $\int Rf(R,\varphi,V_{x})dV_x$ is better suited for revealing finer details of the flux map. 
The first order moment divided by the integrated flux
\begin{equation} \label{eq6}
<V_x>=\frac{\int_{-\infty}^{+\infty}V_x f(y,z,V_x)dV_x}{F(y,z)}
\end{equation}
gives the mean Doppler 
velocity in each pixel, which is a useful indicator of a possible asymmetry 
of the wind velocities. Similarly, $<|V_x|>$ and $<V_x^2>$ provide useful 
information.

Fixing a sky coordinate, $y$ or $z$, or integrating over it corresponds to 
the P-V diagrams commonly used in astronomy. As the dependence on $\varphi$ 
of the flux densities is often only weakly coupled to its dependence on $R$ 
(to the extent that the dependence on stellar latitude of the star properties 
is only weakly coupled to its dependence on $r$) polar coordinates are better 
suited to such analyses, displaying quantities such as 
$\int_{0}^{+\infty}f(R,\varphi,V_x)dR$ and $\int_{0}^{+\infty}Rf(R,\varphi,V_x)dR$.

It is convenient to express the components of the wind velocities 
in the star frame (Figure~\ref{fig2}). Their symmetry properties are detailed 
in Appendix~\ref{seca2}. Precisely, we define a radial component $V_{rad}$, 
an axial component $V_{ax}$ along the star axis and a rotation component about 
the star axis, $V_{rot}$. In terms of the stellar latitude $\alpha$ and stellar 
longitude $\omega$, the $(\xi,\eta,\zeta)$ components of the wind velocity are 
therefore $(\xi,\eta,\zeta)V_{rad}$; $(1,0,0)V_{ax}$; and 
$(0,-\sin\omega,\cos\omega)V_{rot}$ respectively and the Doppler velocity reads:

\begin{equation}\label{eq7}	
V_x=\frac{x}{r}V_{rad}+\cos\theta V_{ax}-\frac{R\cos(\varphi-\psi)\sin\theta}{r\cos\alpha}V_{rot}
\end{equation}
\begin{figure}
\begin{center}
  \includegraphics[width=0.9\columnwidth,trim={0cm 0cm 0cm 0cm},clip]{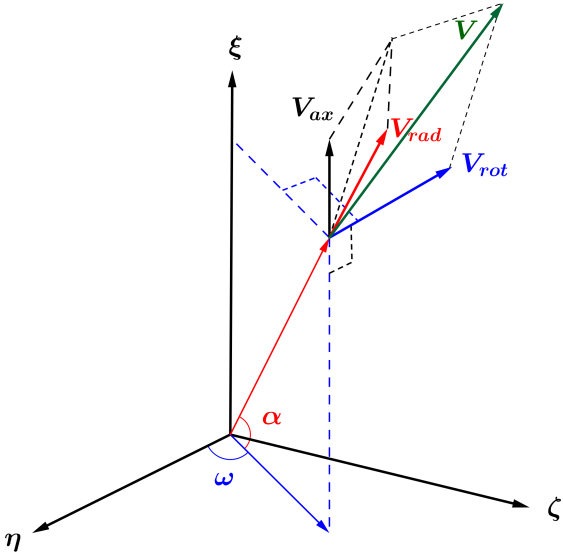}
  \caption{Wind velocity components.}
  \label{fig2}
\end{center}
\end{figure}			

\subsection{Velocity spectra and P-V diagrams}
\label{sec2.3}
For stars that have lived long enough on the AGB for the morphology of their 
gas envelope to be significantly altered, the sky map of the measured flux, 
$F(y,z)$, directly reveals the projection of the star axis on the sky plane. 
However, this is often not the case for AGB stars having mass loss rate at 
the scale of $10^{-7}$ M$_{\odot}$\,yr$^{-1}$ or smaller: they display spherical 
morphology and axial symmetry, if present, can only be revealed from the 
velocity distribution of the wind. 

However, mapping the mean value $<V_x>$ of the gas velocity on the sky plane 
is not sufficient to find the projection of the star axis. Expansion and 
rotation combine with different symmetry properties, resulting in an apparent 
tilt of the projection of the star axis on the sky plane, by an unknown 
amount. It is only in the cases of pure rotation and pure expansion that the 
axis of the star is an axis of symmetry of the sky map. However, in such 
cases, this axis of symmetry is the projection of the star axis in the case 
of pure expansion, but is perpendicular to it in the case of pure rotation. 
It is therefore necessary, in order to evaluate the value of $\psi$, 
the position angle of the projection of the star axis on the sky
plane, to exploit the information contained in the velocity spectra rather 
than integrating over velocities.

\begin{figure*}
\begin{center}
  \includegraphics[width=.9\columnwidth,trim={0cm 0cm 0cm 0cm},clip]{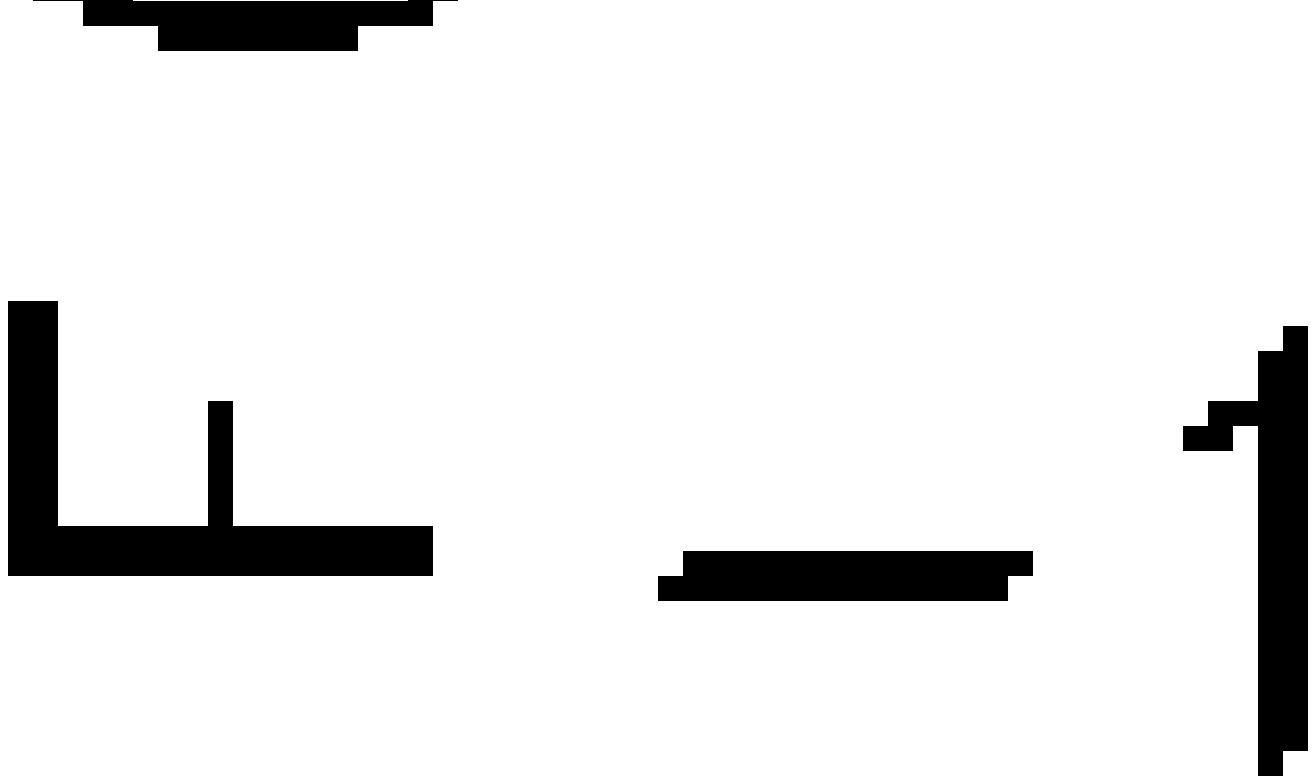}
  \includegraphics[width=.9\columnwidth,trim={0cm 0cm 0cm 0cm},clip]{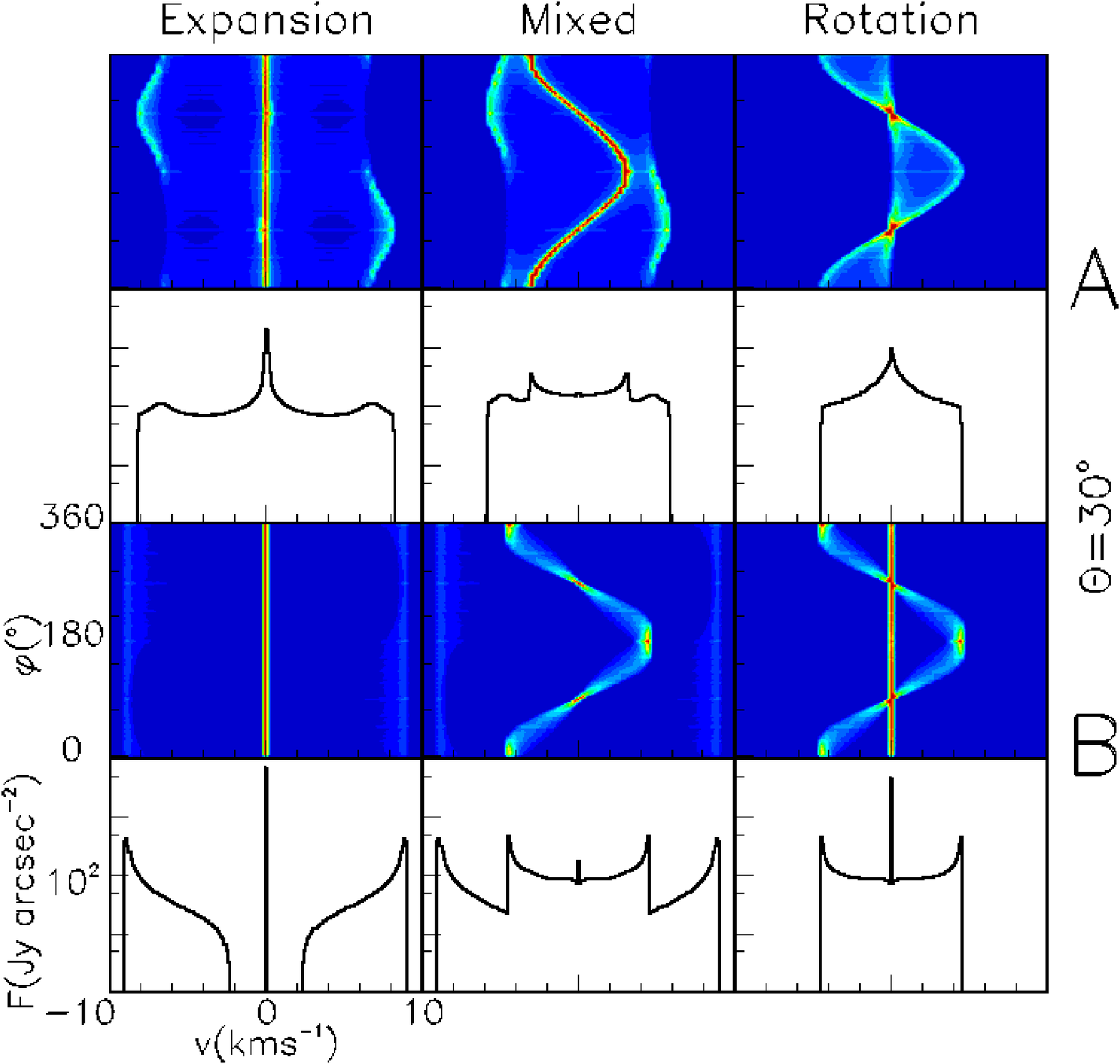}
  \includegraphics[width=.9\columnwidth,trim={0cm 0cm 0cm 0cm},clip]{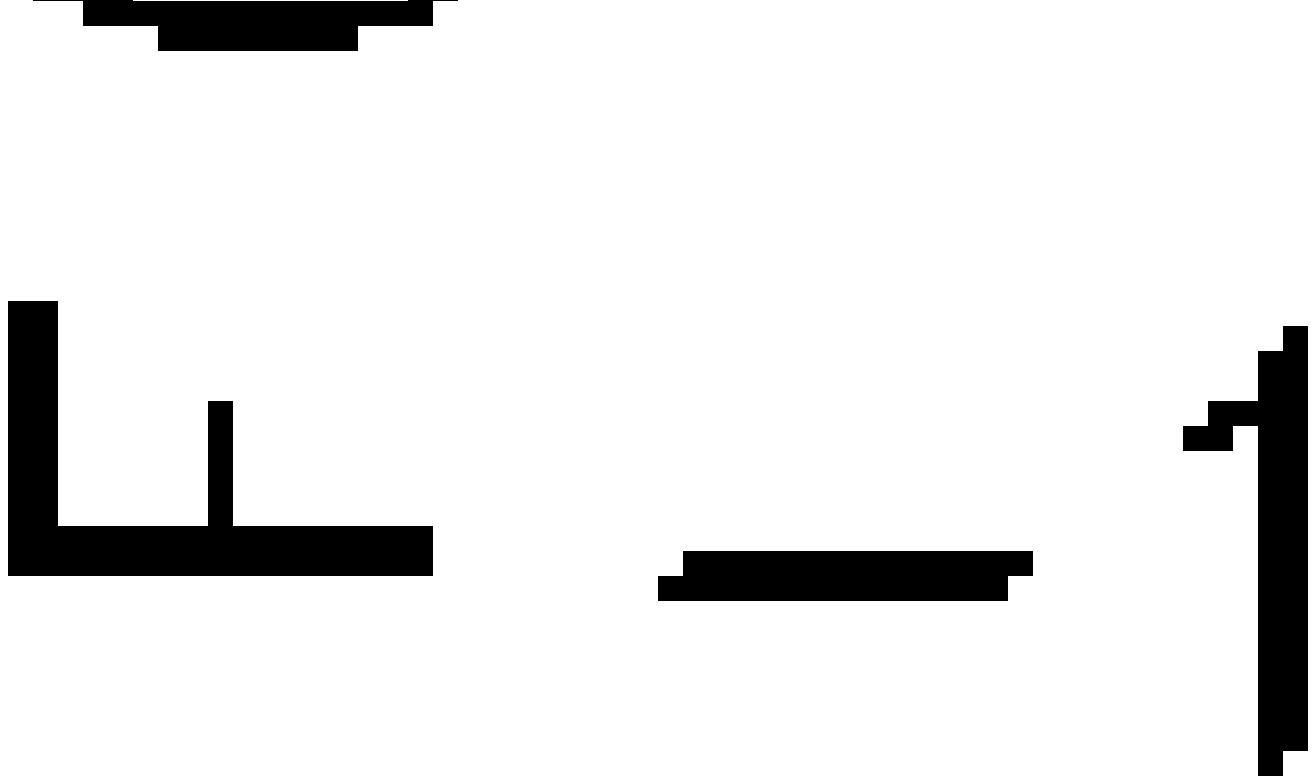}
  \includegraphics[width=.9\columnwidth,trim={0cm 0cm 0cm 0cm},clip]{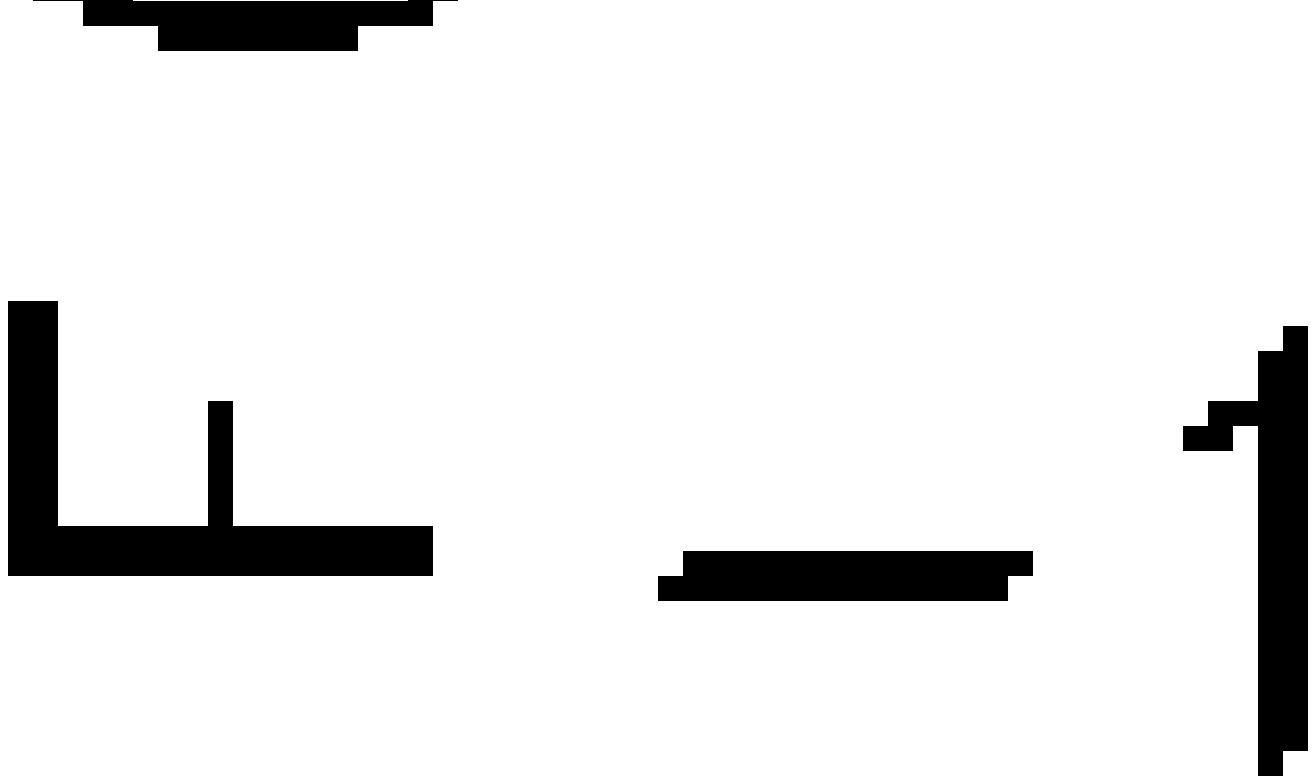}
  \includegraphics[width=.9\columnwidth,trim={0cm 0cm 0cm 0cm},clip]{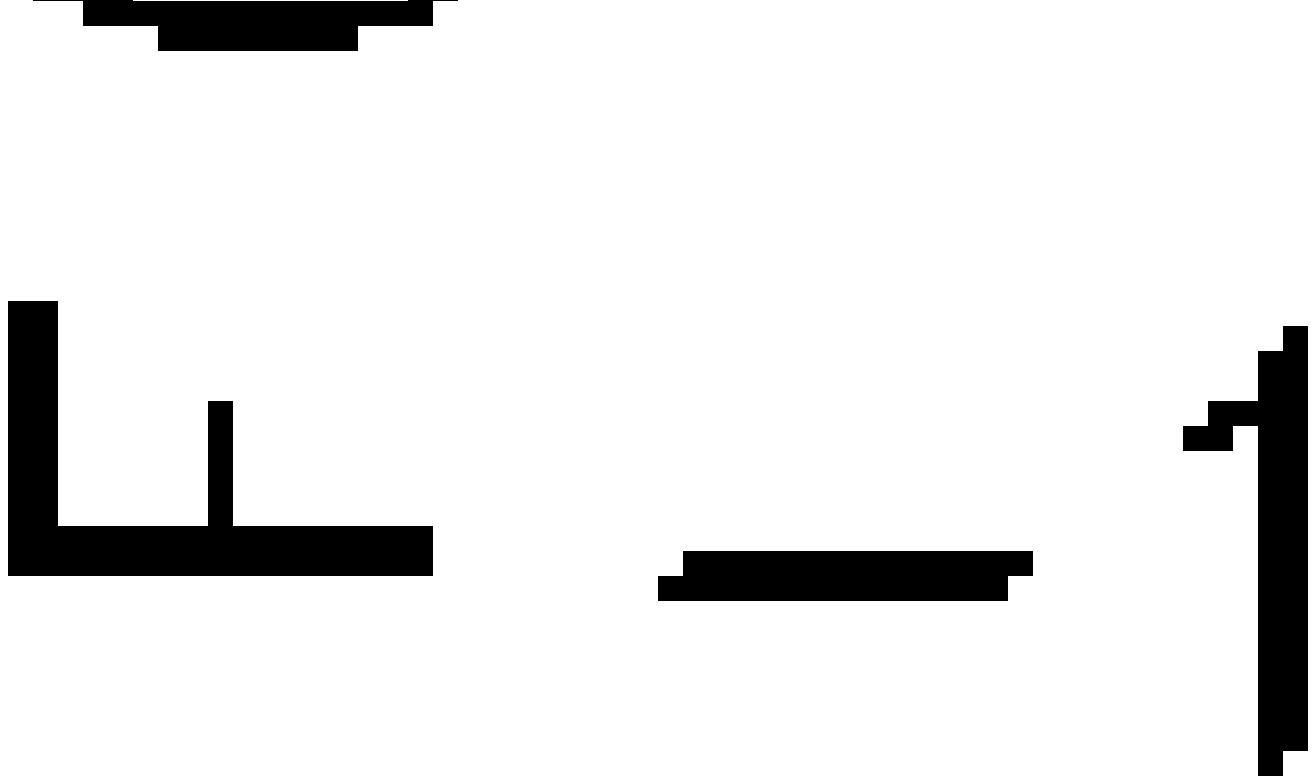}
  \includegraphics[width=.9\columnwidth,trim={0cm 0cm 0cm 0cm},clip]{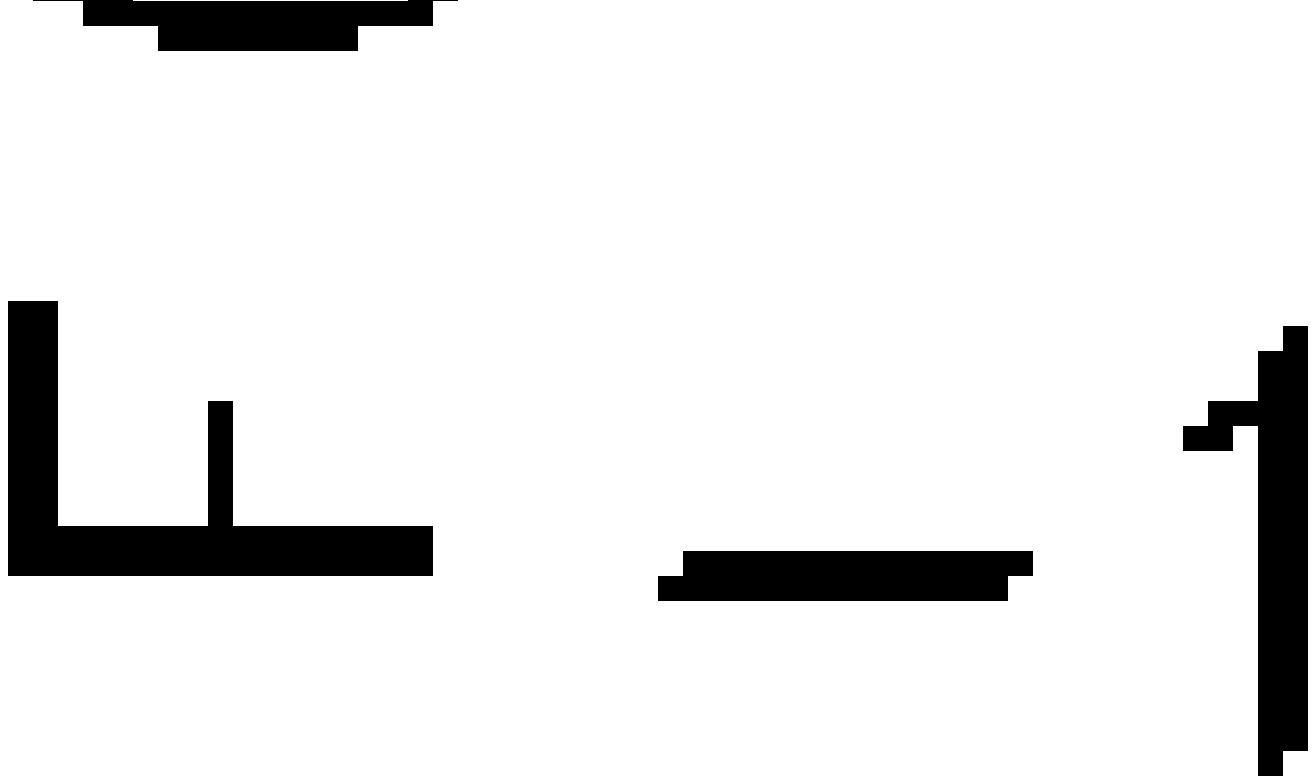}
  \caption{P-V diagrams. The spikes observed at $V_x=0$ in the pure expansion 
and pure rotation cases correspond to the contributions of respectively 
$V_{rot}=0$ and $V_{rad}=0$: in such cases, the outside and respectively 
inside volumes of the cone do not emit, this absence of emission causes 
the spikes. In the mixed case, both the inside and outside volumes emit.}
  \label{fig3}
\end{center}
\end{figure*}

We illustrate this point with a series of simulations, the results of which 
are displayed in Figure~\ref{fig3}. In this figure, the angle $\varphi$ is 
the position angle of the pixel on the sky map. The positive $y$ axis has 
$\varphi=0$, the positive $z$ axis has $\varphi=90^\circ$, etc. The Doppler 
velocity $V_x$ is measured positive when red shifted. Figure~\ref{fig3} 
includes six different sets, each associated with a particular orientation 
$\theta$ of the star axis. The position angle of the projection of the star 
axis on the sky plane is always $\psi=0$, meaning pointing north. Changing 
its value results simply in a rotation of the sky map about the origin, 
meaning a simple translation of the $\varphi$ vs $V_x$ P-V diagrams along 
the $\varphi$ axis. The inclination of the star axis with respect to the line 
of sight ($x$ axis) is measured by the angle $\theta$ indicated on the figure, 
between $15^\circ$ and $90^\circ$ in steps of $15^\circ$. For $\theta=0$ the star 
is seen pole on; for $\theta=90^\circ$ the observer is in the equatorial plane 
of the star.

\begin{figure}
\begin{center}
  \includegraphics[width=.98\columnwidth,trim={0cm 0cm 0cm 0cm},clip]{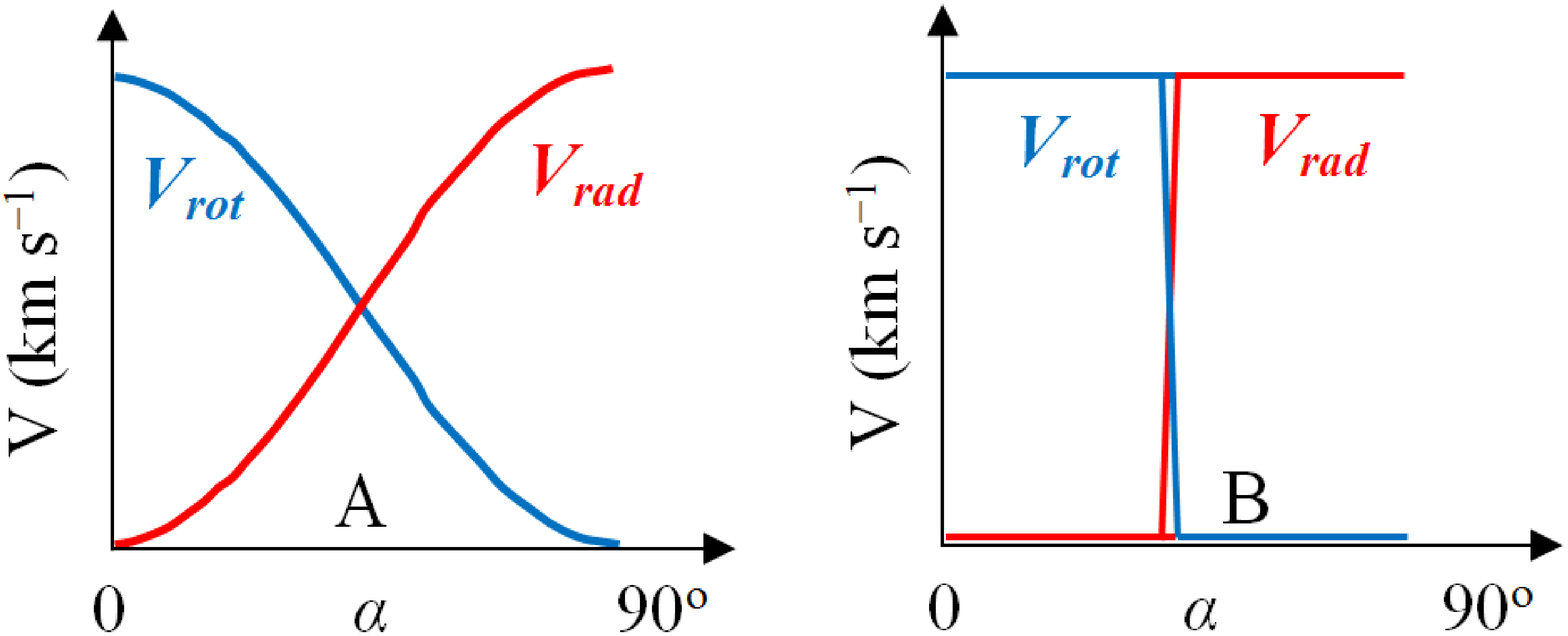}
  \caption{Dependence of gas velocity (rotation in blue, expansion in red) 
on star latitude in cases $A$ and $B$ used in the simulation.}
  \label{fig4}
\end{center}
\end{figure}

Simulations are made for three cases, each corresponding to a column: 
pure expansion on the left, a combination of expansion and rotation in 
the middle and pure rotation on the right. The star properties are supposed 
to be invariant under rotation about the star axis. The effective emissivity 
is taken isotropic with dependence in $r^{-2}$ on the distance $r$ from the 
star. Two different forms, $A$ and $B$, are used to define the dependence of 
the gas velocity on the star latitude, each corresponding to a row in the 
figure. The velocities are otherwise supposed to be independent of the 
star longitude (imposed by rotational invariance) and of the distance $r$ to 
the star (the absence of velocity gradients is a simplification expected 
not to have important consequences on the qualitative comments made here). 
Moreover, they are supposed to be symmetric with respect to the equatorial 
plane, implying central symmetry when combined with rotational invariance, 
meaning that changing $(x,y,z)$ in $(-x,-y,-z)$ leaves the effective emissivity
invariant and changes the sign of the gas velocity. This corresponds to adding 
$180^\circ$ to $\varphi$ and its effect is immediately visible in Figure~\ref{fig3}. 
The expansion velocity $V_{rad}$ is taken purely radial and the rotation 
velocity $V_{rot}$ is taken perpendicular to the star axis. In both cases $A$ 
and $B$, rotation is enhanced near the equator and expansion is enhanced 
near the poles. For each value of $\theta$ the upper row is for case $A$, 
\begin{equation} \label{eq8}
\begin{split}
&V_{rad}=V_0\sin^2\alpha\\
&V_{rot}=V_0\cos^2\alpha
\end{split}
\end{equation}
In the lower row, case $B$, $V_{rad}$ is set to 0 and $V_{rot}$ to $V_0$ 
for $\alpha<45^\circ$ while $V_{rot}$ is set to zero and $V_{rad}$ to $V_0$ 
for $\alpha>45^\circ$ (Figure~\ref{fig4}). Each simulation is illustrated by 
two panels. The upper panel is the P-V diagram, $\varphi$ vs $V_x$, 
with the abscissa running from $-10$ km\,s\,$^{-1}$ to 10 km\,s\,$^{-1}$ 
and the ordinate from 0 to $360^\circ$; the lower panel is the projection of 
the diagram on the velocity axis.

In order to construct an axially symmetric model of the CSE of the star, 
one needs to find the position of the star axis in space, namely the values 
of $\theta$ and $\psi$, before evaluating the values of the effective 
emissivity and gas velocity as a function of the star latitude $\alpha$ and 
the distance $r$ to the star. Once $\psi$ is known, one can rotate the data 
on the sky map in order to have $\psi=0$, namely the projection of the 
star axis pointing north. Then the star properties display simple symmetries 
with respect to transformation $S_1$, changing $(x,y,z)$ in $(-x,y,-z)$ 
and transformation $S_2$, changing $(x,y,z)$ in $(x,-y,z)$; the Doppler velocity 
is even under $S_1$ and odd under $S_2$ for pure rotation and even 
under $S_2$ and odd under $S_1$ for pure expansion. These symmetries are 
clearly seen in Figure~\ref{fig3} in the cases of pure expansion 
(left columns) or pure rotation (right columns). However, if one does not know 
the value of $\psi$ (and one has no reason to know it a priori), the symmetry 
obeyed by the Doppler velocity as a function of $\varphi$ does not tell the 
value of $\psi$: the values associated with pure expansion and pure rotation 
differ by $90^\circ$. It is then essential to make a distinction between the 
effects of expansion and rotation on the Doppler velocity and the P-V diagrams 
before having a chance to measure $\psi$. Assuming that we know $\theta$ and 
$\psi$, and given the dependence on $r$ and $\alpha$ of the effective 
emissivity and of the gas velocity, one can calculate, for each panel of 
Figure~\ref{fig3}, the flux density $f(y,z,V_x)=\rho(x,y,z)dx/dV_x$ for each 
pixel $(y,z)$, or equivalently $(R,\varphi)$. Further details are given in 
Appendix~\ref{seca3}.

\subsubsection{The case $\theta =0^\circ$}
\label{sec2.3.1}
In this case, the flux density is independent of $\varphi$: the whole 
information is contained in the projections on the $V_x$ axis of the 
P-V diagrams. Moreover, the rotation velocity being perpendicular to 
the line of sight does not contribute to the Doppler velocity: for pure 
rotation, $V_x$ identically cancels. A consequence is that the mixed case 
is identical to the pure expansion case.

When attempting to construct a model, there exists no constraint imposed on the forms chosen for the radial and axial components of the velocity vector (see Sub-section~\ref{sec2.2}), respectively $V_{rad}(r,\alpha)$ and $V_{ax}(r,\alpha)$ 
(not to mention $V_{rot}$, which can take any value!). In particular, the 
hypothesis of rotational invariance is of no use in this case: all it says 
is that the measured flux density $f(R,\varphi,V_x)$ must be independent 
of $\varphi$. Either it is, in which case the data are consistent with the 
hypothesis, or it is not, in which case they are not. 

In summary, when $\theta =0^\circ$, there is nothing to be learned about rotation 
and about the dependence of the expansion velocity on the star latitude and 
the distance from the star. But when making assumptions, on the basis of 
physics arguments, about the form they take, one can reconstruct the values 
of the effective emissivity and expansion velocity at any point in space. 
If one assumes that $\rho$, $V_{rad}$ and $V_{ax}$ depend only on $r$ the 
integral equation is easily solved to give $\rho(r)$ for any $r$ value. 
Then, assuming particular forms for the $r$ dependence of $V_{rad}(r)$ and 
$V_{ax}(r)$, one can also calculate them for any $r$ value. In such a case, 
one obtains a spherical model of the CSE.

\subsubsection{The case $\theta =90^\circ$}
\label{sec2.3.2} 
In this case, as $R < r$, the angle $\varphi$ measures the maximal value 
taken by the stellar latitude $\alpha$ in a given pixel (for $x=0$): 
\begin{equation} \label{eq9}
r\sin\alpha =z=R\sin\varphi
\end{equation}

In the case of pure expansion, as $V_x=(x/r)V_{rad}$, the Doppler velocity 
tends to $V_{rad}$ when $x$ is large. Near the $y$ axis, $\varphi =0^\circ$, 
one is probing low latitudes where $V_{rad}$ cancels.  On the contrary, 
near the $z$ axis, $\varphi =90^\circ$, one is probing large latitudes where 
$V_{rad}$ takes its maximal values. The maximal values of $V_x$ are therefore 
reached around $\varphi =90^\circ$ and the minimal values around $\varphi =0$. 
The precise value taken by $V_x$ at maximum depends on the dependence of 
$V_{rad}$ over $\alpha$, the larger the more concentrated near the poles 
are high $V_{rad}$ values. 

In the case of pure rotation, as 
\begin{equation} \label{eq10}
V_x=-\frac{R\cos\varphi}{r\cos\alpha}V_{rot}
\end{equation}
the $\cos\varphi$ term modulates the maximal value of $V_x$ as a function 
of $\varphi$: it oscillates between $-R(r\cos\alpha)^{-1}V_{rot}$ and 
$+R(r\cos\alpha)^{-1}V_{rot}$ when $\varphi$ goes from 0 to $180^\circ$, 
cancelling at $\varphi =90^\circ$. When $\theta =90^\circ$, contrary to the 
previous case ($\theta =0^\circ$), rotational invariance cannot be verified 
but, if it is assumed to be obeyed, it strongly constrains the data. 
It implies that for a given value of $z$, the effective emissivity is a 
function of only $r$ and $\sin\alpha =z/r$, namely of $r$ alone. 
Therefore, for each value of $z$, one can write 
\begin{equation} \label{eq11}
F(y,z)=\int \rho (r,z)dx=\int \rho (r,z)r(r^2-R^2)^{-1/2}dr
\end{equation} 
an integral equation easily solvable by iteration. This property has been 
known for a long time and is being applied extensively in deprojection 
algorithms used in optical astronomy \citep[][and references therein]{Wenger2013}. 
Rotational invariance constrains the dependence of $F(y,z)$ 
on $y$ for each value of $z$. Similarly, for each value of $z$, $V_{rad}$ 
and $V_{rot}$ depend only on $r$, so does therefore $dx/dV_x$. 
Assuming particular forms for the $r$-dependence of $V_{rad}$ and $V_{rot}$, 
one can then evaluate their values at each point in space. More precisely, 
for each value of $z$, as long as the forms chosen for $V_{rad}$ and $V_{rot}$ 
depend on a small enough number of parameters compared with the number 
of pixels, one can calculate the values taken by these parameters and 
obtain $V_{rad}$ and $V_{rot}$ as functions of $r$ and $z=r\sin\alpha$, 
namely as functions of $r$ and $\alpha$. 

In summary, when $\theta=90^\circ$, one cannot verify the validity of the 
hypothesis of rotational invariance (even if one can infer its absence when 
$F(y,z)$ and $F(-y,z)$ are not equal). But, assuming it, one can evaluate 
the effective emissivity at any point in space. One can also evaluate the 
values taken by $V_{rad}$ and $V_{rot}$ at any point in space once arbitrary 
forms, chosen according to physics arguments, have been adopted 
for their $r$-dependence, but $V_{ax}$ can take any arbitrary value.

\subsubsection{Pure rotation and pure expansion}
\label{sec2.3.3}
In a given pixel, both pure rotation and pure expansion typically generate 
double-horned spectral distributions. Indeed, if $V_{max}$ is the maximal 
value taken by $|V_x|$ in this pixel, $dN/dV_x=(dN/dx)(dx/dV_x)$ 
is infinite at $V_x=V_{max}$ where \mbox{$dV_x/dx=0$}. A simple illustration is 
obtained using \mbox{$V_{rot}=V_0\cos\alpha$} and \mbox{$V_{rad}=V_0$}. The corresponding 
spectral distributions are respectively proportional to 
$(V_{max}^2-V_x^2)^{-1/2}$ and $(V_{max}^2-V_x^2)^{-3/2}$.  The limits of the 
spectrum are $\pm V_{max}$, with $V_{max}=V_0\cos\varphi \sin\theta$ for 
pure rotation and $V_{max}=V_0$ for pure expansion. In both cases, the spectrum 
grows from its value at $V_x=0$ to infinity at the limits. Details are given 
in Appendix~\ref{seca4}.

\subsubsection{Mixing rotation and expansion}
\label{sec2.3.4}
\begin{figure*}
\begin{center}
  \includegraphics[width=.4\linewidth,trim={0cm 0cm 0cm 0cm},clip]{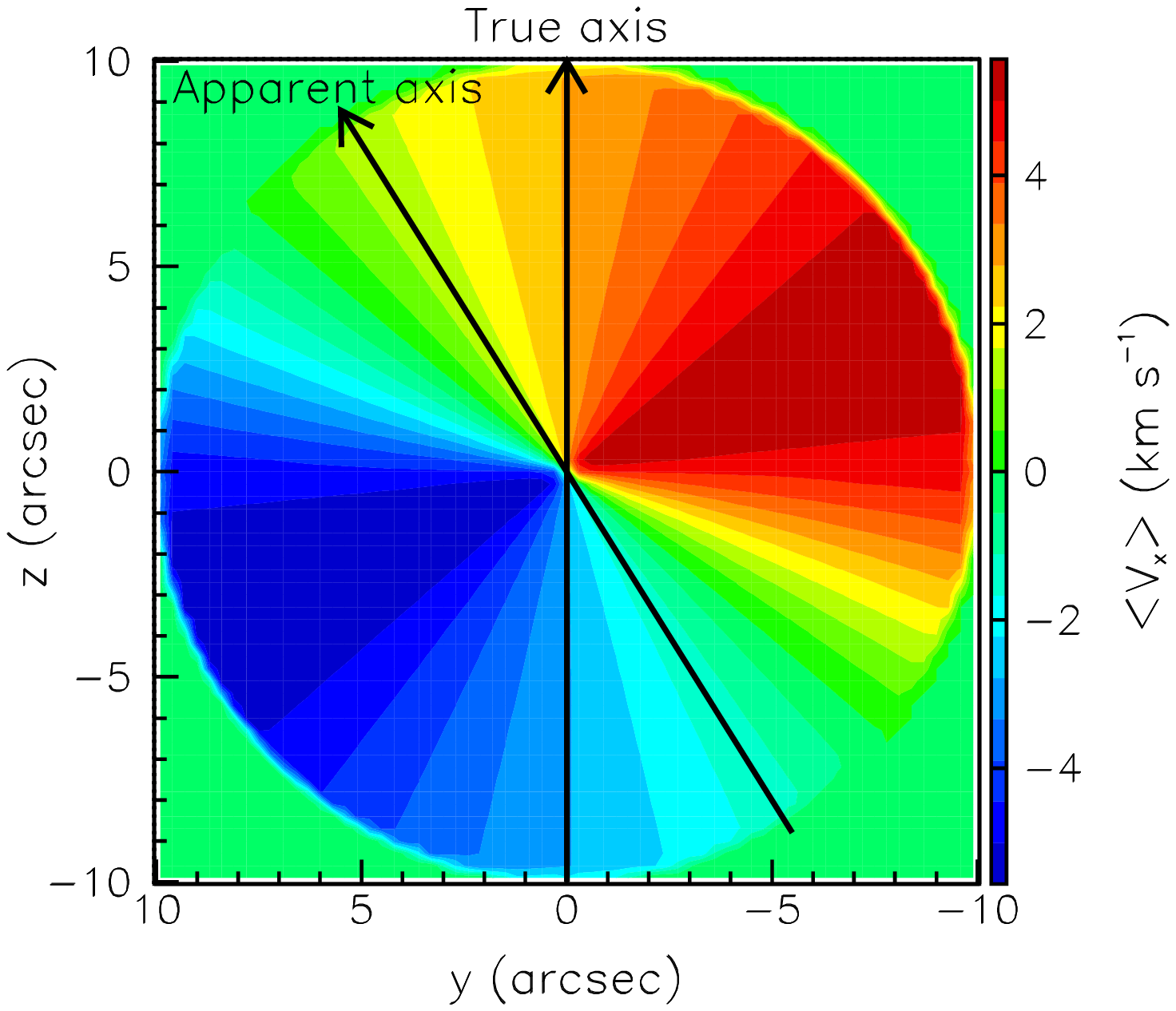}
\hspace{1cm}
  \includegraphics[width=.4\linewidth,trim={0cm 0cm 0cm 0cm},clip]{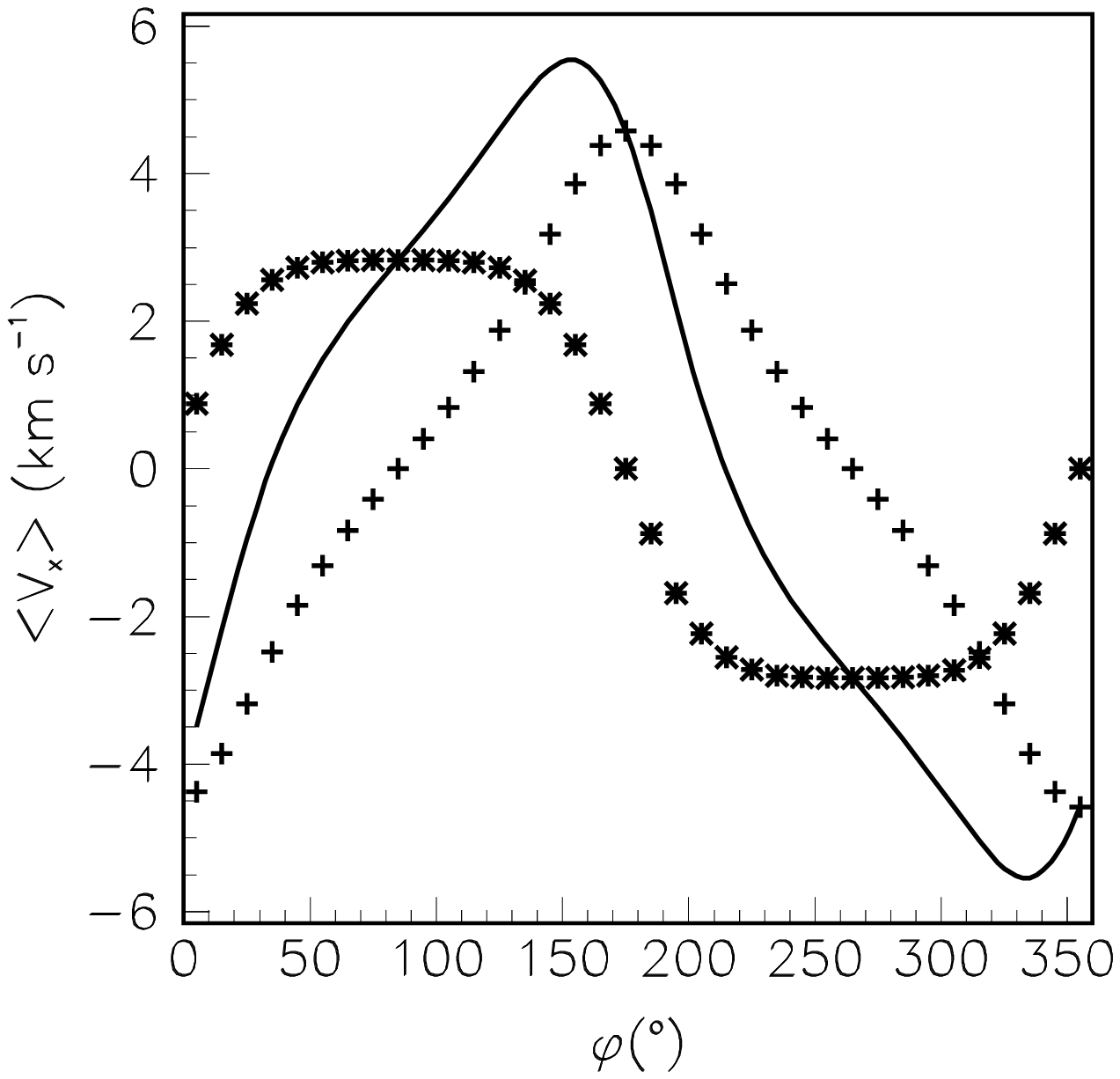}
  \caption{Mean value of the Doppler velocity, $<V_x>$ in case $A$ 
for $\theta =45^\circ$, $\psi=0$ and mixed expansion and rotation 
(middle column of the set of P-V diagrams in Figure~\ref{fig3}).
Left: its sky map; Right: its dependence on $\varphi$ (full line) and that 
of the rotation (+) and expansion (*) components.}
  \label{fig5}
\end{center}
\end{figure*}

The simplest case is case $B$: $x$ probes either inside the cone where it 
sees only expansion or outside where it sees only rotation. Hence, for case B,
the middle columns of Figure~\ref{fig3} are simply the superposition of 
the left and right columns. This remains qualitatively true when the regions 
probed by $x$ are dominated by a single regime, rotation or expansion, 
respectively near the equator or near the poles. However, in general, for a 
given pixel, $V_x$ receives significant contributions from both rotation and 
expansion. As a result, when mapping $<V_x>$, one sees a symmetry that 
reflects both the value of $\psi$ and the relative importance of rotation 
and expansion: Figure~\ref{fig5} illustrates the difficulty of finding the 
axis of the star when nothing is known a priori about their relative 
contributions. The next sections (\ref{sec2.3.5}, \ref{sec2.4} and 
\ref{sec2.5}) address issues related to this problem, the nature of 
which deserves some preliminary comments.

A first remark is that we are concerned here with the kinematics
of the CSE, not its morphology, motivated by the fact that many
low mass-loss rate AGB stars display an essentially spherical
morphology. If such is not the case, namely if axisymmetry is
already revealed by the morphology, so much the better: the
projection of the star axis on the sky plane will be immediately
identified. But the general problems of measuring the inclination $\theta$
of the star axis with respect to the line of sight and of telling
expansion from rotation will remain. For this reason, we assume
in what follows that the effective emissivity is spherically
symmetric.

A second remark is related with understanding what is causing the
apparent tilt of the star axis in Figure ~\ref{fig5}. For such a tilt 
to take place, as already mentioned, two conditions must be obeyed: both
rotation and expansion must be present and $\theta$ must differ from
both $0^\circ$ and $90^\circ$. But a third condition is also mandatory:
expansion must deviate from isotropic, namely be bipolar
(isotropic expansion contributes equally to positive and negative
values of $V_x$, causing the mean to cancel). All three conditions
make sense from a physics point of view and detecting the
possible simultaneous presence of rotation and expansion is a
difficult but important task in AGB astrophysics. Taking the $z$ axis
as star axis, the velocity fields of expansion and rotation have
different symmetries: rotation is symmetric about the $y$ axis and
antisymmetric about the $z$ axis, expansion is symmetric about the
$z$ axis and antisymmetric about the $y$ axis. Namely, in the case of
pure rotation, the projection of the star axis on the sky plane is
perpendicular to the symmetry axis of the map, while in the case
of pure expansion it is the symmetry axis of the map. In general,
in a given pixel, $<V_x>$ receives contributions from both expansion
and rotation, respectively $V_1=<xV_{rad}/r>$ and 
$V_2=R\cos\varphi\sin\theta<V_{rot}/r\cos\alpha>$. 
Assuming that both $V_{rad}$ and $V_{rot}$ are independent of $r$ and 
even functions of $\sin\alpha$, the problem is scale
invariant and $V_1$, $V_2$ and their sum do not depend on $R$, but only
on $\varphi$. This is immediately apparent on Figure~\ref{fig5}. 
While $V_1$ is symmetric about $\varphi=90^\circ$, $V_2$ is symmetric about 
$0^\circ$. But their sum, in general, has no reason to display any symmetry. 
Indeed the apparent symmetry of the map displayed in Figure~\ref{fig5} 
is only approximate and the symmetry is in fact destroyed rather than the
symmetry axis being simply tilted.

A third remark is that while pure rotation splits the sky map in
two regions, one blue-shifted on one side of the projection of the
star axis on the sky plane, the other red-shifted on the other side,
pure expansion can generate both blue-shifted and red-shifted
velocities in a same pixel if the polar outflows are broad enough.
This is clearly seen in Figure~\ref{fig3}, particularly for large values 
of $\theta$.

\subsubsection{Evaluating the value of the position angle $\psi$ of the projection of the symmetry axis on the sky plane}
\label{sec2.3.5}
When $\psi=0$, in the vicinity of the $y$ axis, changing $z$ in $-z$ 
leaves the velocity spectrum unchanged in the case of pure rotation but 
reflects it about the origin in the case of pure expansion. Similarly, 
in the vicinity of the $z$ axis, changing $y$ in $-y$ leaves the velocity 
spectrum unchanged in the case of pure expansion but reflects it about 
the origin in the case of pure rotation. Appendix~\ref{seca5} introduces 
a function $\chi^2_{\psi}$ of the position angle constructed on this remark, 
which is meant to reach its minimum at $\psi$. However, in case A, 
when $\theta$ is sufficiently different from 0 and 90$^\circ$, and when 
both expansion and rotation contribute significantly to $V_x$, $\chi^2_{\psi}$ 
is found to be less dependent on $\psi$ and the identification of the minimum 
is difficult. It is interesting to remark that a clear minimum is always 
found in case $B$. In this case, one is probing either the interior of 
the cone and sees only expansion, or the exterior of the cone, and sees 
only rotation: rotation and expansion never compete at a same point in space. 

In practice, when there are reasons to suspect the co-existence of
significant rotation and expansion velocity components, the best
will be to use a model making optimal use of the known physical
properties of the system and including $\psi$ among the parameters to
be adjusted by chi square minimization. An example is given in
Section 3.1.

\subsection{Telling rotation from expansion}
\label{sec2.4}
Having found the value of $\psi$ is not sufficient to tell expansion 
from rotation. Even in a pure case, rotation around the y axis and expansion 
along the $z$ axis will both produce P-V diagrams odd under $S_1$ and 
even under $S_2$. A discriminator between rotation and expansion would 
therefore be a useful tool. After rotation of the data on the sky plane 
making $\psi$ to cancel, in the case of pure radial expansion 
$xV_x=(x^2/r)V_{rad}$ is positive and in the case of pure rotation 
$yV_x=-y^2\sin\theta(r\cos\alpha)^{-1}V_{rot}$ has the sign of $-V_{rot}$. 
Appendix~\ref{seca6} makes use of this property to construct a 
function of $\varphi$, $A(\varphi)$, as a possible discriminator between 
rotation and expansion. The values of $A(\varphi)$ at $\varphi = 0^\circ$ 
and $\varphi = 90^\circ$ or its integral over $\varphi$ are generally good 
indicators of the relative importance of rotation over expansion. 
However, when dealing with a small perturbation to a global isotropic 
expansion, it is difficult to assert its precise nature, rotation or 
bipolar expansion. Here again, as in the preceding sub-section, 
the use of a model adjusted to best fit the
data will often be the most efficient method. It will include what
is known of the physical properties of the system, in particular the
likely form taken by the radial dependence of the rotation and 
expansion velocities. The radial dependences of the rotation
velocity (typically Keplerian or decreasing faster at large
distances) and of the expansion velocity are indeed important
quantities to evaluate. We address briefly the issue in Section 3, in
the context of RS Cnc and L1527 but a general treatment is
beyond the scope of the present article. Our choice in the present
section (constant rotation and expansion velocities) is motivated
by simplicity and by the fact that it has little impact on the general
qualitative ideas developed in the article because of the rapid
radial decrease of the effective emissivity. In practice,
preconceived ideas about the radial dependences of the rotation
and expansion velocities, based on physics arguments, will help
with the solution of the problem.

\subsection{Inclination $\theta$ of the star axis with respect to the line of sight}
\label{sec2.5}
Having evaluated the value of $\psi$ and assuming for simplicity that 
$\psi=0$, one needs to evaluate the value of the inclination $\theta$ 
of the star axis with respect to the line of sight. For the star axis 
to be well defined, there must be a large enough asymmetry of the 
effective emissivity and/or the wind velocity: spherical distributions 
prevent the definition of such an axis. As a result, what can be 
generally measured is the product of a measure of the inclination, 
typically $\sin\theta$, by a measure of the elongation of the 
effective emissivity and/or velocities along the star axis. 
This result, which is well known for morphology \citep[][\citeauthor{Magnor2004} \citeyear{Magnor2004} \& \citeyear{Magnor2005}]{Steffen2011,Leahy1991,Palmer1994} is also valid 
for kinematics. As detailed in Appendix~\ref{seca7}, this implies 
strong correlations relating the inclination angle to the elongation 
of the outflow on the star axis, namely the ratio of polar to 
equatorial wind velocities. 

\subsection{Radial dependence of the CSE properties}
\label{sec2.6}
One expects the $R$-dependence of the integrated flux to carry 
much information about the $r$-dependence of the effective emissivity. 
Indeed, from pure dimensional considerations, one would expect a $r^{-n}$ 
dependence of the effective emissivity to produce a $R^{-(n-1)}$ dependence 
of its integral on the sky map. As detailed in Appendix~\ref{seca8}, this 
is generally the case. Similarly, the sky map of the mean of the absolute 
value of the Doppler velocity is found to reveal directly the presence of 
velocity gradients. 
 
These results illustrate further the advantage of displaying the measured 
flux densities in terms of polar rather than Cartesian coordinates on the 
sky plane. Their dependence on $R$ is related simply to the dependence on 
$r$ of the effective emissivity and wind velocity, while their dependence 
on $\varphi$ is related to the dependence of the effective emissivity and 
wind velocity on the orientation of the star axis and on star latitude. 
Without having recourse to a model, much information about the morphology 
and kinematics of the CSE can be obtained from simple quantities 
constructed directly from the expression of the flux densities 
in polar coordinates.

\section{Application to real observations}
\label{sec3}
\subsection{A typical AGB star: RS Cnc}
\label{sec3.1}
\begin{figure*}
\begin{center}
  \includegraphics[width=.245\linewidth,trim={0cm 0.cm 0.1cm 1.2cm},clip]{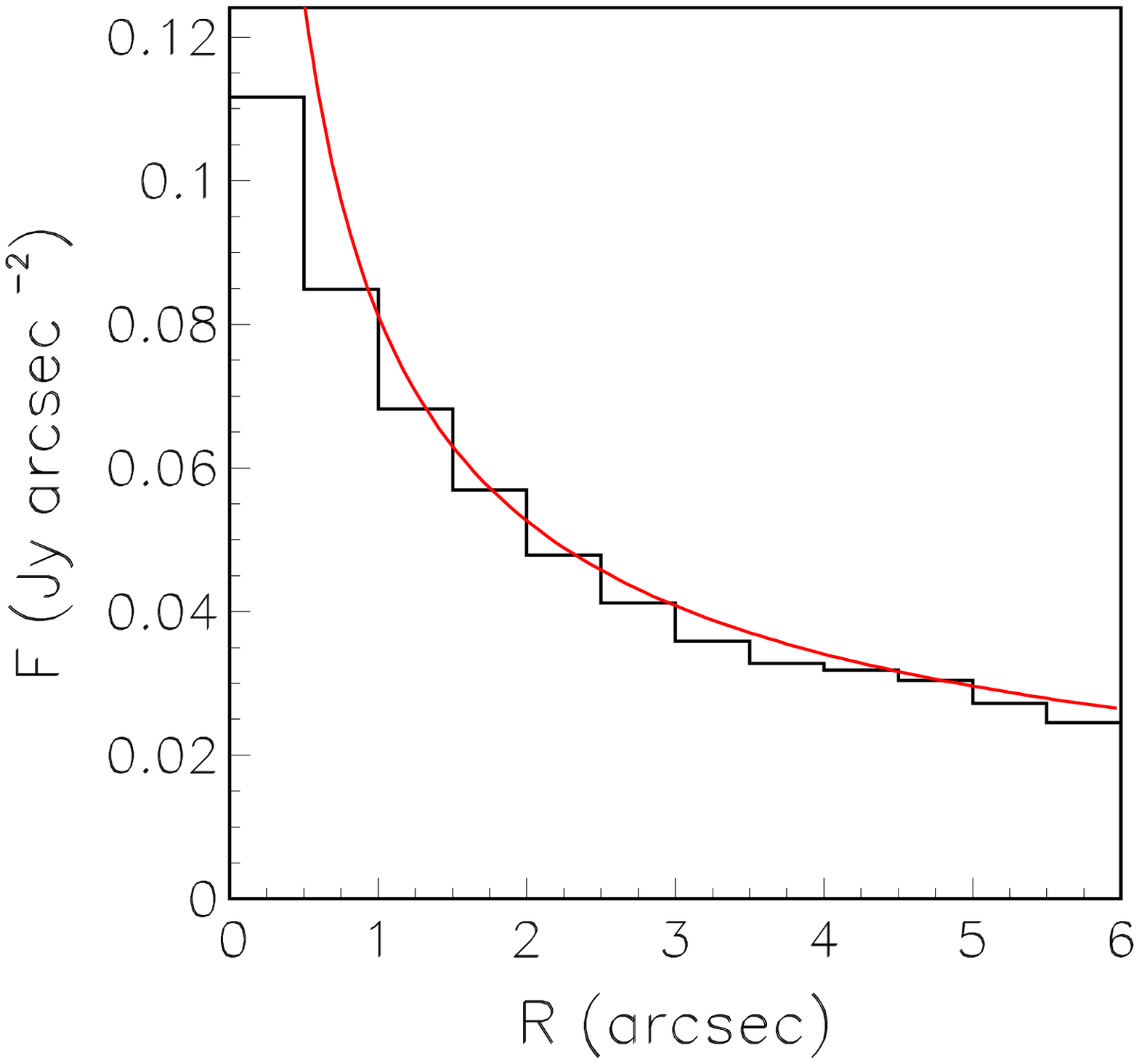}
  \includegraphics[width=.245\linewidth,trim={0cm 0.cm 0.1cm 1.2cm},clip]{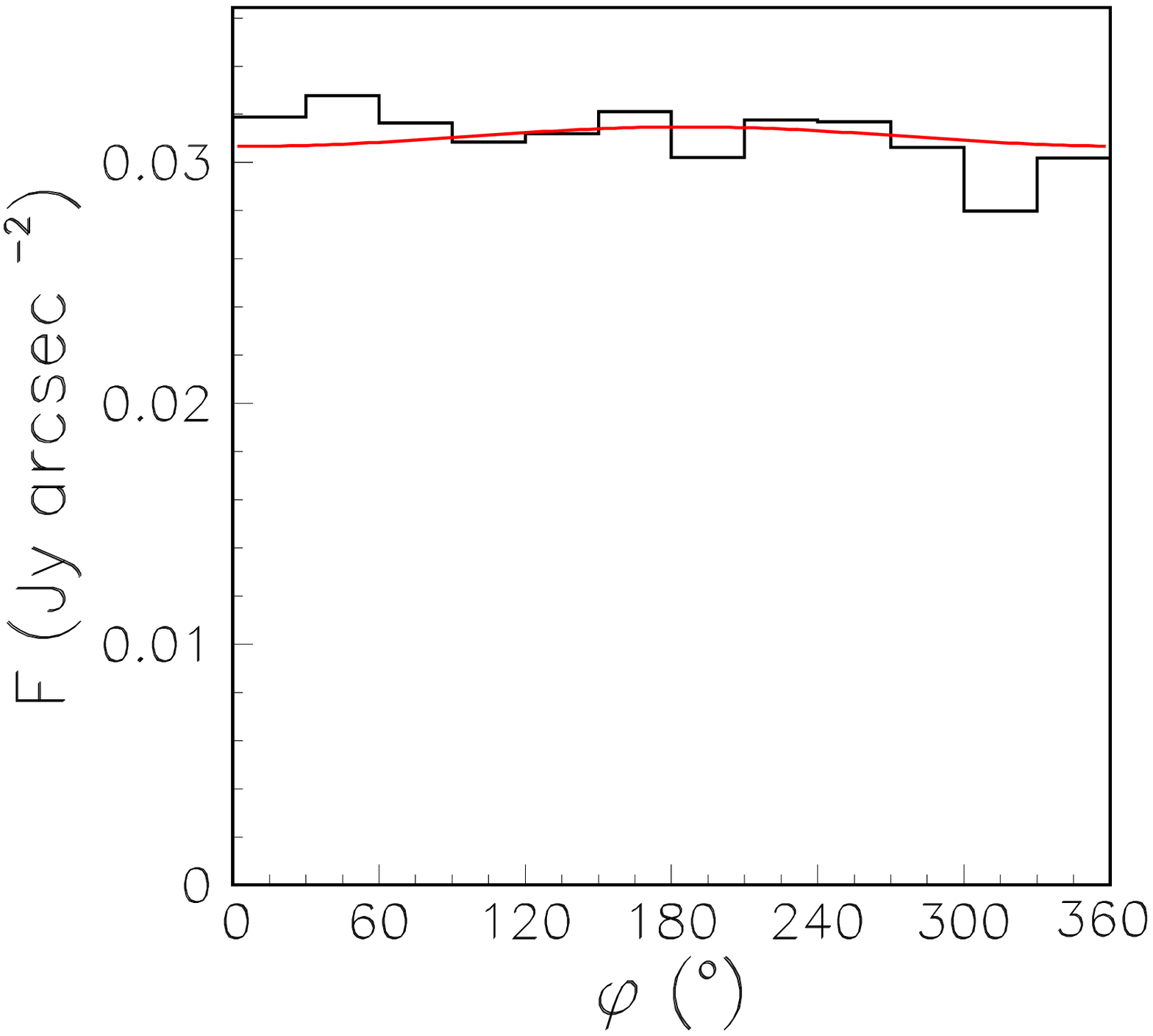}
  \includegraphics[width=.245\linewidth,trim={0cm 0.cm 0.1cm 1.2cm},clip]{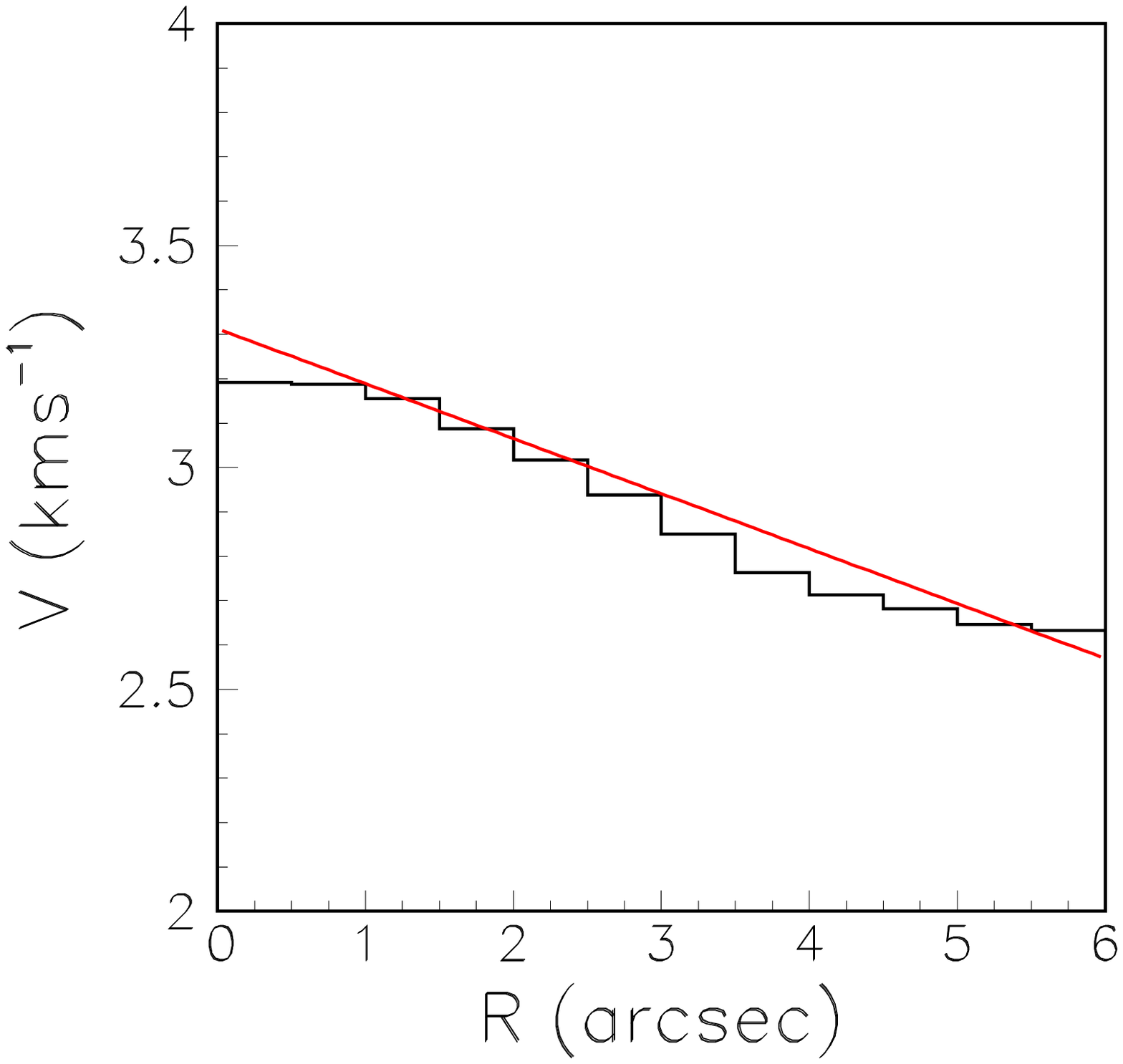}
  \includegraphics[width=.245\linewidth,trim={0cm 0.cm 0.1cm 1.2cm},clip]{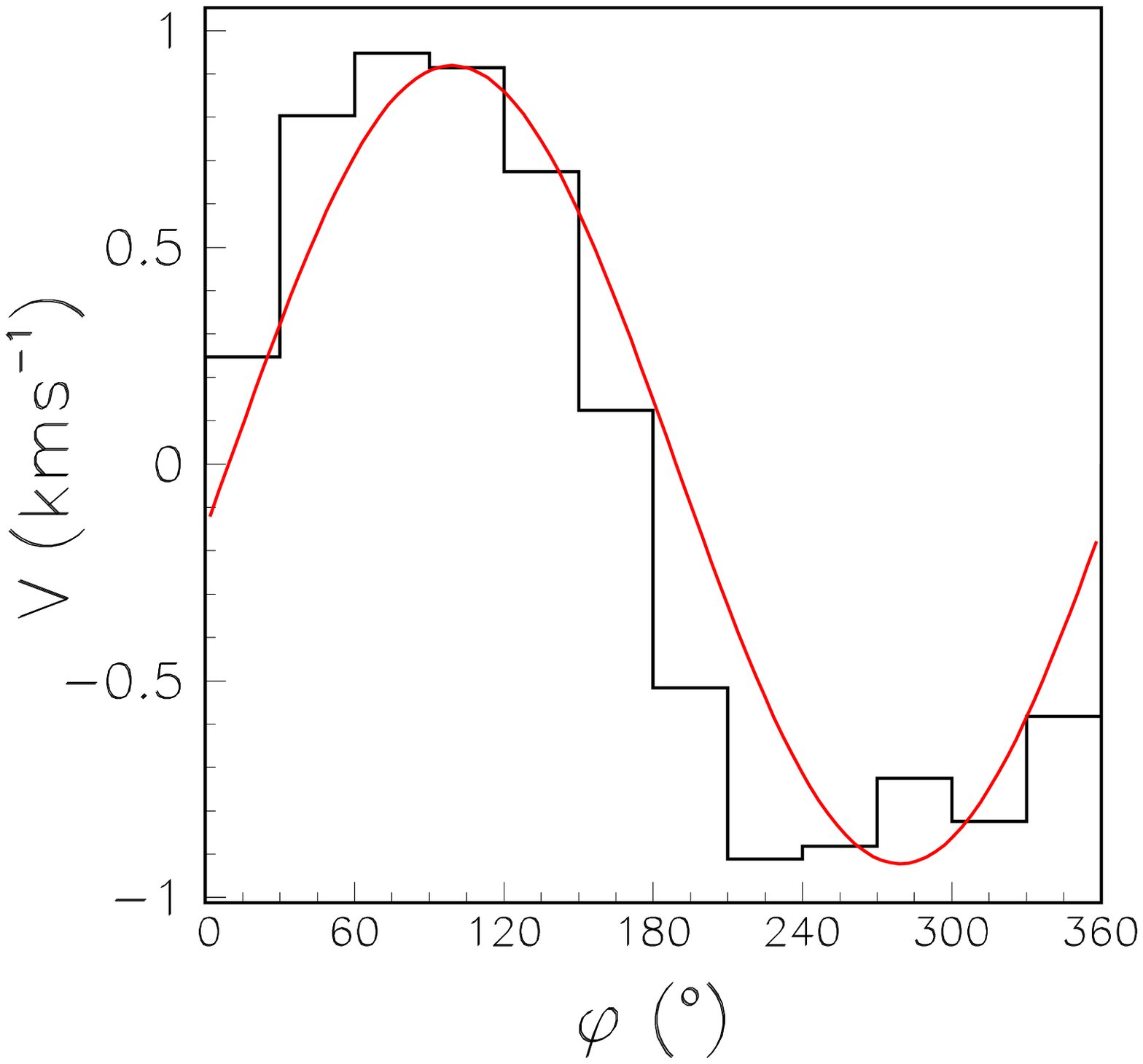} 
  \caption{Measured \mbox{CO(1-0)} flux of RS Cnc. Left: $R$-dependence 
of $F(y,z)$ averaged over $\varphi$; the fit (in red) is a power law of 
index $-$0.63. Central left: $\varphi$-dependence of $F(y,z)$ averaged 
over $R$; the fit (in red) is a sine wave of amplitude 1.3\% of the mean 
value. Centre right: $R$-dependence of $<|V_x|>$ averaged over $\varphi$; 
the fit (in red, excluding the first bin) is linear, with a slope of 
$-$0.12 km\,s$^{-1}$ per arcsecond. Right: $\varphi$-dependence of $<V_x>$ 
averaged over $R$; the fit (in red) is a sine wave of amplitude 
0.92 km\,s$^{-1}$ and phase of 6$^{\circ}$.}
  \label{fig6}
\end{center}
\end{figure*}

\begin{figure*}
\begin{center}
  \includegraphics[width=.244\linewidth,trim={0.4cm 0cm 0cm 0cm},clip]{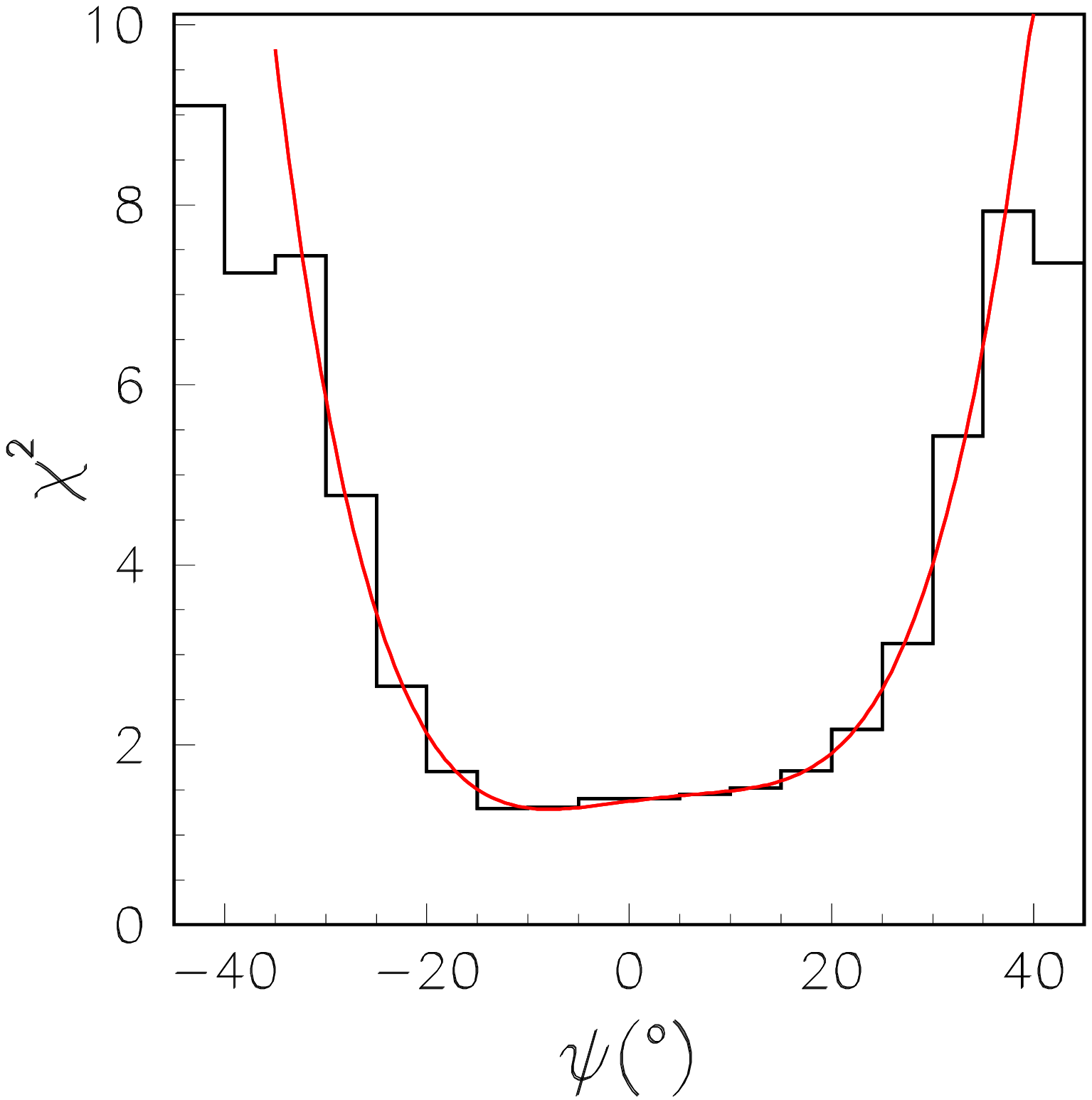}
  \includegraphics[width=.242\linewidth,trim={0cm 0cm 0cm 0cm},clip]{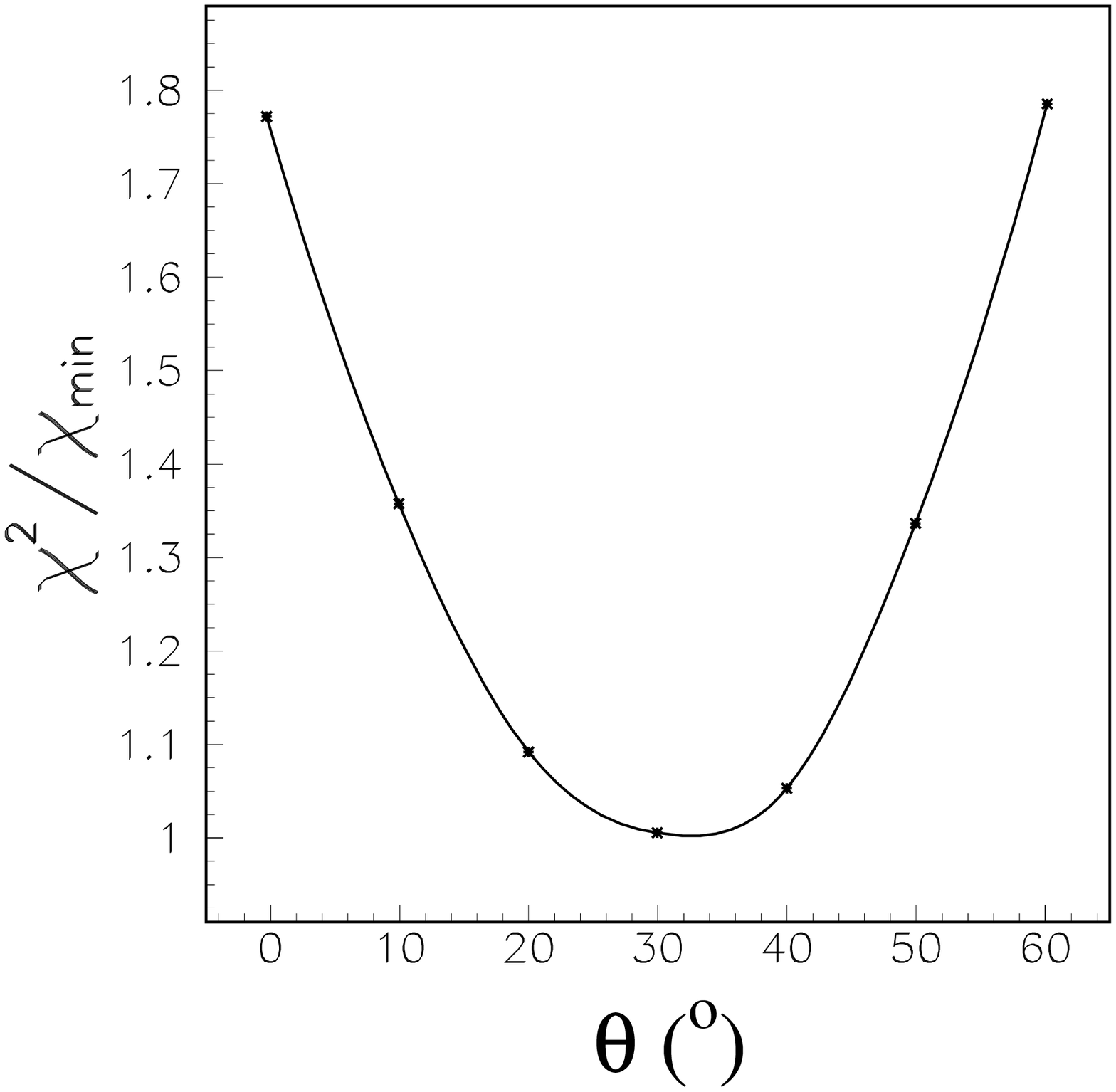}
  \includegraphics[width=.254\linewidth,trim={0cm 0cm 0cm 0cm},clip]{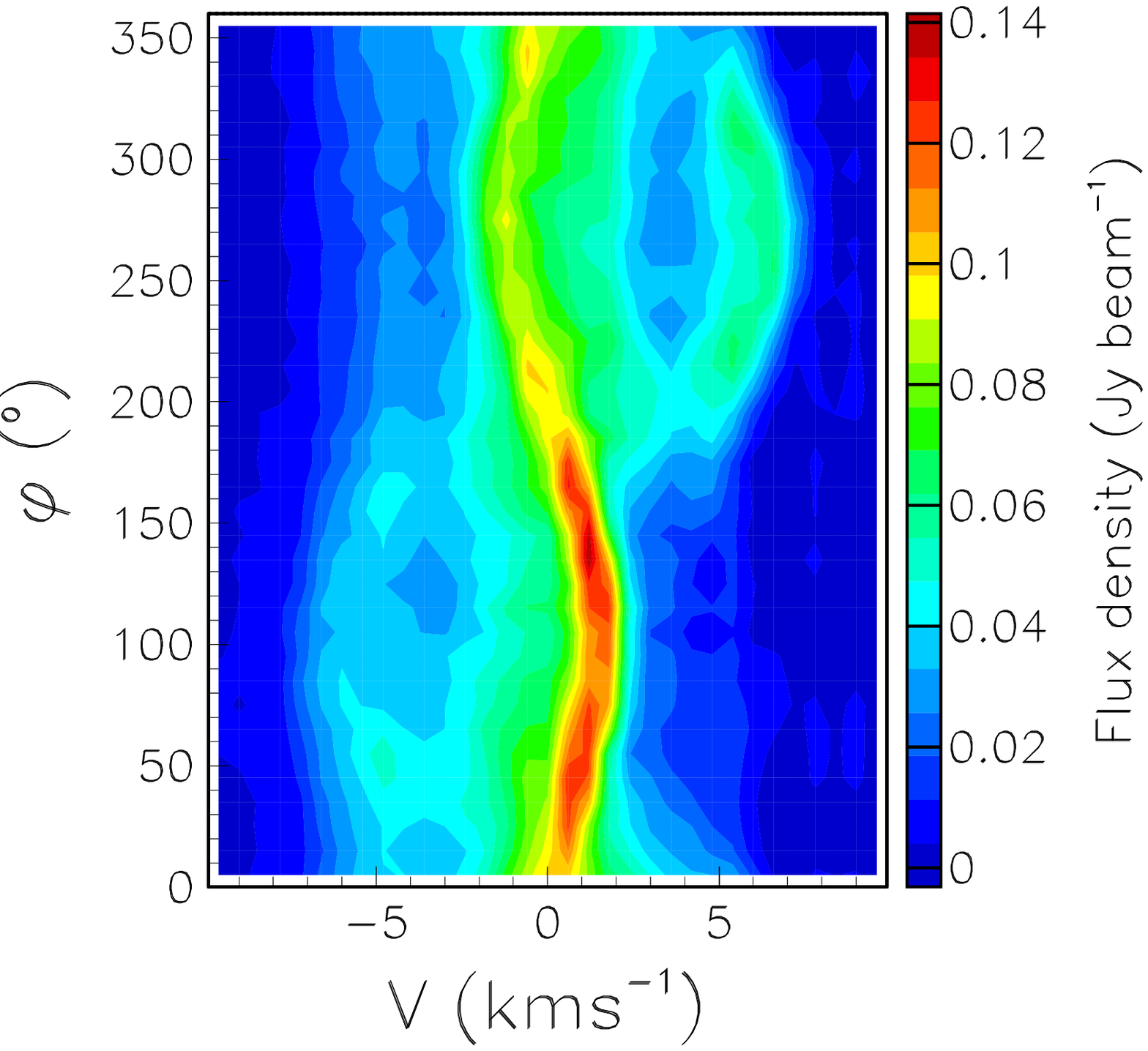}
  \includegraphics[width=.242\linewidth,trim={0cm 0cm 0cm 0cm},clip]{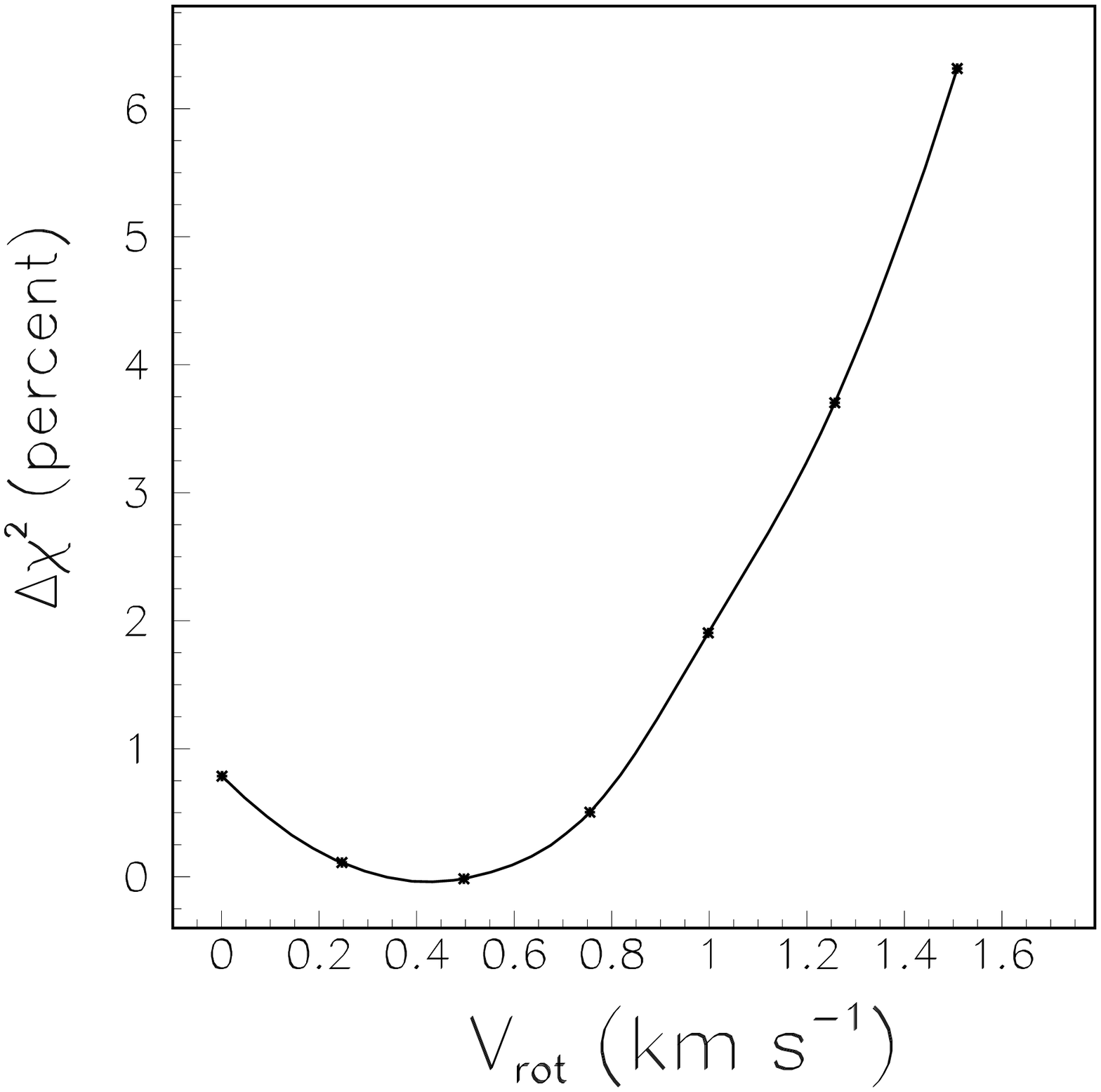}
  \caption{RS Cnc. From left to right: dependences on $\psi$ and on $\theta$
 of the best fit $\chi^2$ normalized to its minimal value; P-V diagram, 
$\varphi$ vs $V_x$; dependence of the best fit $\chi^2$ (in percent 
excess with respect to its minimal value) on the rotation velocity.}
  \label{fig7}
\end{center}
\end{figure*}

As a first application of the results of the preceding sections, we 
consider a typical AGB star, RS Cnc that has been studied in some detail, 
in particular from observations of \mbox{CO(1-0)} emission using the 
Plateau de Bure Interferometer \citep{Hoai2014,Nhung2015a}.

Figure~\ref{fig6} displays the dependence on $R$ and $\varphi$ of 
the measured flux multiplied by $R$. The $R$-dependence is well 
described by a power law of the form $R^{-0.63}$ implying a 
\mbox{$r$-dependence} of the effective emissivity of the approximate 
form $r^{-1.6}$. The $\varphi$-dependence is very uniform, with 
excursions from the mean not exceeding $\sim 6\%$; Figure~\ref{fig6} 
displays also the dependence on $R$ and $\varphi$ of the mean Doppler 
velocity, which reveals a very clear north-south asymmetry.

The $R$-dependence suggests the presence of a significant negative velocity gradient, at variance with the results of the best fits published earlier \citep{Nhung2015a}, which imply a positive velocity gradient both at the equator and near the poles. This is a good illustration of a general situation and deserves therefore a detailed analysis. The model of \cite{Nhung2015a} assumes a stationary regime: a radial wind velocity decreasing with $r$ implies a product $dr^2$ ($d$ is the gas density) increasing with $r$ at the same rate. However, the dependence on $r$ of the effective emissivity includes in addition a temperature factor (that accounts for the population of the emitting quantum state and the probability of emission) and a UV dissociation factor. In the model of \cite{Nhung2015a}, the radial dependence of the temperature is discussed in detail, using in particular the ratio of \mbox{CO(2-1)} and \mbox{CO(1-0)} emission, and the UV dissociation factor is taken from the literature. Both decrease strongly with $r$. As a result, the product $\rho r^2$ ($\rho$ is the effective emissivity) decreases with $r$ because the contributions of the temperature and of UV dissociation win over the contribution of stationarity. The hypothesis of stationarity links the radial dependence of $\rho r^2$ to that of $<|V_x|>$, which receives therefore three contributions: a 1.9\%/arcsecond decrease from temperature, a 3.4\%/arcsec decrease from UV dissociation and 1.8\%/arcsec increase from the velocity gradient, adjusted to best fit the observed 3.5\%/arcsec negative velocity gradient observed in the data. The result of \cite{Nhung2015a} is therefore perfectly consistent with the observation of a negative velocity gradient made here, the reason being that the two velocity gradients measure in fact different quantities, the result of \cite{Nhung2015a} resting heavily on the hypothesis of a stationary flow.
 
The question then arises of the validity of the assumption of stationarity used in the model: the radial scale spanned by Figure~\ref{fig6} means some 10$^5$ years, the wind velocity might have increased somewhat over such a long period, implying higher gas densities at large distances from the star and violating the stationarity hypothesis. To answer the question we compare the quality of two fits: one assuming stationarity and allowing for gradients on the velocities, the other assuming instead $r$-independent velocities but allowing for gradients on the fluxes of matter. The best fit is obtained for the scenario that assumes stationarity, but the other fit gives a value of $\chi^2$ that is only 12\% larger. This result illustrates the difficulty to ascertain the validity of the stationarity hypothesis with confidence from the presently available observations: a dependence on age of the mass-loss rate reproduces the data nearly as well without requiring the presence of a positive velocity gradient.

RS Cnc is also a good test case to illustrate the problems associated 
with the measurement of the orientation of the star axis and the 
relative importance of expansion and rotation. 

The P-V diagram displayed in Figure~\ref{fig7}, $\varphi$ vs $V_x$, 
shows arcs near $V_x=0$ and at $|V_x|>4$ km\,s$^{-1}$ in phase 
opposition at a same value of $\varphi$, namely populating both 
positive and negative Doppler velocity intervals, showing, 
at least qualitatively, that one is dealing with expansion 
rather than with rotation. Moreover, the absence of significant features in quadrature, i.e. shifted in phase by $\pm 90^\circ$ with respect to these arcs, indicate that a possible contribution of rotation, if present, must be quite small (again qualitatively). The asymmetry parameter $A(\varphi)$, defined in A6, reaches minimal values at the 10\% level around $\varphi=0^\circ$ and $\varphi=180^\circ$ where the contribution of pure expansion cancels. To better quantify a possible contribution of rotation, we use the model described in \cite{Hoai2014} and \cite{Nhung2015a}. We allow for a rotation velocity having a Gaussian dependence on $\sin\alpha$, the sine of stellar latitude, and decreasing with distance as a power of $r$. Calling $V_{rot}$ the equatorial rotation velocity at $r=1''$, $\sigma_{rot}$ the $\sigma$ of the Gaussian and $n_{rot}$ the power index, we obtain a best fit for $V_{rot}=0.4$ km\,s$^{-1}$, $\sigma_{rot} =0.3''$ and $n_{rot}=-1.0$. The only other parameters allowed to vary in the fit are the values of the radial expansion velocities. Figure~\ref{fig7} displays the dependences on $\psi$ and $\theta$ of the best fit $\chi^2$, both angles being measured with an uncertainty of $\sim \pm 25^{\circ}$ 
and the values of the
parameters describing the elongation along the star axis are strongly correlated with $\theta$. Then, fixing $\sigma_{rot}$  and $n_{rot}$ at their best fit values, we obtain the dependence of the best fit $\chi^2$ on $V_{rot}$ displayed in Figure~\ref{fig7}, suggesting that a rotation velocity of up to $\sim$1 km\,s$^{-1}$ at $r=1''$  could be accommodated by the data. Observations, of better sensitivity than presently available would be necessary to conclude reliably on the possible presence of rotation.

\subsection{The $\theta=0$ case: EP Aquarii}
\label{sec3.2}
\begin{figure*}
\begin{center}
  \includegraphics[width=.26\linewidth,trim={0cm 0cm 0cm 0cm},clip]{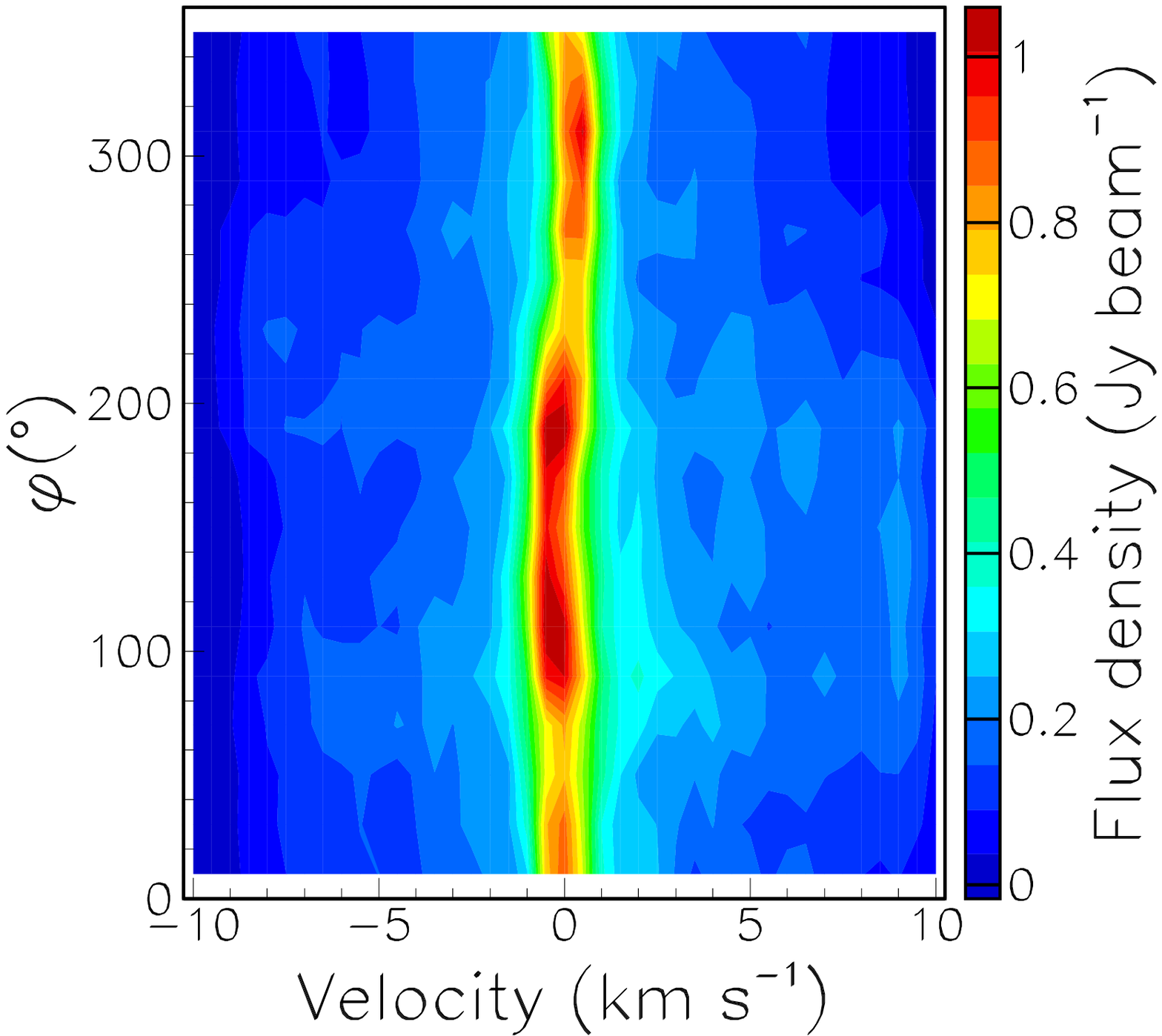}
  \includegraphics[width=.24\linewidth,trim={0cm 0cm 0cm 0cm},clip]{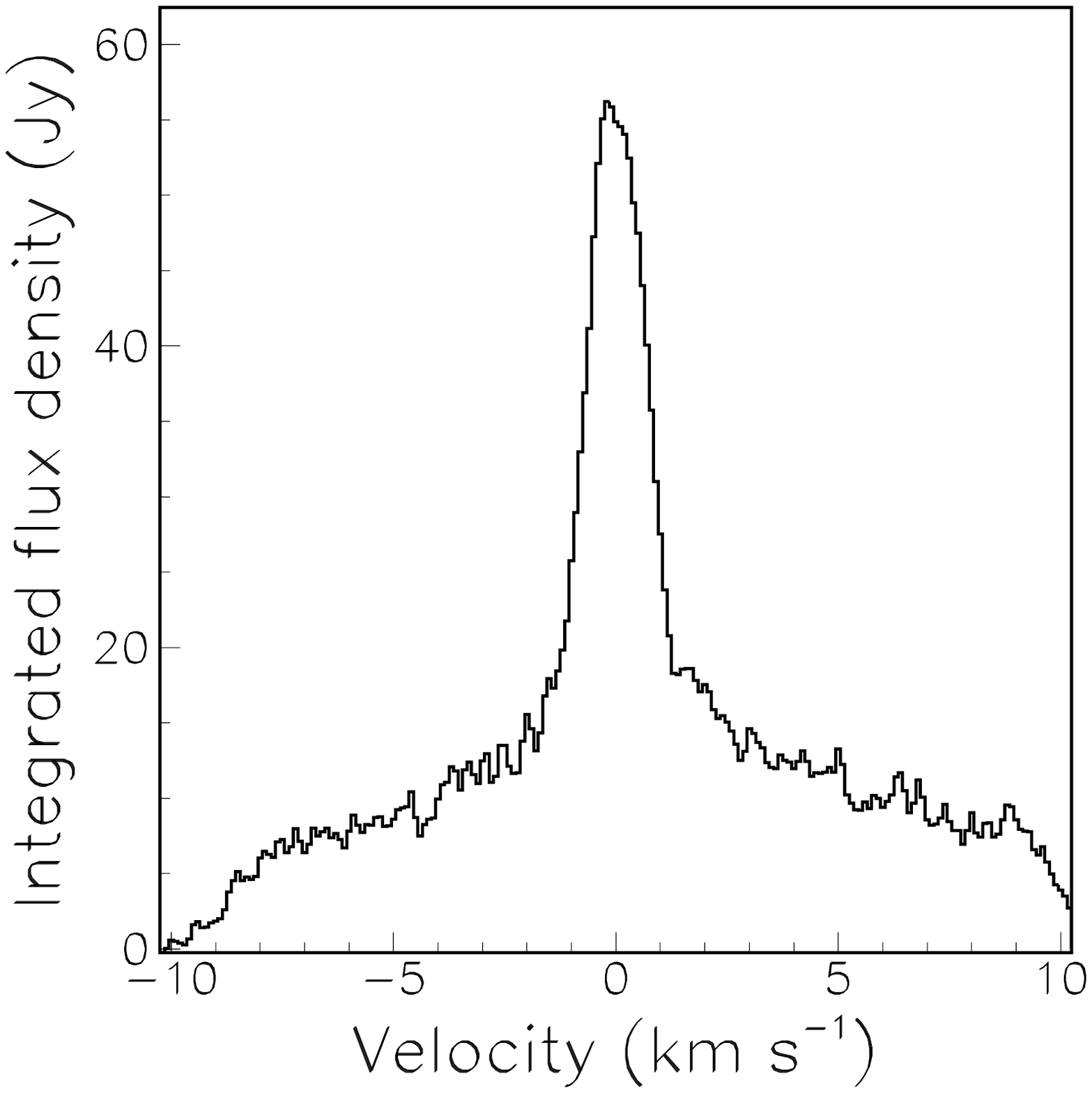}
  \includegraphics[width=.24\linewidth,trim={0cm 0cm 0cm 0cm},clip]{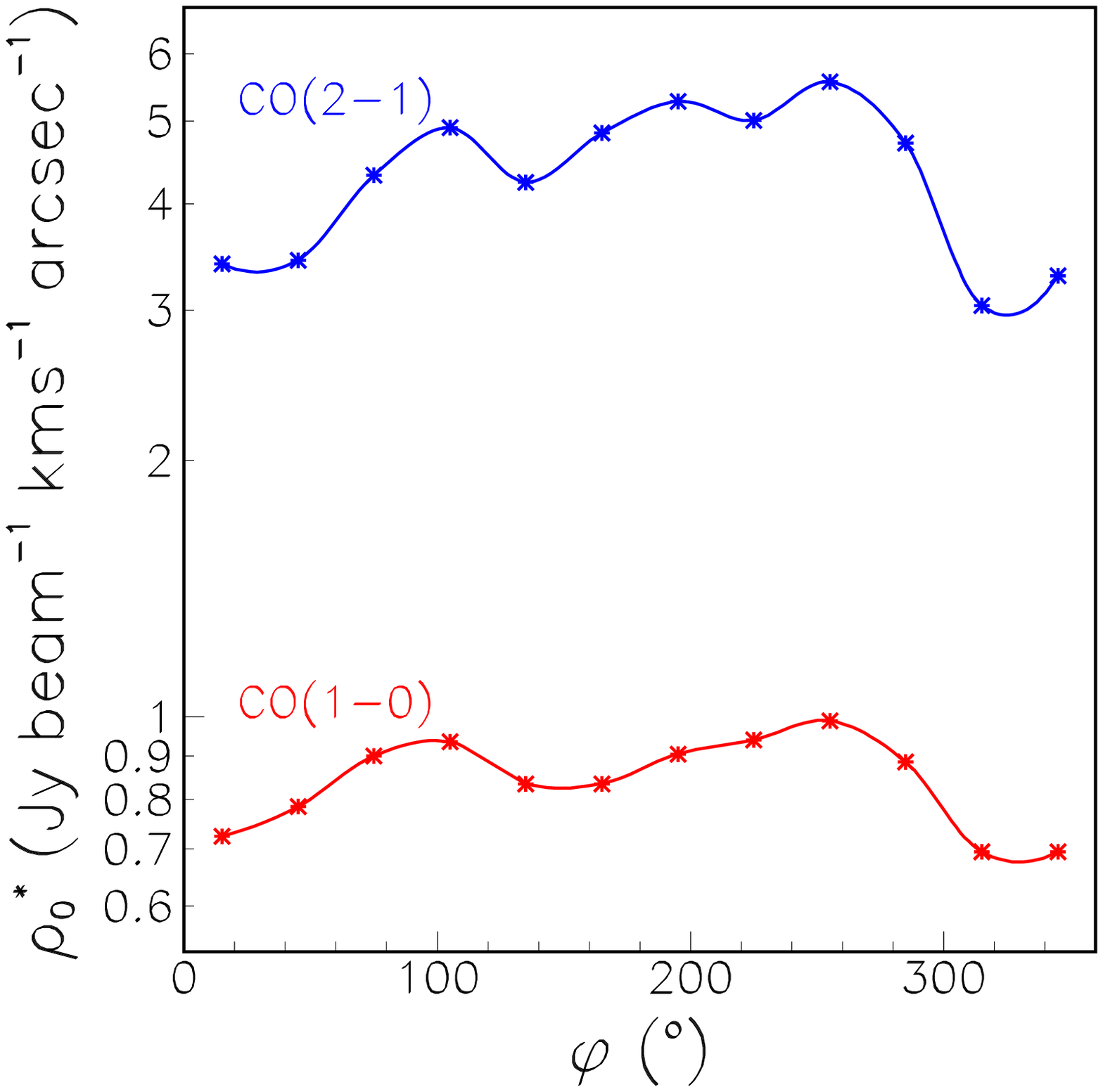}
  \includegraphics[width=.24\linewidth,trim={0cm 0cm 0cm 0cm},clip]{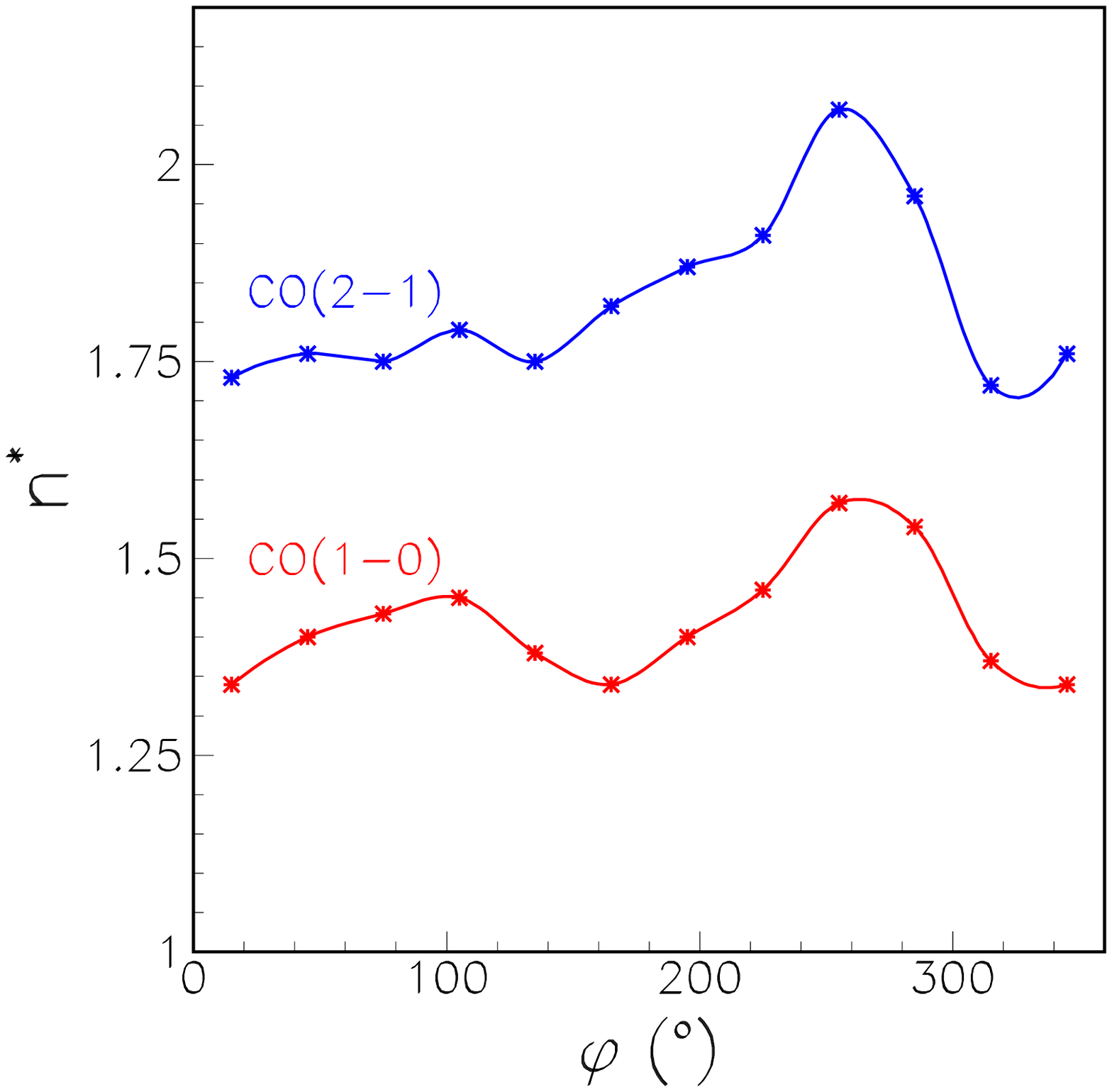}
  \caption{EP Aqr, from left to right: P-V diagrams in the $\varphi$ vs $V_x$ plane of the \mbox{CO(2-1)} data; its projection of the $V_x$ axis; dependence on $\varphi$ of $\rho^*_0$ (centre right) and $n^*$ (right) for \mbox{CO(1-0)} (red) and \mbox{CO(2-1)} (blue) observations.}
  \label{fig8}
\end{center}
\end{figure*}

EP Aqr is a textbook case of an AGB star having its axis nearly parallel to the line of sight. Its \mbox{CO(1-0)} and \mbox{CO(2-1)} emissions have been previously analysed both in terms of spherical winds and of bipolar outflows \citep{Winters2007,Nhung2015b}. We refer the reader to the respective analyses and, as in the case of RS Cnc, we limit the present discussion to issues that are of direct relevance to the considerations developed in the previous sections, in particular aiming at a unified picture of the former very different analyses. Figure~\ref{fig8} shows the \mbox{CO(2-1)} P-V diagram in the $\varphi$ vs $V_x$ plane and its projection on the Doppler velocity axis. As expected, it is typical of a star having its axis parallel to the line of sight. The dependence on both $R$ and $\varphi$ of the measured flux multiplied by $R$ is dominated by important inhomogeneities that extend up to large radii. The $R$-dependence is steeper for \mbox{CO(2-1)} than for \mbox{CO(1-0)} as expected, with respective power indices $-1.2$ and $-0.9$. The sky maps of the mean Doppler velocity display a small north-west/south-east asymmetry, interpreted by \cite{Nhung2015b} as resulting from the small inclination of the star axis on the line of sight.

In the $\theta=0$ approximation the stellar longitude $\omega$ is equal to the sky position angle $\varphi$ of the pixel and the stellar latitude $\alpha$ is related to $x$ by the relation $x=R\tan\alpha$. The relation $f(R,\varphi,V_x)dV_x=\rho(r,\alpha,\omega)dx$ can then be dealt with in each $\varphi=\omega$ interval separately. As rotation cannot be detected, we assume a radial wind of velocity $V_{rad}$. Using the relations 
\begin{equation} \label{eq12}
\begin{split}
&V_x=V_{rad}\sin\alpha \\
&x=r\sin\alpha\\
&R=r\cos\alpha
\end{split}
\end{equation}
one can associate in each pixel a value of $x$ to each value of $V_x$ if we know $V_{rad}$: $x$ is solution of the equation 
\begin{equation} \label{eq13}
rV_x=\sqrt{x^2+R^2}V_x=xV_{rad}
\end{equation}
namely 
\begin{equation} \label{eq14}
x=\frac{V_x}{\sqrt{V_{rad}^2-V_x^2}}R
\end{equation}
knowing $x$ one also knows \mbox{$r=\sqrt{x^2+R^2}$} and $\sin\alpha=x/r$. 

\begin{figure*}
\begin{center}
  \includegraphics[width=.255\linewidth,trim={0cm 0cm 0cm 0cm},clip]{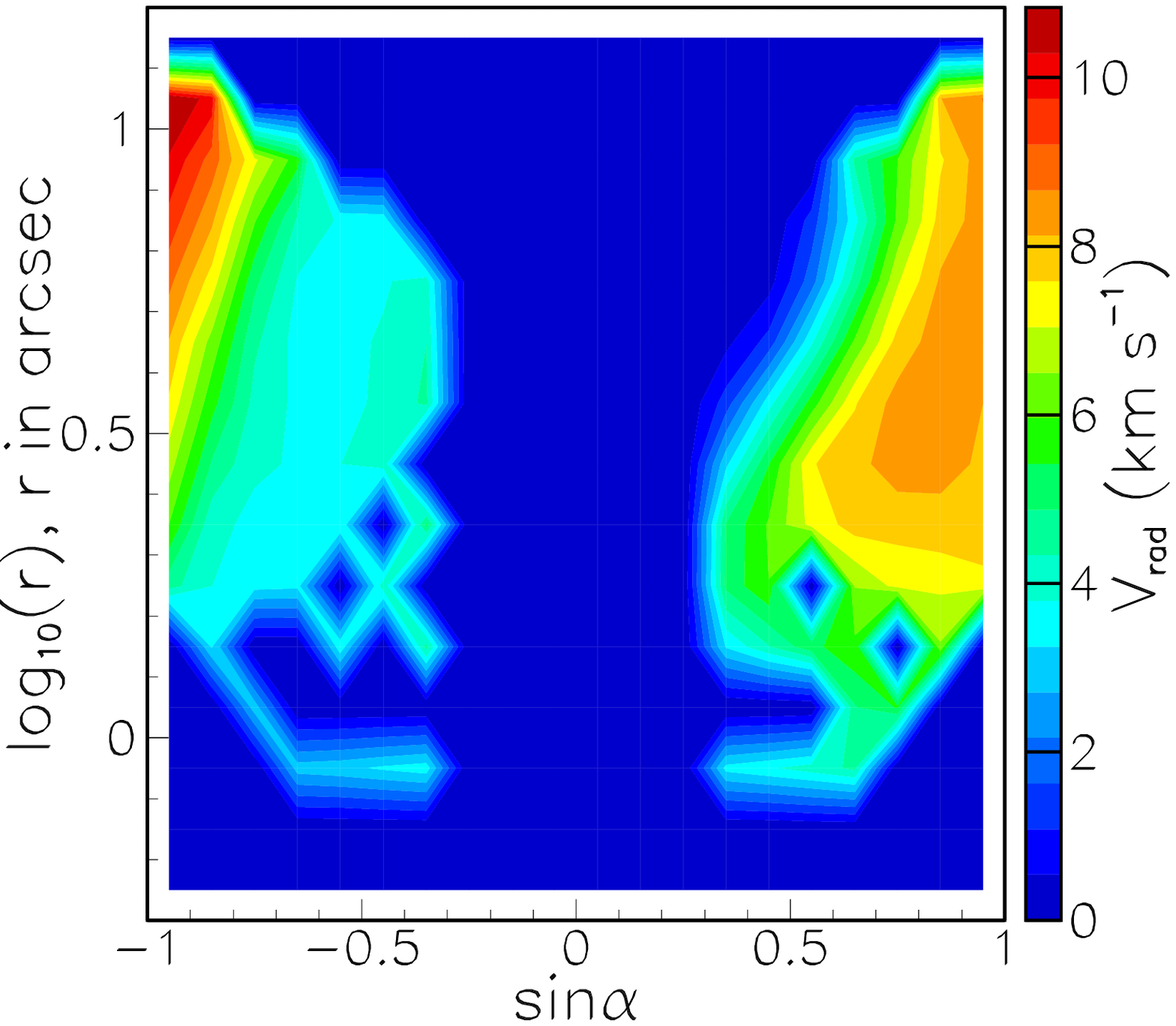}
  \includegraphics[width=.255\linewidth,trim={0cm 0cm 0cm 0cm},clip]{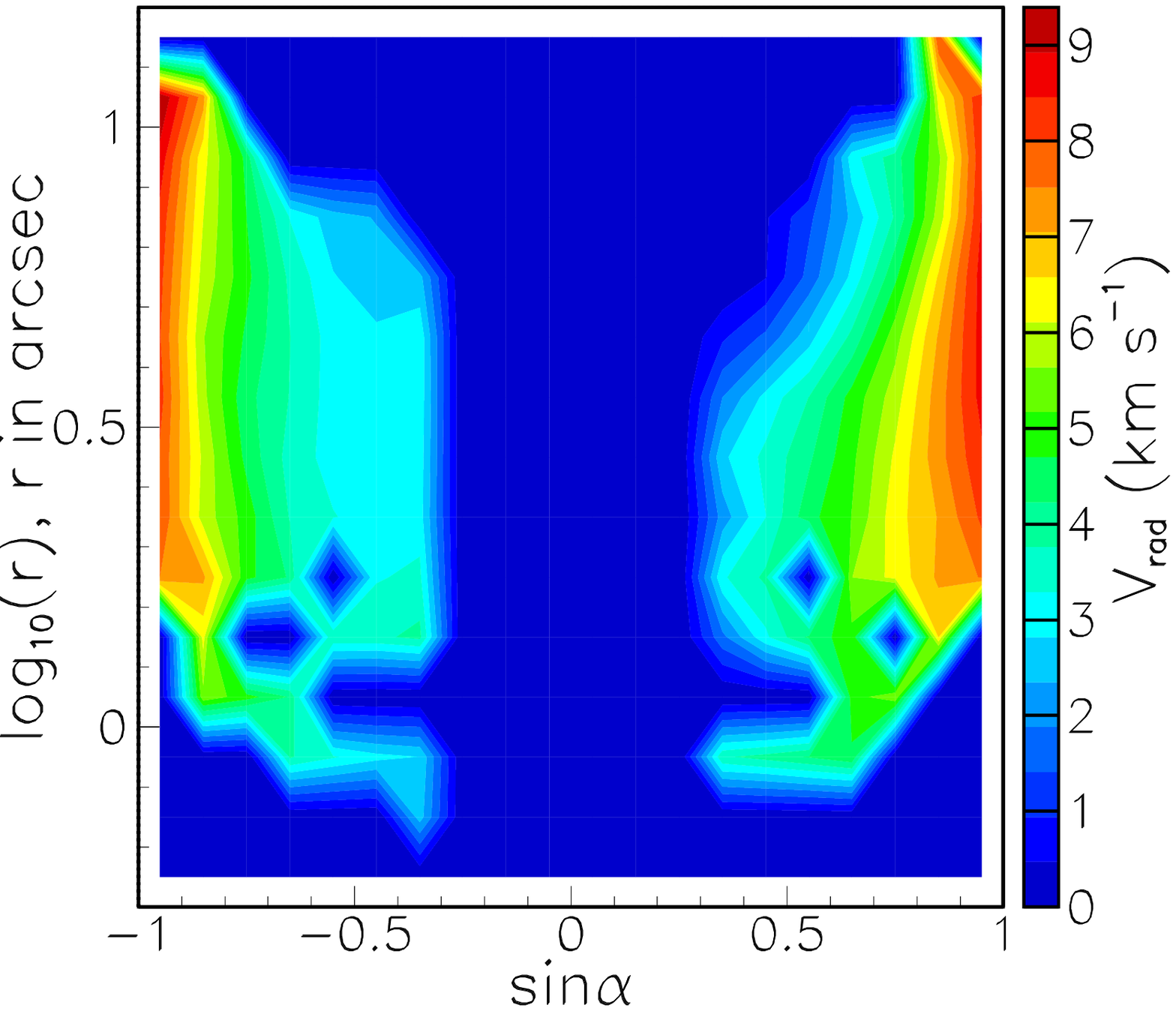}
  \includegraphics[width=.235\linewidth,trim={0cm 0cm 0cm 0cm},clip]{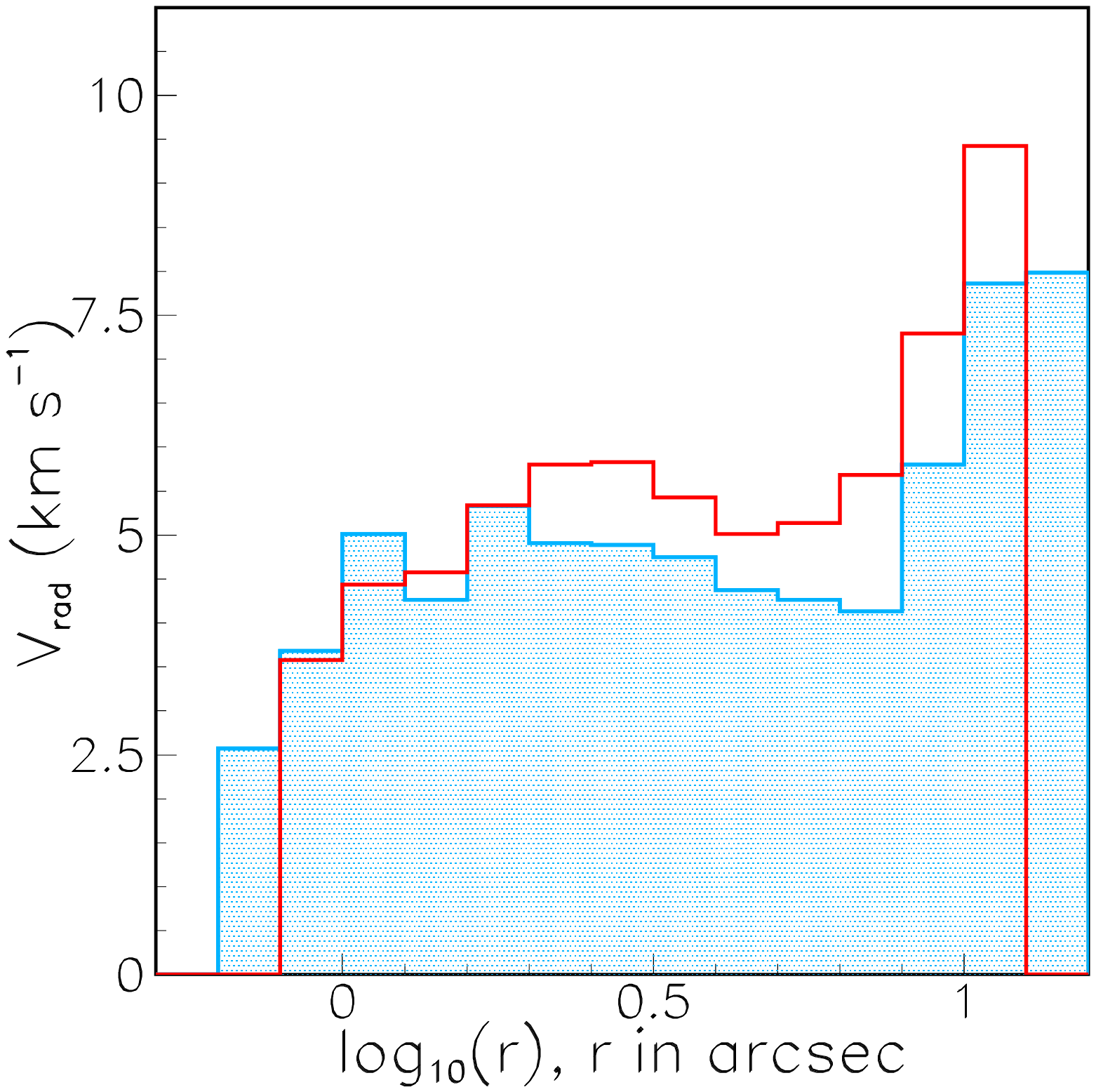}
  \includegraphics[width=.235\linewidth,trim={0cm 0cm 0cm 0cm},clip]{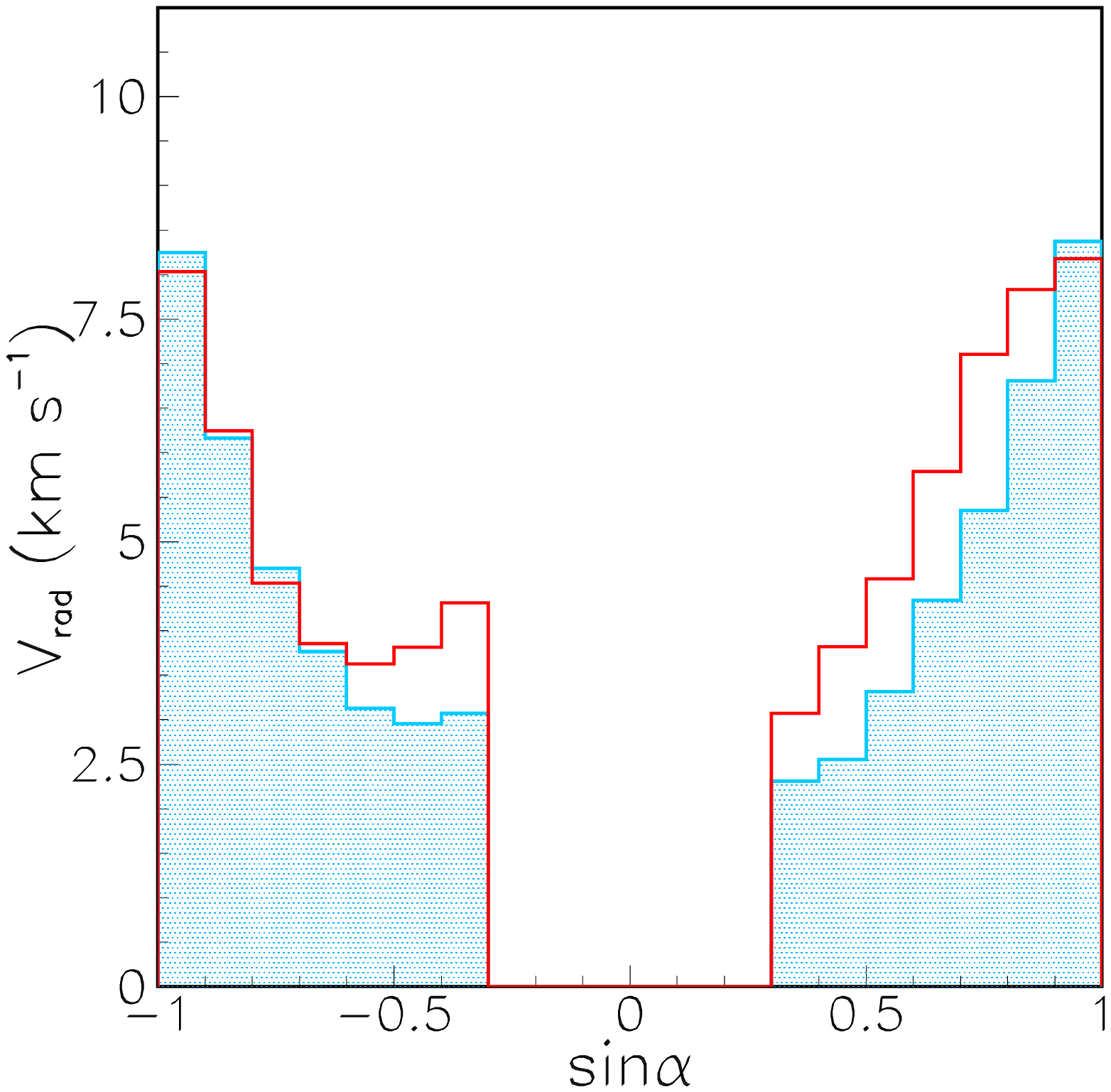}
  \caption{EP Aqr. Distribution of the radial expansion velocity $V_{rad}$, averaged over $\varphi$. Left: in the $(\log _{10}(r),\sin\alpha)$ plane for \mbox{CO(1-0)} and \mbox{CO(2-1)}. Right: as a function of $\sin\alpha$ (centre right) and $\log _{10}(r)$ (extreme right) for \mbox{CO(1-0)} (red histogram) and \mbox{CO(2-1)} (filled blue).}
  \label{fig9}
\end{center}
\end{figure*}

An estimate of the effective emissivity averaged over stellar latitude (corresponding to spherical symmetry of the effective emissivity), $\rho^*(r)$, can be obtained from the flux measured in each pixel, $F(y,z)=\int \rho(r,\alpha,\omega)dx$. In each $\varphi=\omega$ interval, one obtains $\rho^*(r)$ by solving the integral equation 
\begin{equation} \label{eq15}
F(R)=\int \rho^*(r)dx=2\int \rho^*(r)r(r^2-R^2)^{-1/2}dr
\end{equation}
where the second integral runs from $R$ to 13$''$, the mean distance at which CO molecules are UV-dissociated. 
Assuming in each $\varphi$ bin a form ($r$ measured in arcseconds).
\begin{equation} \label{eq16}
\rho^*(r)=\rho^*_0r^{-n^*}
\end{equation}
good fits are obtained and the dependence on $\varphi$ of $\rho^*_0$ and $n^*$ is displayed in the right panels of Figure~\ref{fig8}. The $r$-dependence of the effective emissivity is steeper for \mbox{CO(2-1)} than for \mbox{CO(1-0)} and its slope reaches a maximum in the north direction for both lines. The relation $f(R,\varphi,V_x)\Delta V_x=\rho(r,\alpha,\omega)\Delta x$ where $\Delta x$ is the $x$ interval spanned by a velocity bin $\Delta V_x$ can then be used in each $(R,\varphi)$ bin to evaluate the radial expansion velocity, $V_{rad}$, as a function of $x$ in the approximation where the effective 
emissivity is taken in each $\varphi$ bin equal to $\rho^*_0r^{-n^*}$ as evaluated in the previous paragraph. For the purpose of illustration, we do so over the whole $\varphi$ range but it can as well be done in each $\varphi$ bin separately when the data are of sufficient quality. In order to exclude the equatorial region, where the evaluation of $V_{rad}$ is less reliable, we limit the stellar latitude to the range $|\sin\alpha|>0.3\ (|\alpha|>17.5^\circ)$, with the effect of excluding low values of $r$ (typically smaller than 0.8$''$). Figure~\ref{fig9} shows the resulting distribution of $V_{rad}$, averaged over $\varphi$, in the stellar meridian plane. It gives evidence for a strong polar enhancement, at variance with a weak radial dependence up to $r\sim 8''$. It provides a unified picture of the two models that have been proposed earlier, the model of \cite{Winters2007} corresponding to $V_{rad}$ dependent on $r$ but independent on $\alpha$, that of \cite{Nhung2015b} corresponding to $V_{rad}$ dependent on $\alpha$ but independent on $r$. The projections of the $V_{rad}$ distributions on the $\sin\alpha$ and $r$ axes illustrate the preference for the latter over the former.

\subsection{The radial expansion representation}
\label{sec3.3}
The analysis presented in the preceding section can be applied to any observation: it is always possible to calculate, given a value of $V_{rad}$, for each pixel and each Doppler velocity, the quantity 
\begin{equation} \label{eq17}
x=rV_x/V_{rad}=\frac{V_x}{\sqrt{V_{rad}^2-V_x^2}}R
\end{equation}
and obtain this way a \mbox{3-D} representation of the observation, with each point $(x,y,z)$ given an effective emissivity
\begin{equation} \label{eq18}
\rho (x,y,z)=f(y,z,V_x)dV_x/dx=\frac{x(V_{rad}^2-V_x^2)}{r^{2}V_x}f(y,z,V_x)
\end{equation}

\begin{figure*}
\begin{center}
  \includegraphics[width=1.\columnwidth,trim={0cm 0cm 0cm 0cm},clip]{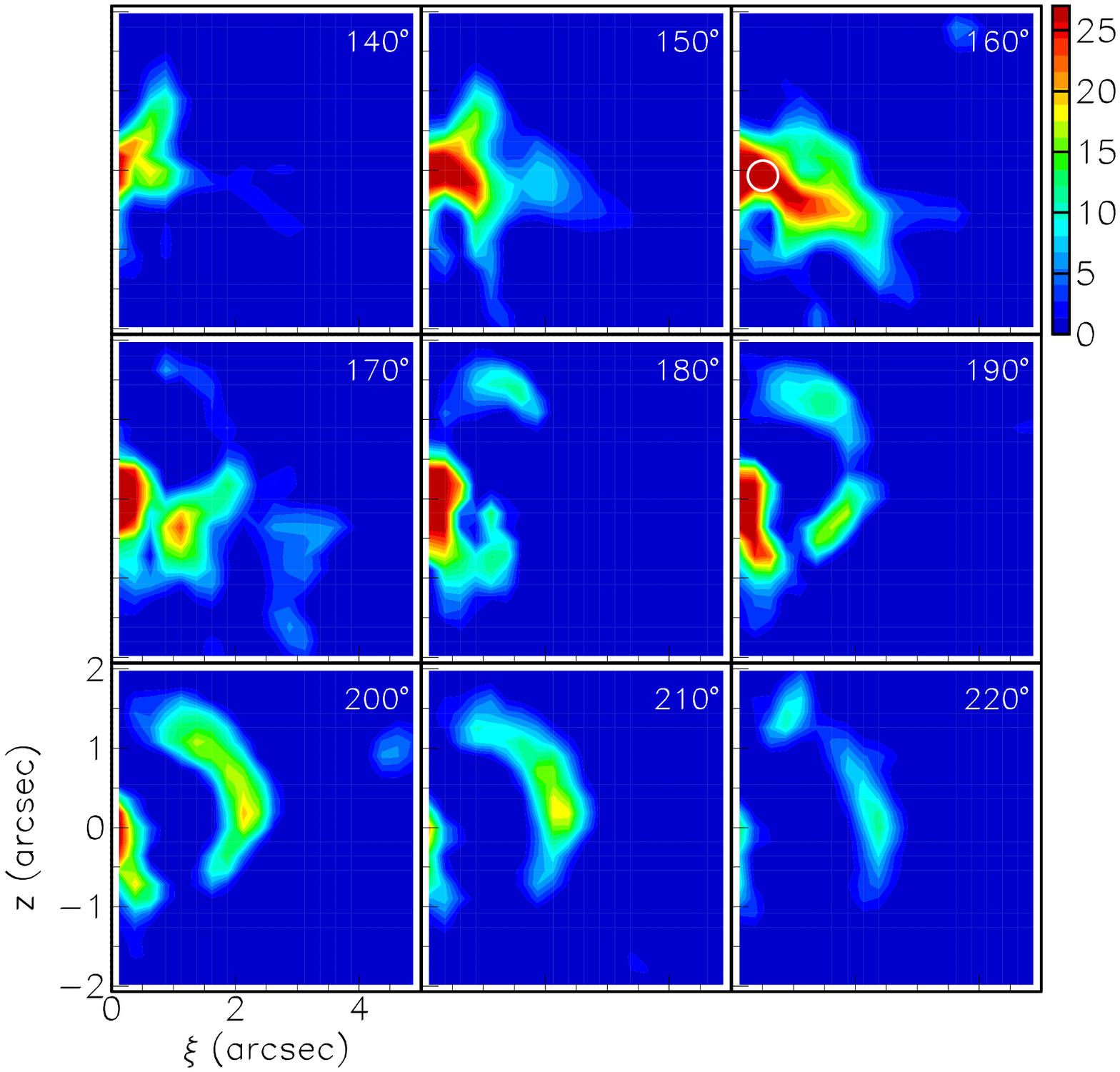}
  \includegraphics[width=1.\columnwidth,trim={0cm 0cm 0cm 0cm},clip]{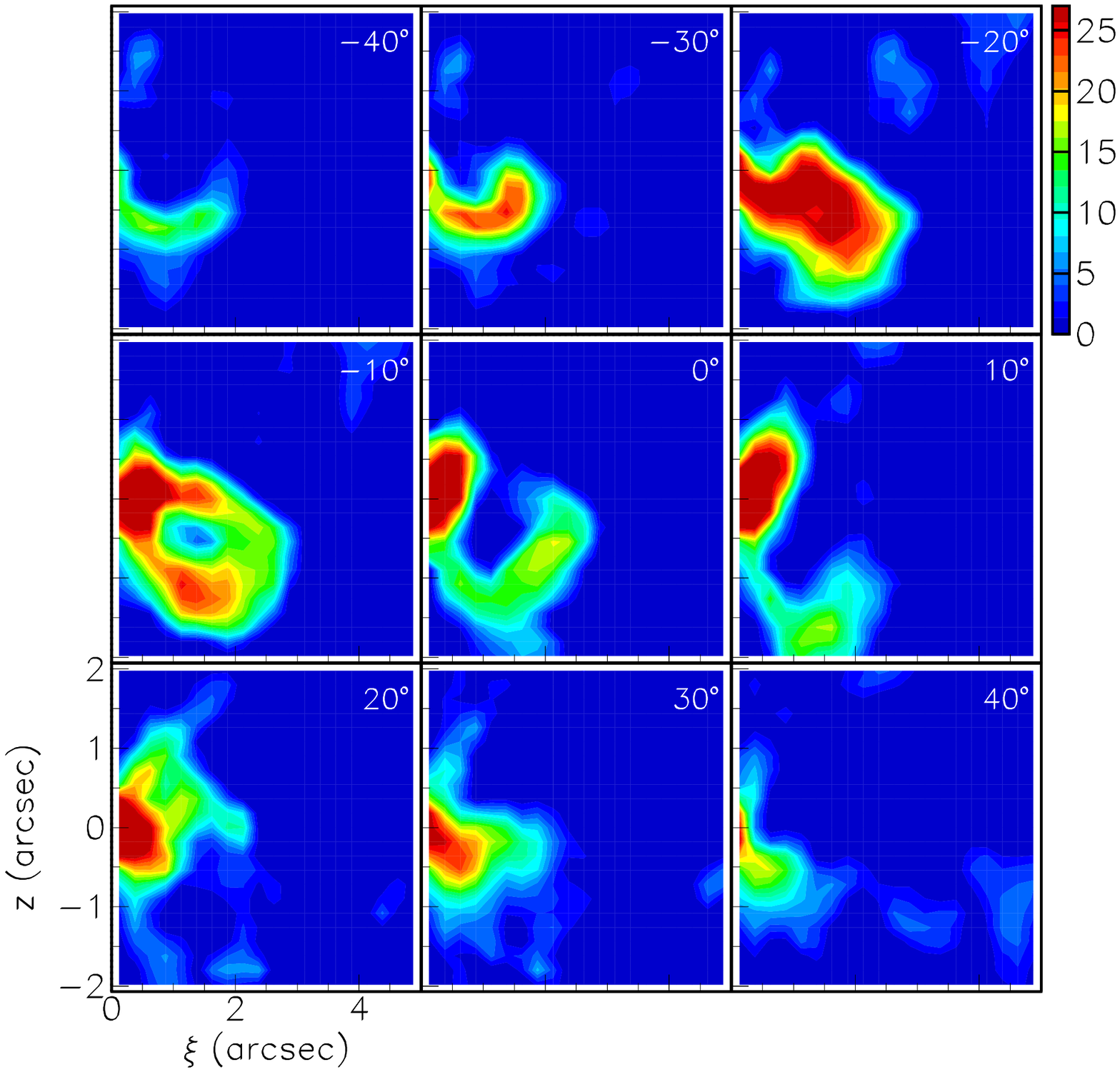}
  \caption{Western (left) and eastern (right) outflows of the central part of the CSE of Mira Ceti: maps of the effective emissivity reconstructed in space under the assumption of a pure radial expansion at constant velocity of 7 km\,s$^{-1}$. The ordinate is $z$ and the abscissa is $\xi =\sqrt{x^2+y^2}$. Each panel is for a 10$^\circ$ wide interval of $\omega$ measured clockwise from the $y$ axis. The label in each panel gives the value of $\omega$ at the centre of the interval. The colour scale is in units of Jy\;km;s$^{-1}$\;arcsec$^{-3}$}
  \label{fig10}
\end{center}
\end{figure*}

Such a representation will have something to do with reality if the kinematics is indeed dominated by a radial expansion with constant velocity $V_{rad}$. In practical cases, this may often be approximately the case, at least locally in some domain of the $(y,z,V_x)$ space. In such cases, it is usually possible to have an estimate of the range in which $V_{rad}$ may reasonably fall (clearly, $V_{rad}$ must exceed $|V_x|$). It is then useful to have some idea of the distortion implied by the adopted representation. in particular when $V_x$ approaches $\pm V_{rad}$. In such a case, $x$ goes to infinity and $\rho (x,y,z)r^2$, and a fortiori $\rho (x,y,z)$, cancel. The representation becomes meaningless in this case: obviously, if there is no expansion to start with, there is no chance to learn about the morphology. However, when $V_{rad}$ exceeds $|V_x|$ significantly, the representation is well behaved and the topology is conserved: a detached shell remains a detached shell, an isolated clump remains an isolated clump, a cavity remains a cavity. While such a representation must be interpreted with care, remembering that it is ill-behaved when $|V_x|$ approaches $V_{rad}$, and keeping in mind the possible presence of important distortions, it provides a convenient 3-D visualization of the morphology, particularly useful in complex cases. Such an example is illustrated in Figure~\ref{fig10}, displaying maps of the effective emissivity of Mira Ceti in $(\xi,z)$ planes, with $\xi =\sqrt{x^2+y^2}$, for various values of the angle $\omega$ between the plane and the $y$ axis. It shows clearly the presence of arcs of detached shells.

\subsection{The $\theta =90^\circ$ case: the Red Rectangle}
\label{sec3.4}
The Red Rectangle is a post-AGB star that has been extensively studied 
at optical, infrared and far infrared wavelengths. Its CO emission has 
been recently analysed along the lines presented in the present paper 
using ALMA data \citep{TuanAnh2015}. It has its axis nearly parallel to 
the sky plane and displays density and velocity configurations similar to 
case B considered in the preceding sections: two distinct regions with a 
rather sharp separation between them, a conical bipolar outflow and a 
rotating equatorial volume. We refer the reader to \cite{TuanAnh2015} 
for details and limit our comments to a brief summary of the main results. 
In the $\theta =90^\circ$ approximation, the stellar latitude $\alpha$ 
is related to $z$ by the relation $z=r\sin\alpha$ and the stellar 
longitude $\omega$ is related to $y$ by the relation 
$y=r\cos\alpha \cos\omega$. Rather than dealing with 
independent $\varphi=\omega$ intervals as in the preceding section, 
we deal instead with independent $z$ intervals.
 Namely, in each pixel, once we know $x$ we know 
\begin{equation} \label{eq19}
\begin{split}
&\sin\alpha=z/\sqrt{x^2+R^2}\\
&\tan\omega=x/y
\end{split}
\end{equation}
The measured flux is \mbox{$F(y,z)=\int \rho(x,y,z)dx$}: the effective 
emissivity expressed as a function of $r$ and $\sin\alpha$ can be 
obtained from this integral equation if we assume that it is 
invariant under rotation about the star axis. Then, one can write 
\begin{equation} \label{eq20}
F(y,z)=\int \rho (r,\sin\alpha)dx=2\int \rho (r,z/r)r (r^2-R^2)^{-1/2}dr 
\end{equation}
and solve the integral equation. Deprojections of optical images
have been presented earlier using this same feature \citep[see for
example][and references therein]{Wenger2013}. As two different 
CO lines have been observed, CO(6-5) and CO(3-2), it is possible 
to disentangle the effects of temperature and density and evaluate 
them in space as shown in Figure 11. Using the result of
such deprojection, and the structure that it suggests, as guidance to
define flux lines, it is then possible to construct a simple model of
the wind kinematics inspired from the observed morphology. It
gives evidence for two well distinct volumes that host very
different kinematical regimes and describes simply the radial
dependence of both the expansion and rotation velocities.
 
\begin{figure*}
\begin{center}
  \includegraphics[width=.25\linewidth,trim={0cm 0.2cm 0cm 0cm},clip]{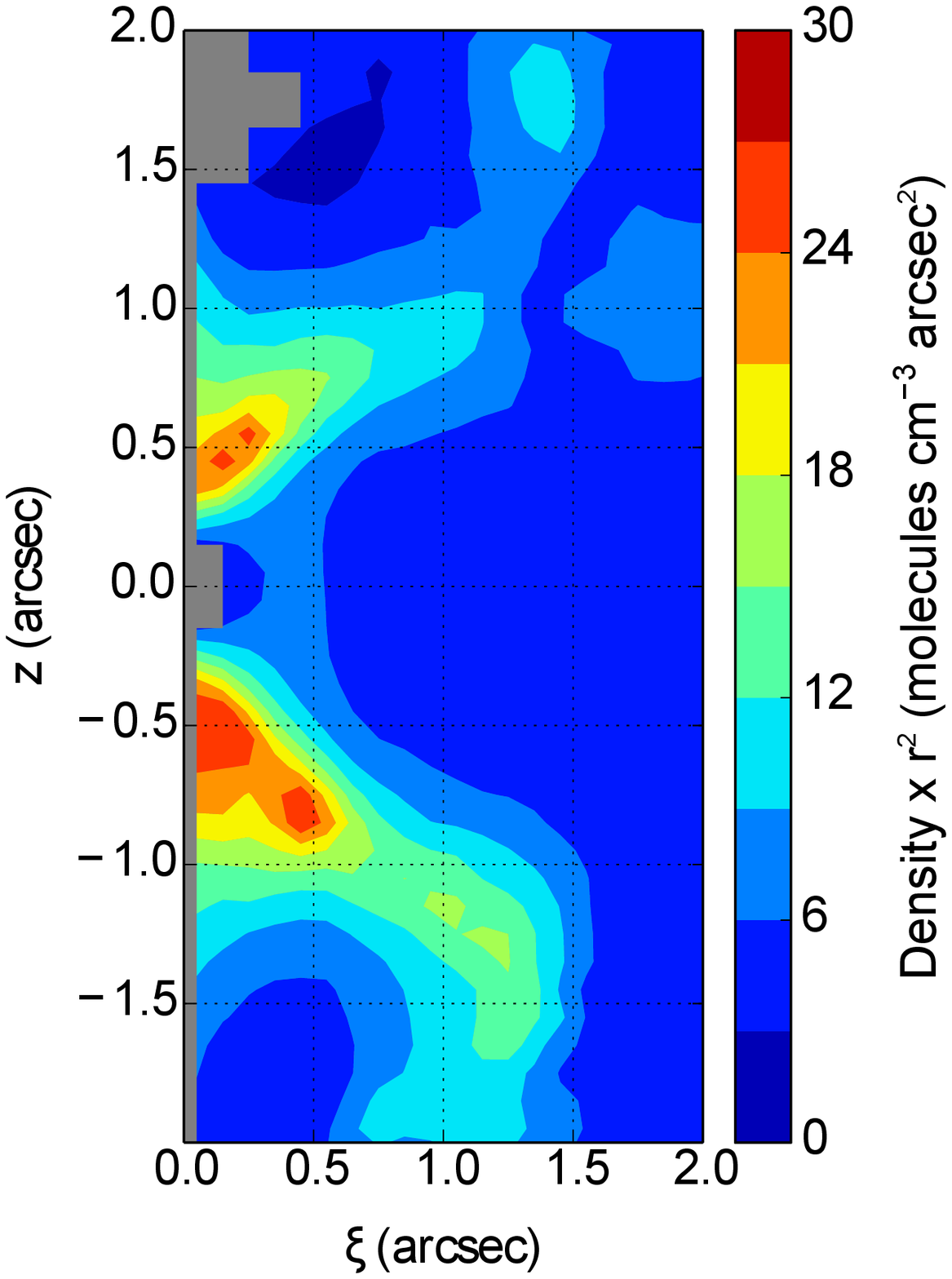}
  \hspace{0.2cm}
  \includegraphics[width=.255\linewidth,trim={0cm 0.2cm 0cm 0cm},clip]{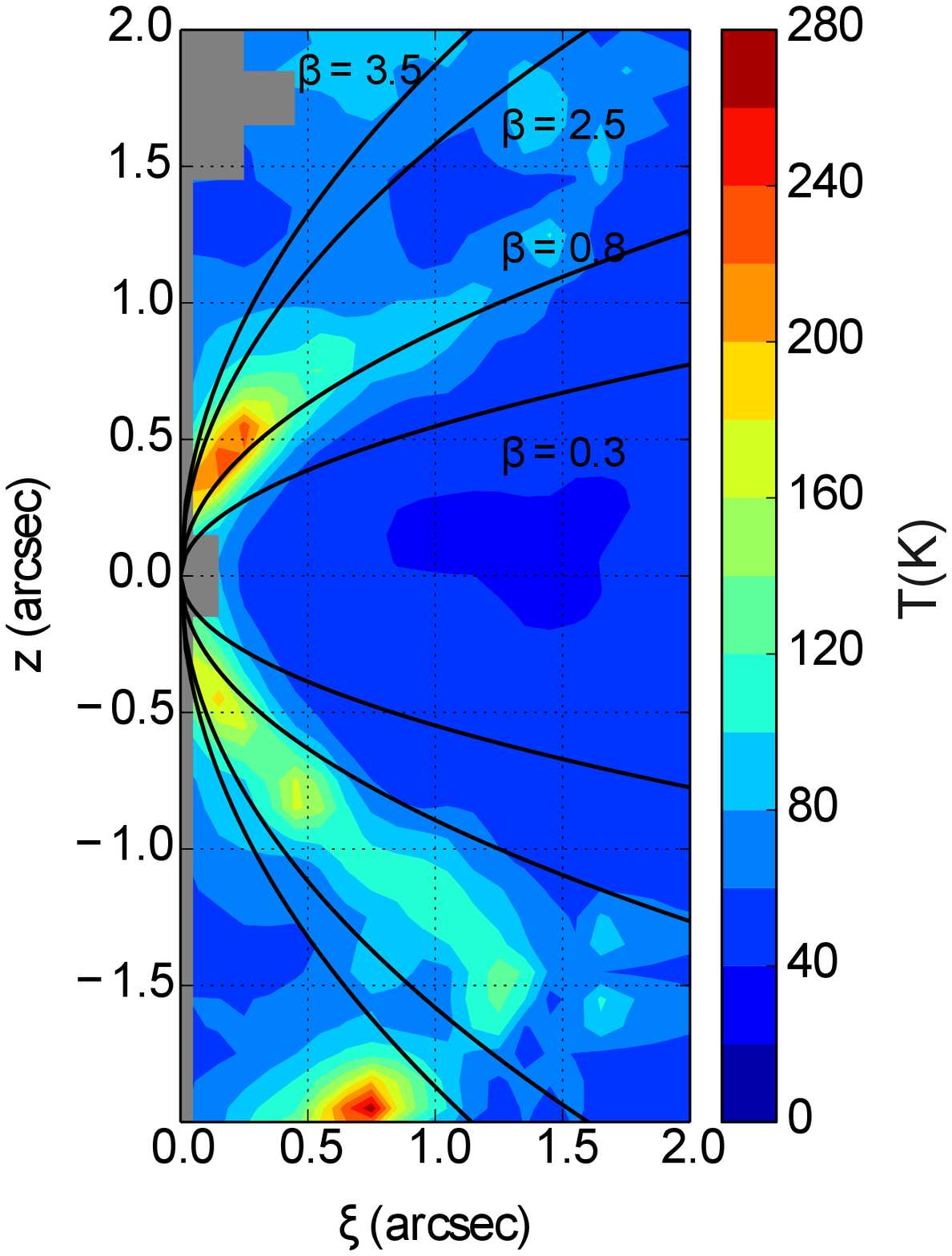}
  \includegraphics[width=.2\linewidth,trim={6.cm -1.6cm 5.2cm 1cm},clip]{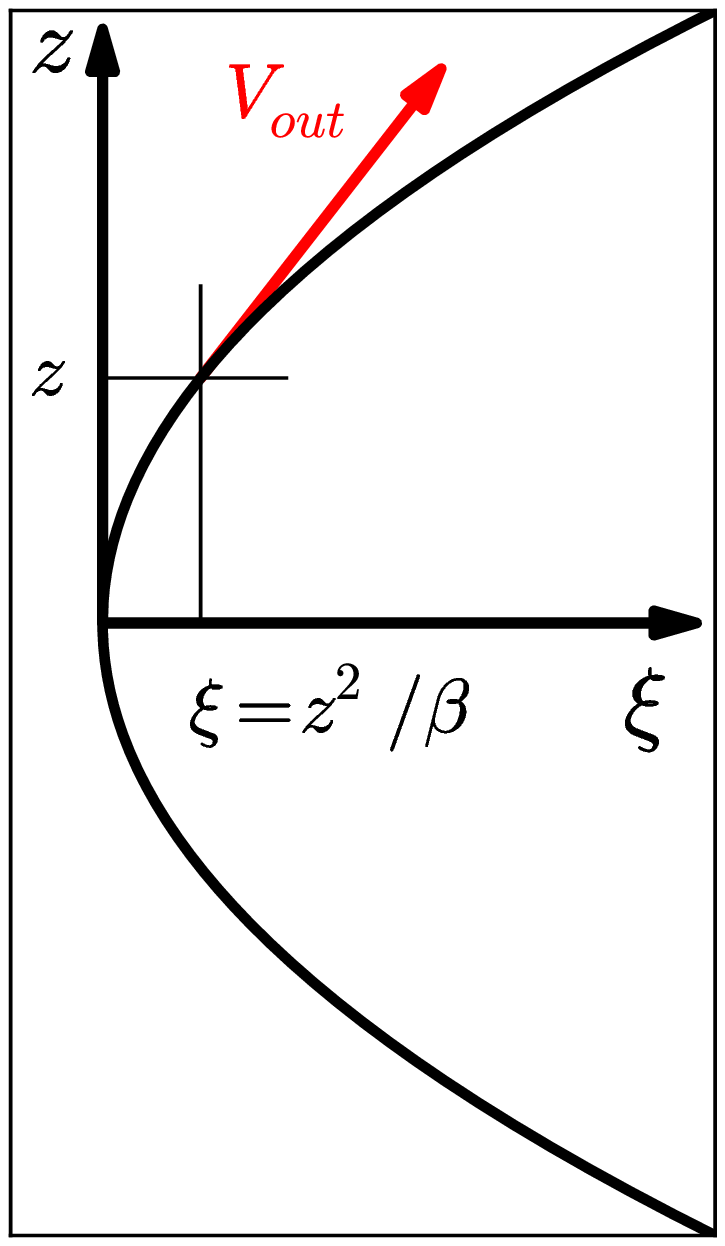}
  \includegraphics[width=.2\linewidth,trim={6.cm -1.6cm 5.2cm 1cm},clip]{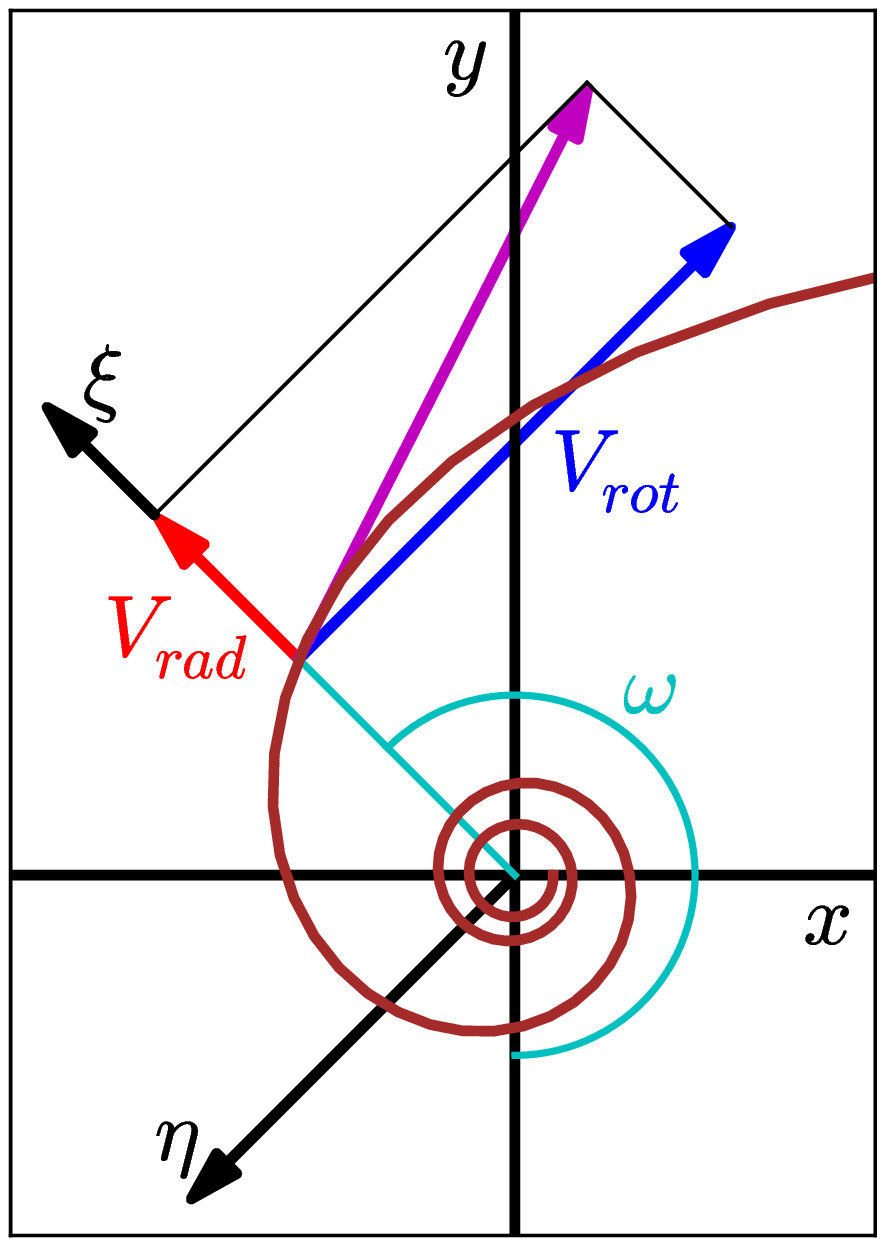}
  \caption{Red Rectangle. Left panels: maps in the half-meridian plane 
of the star of temperatures (extreme left) and CO density (centre left, 
in molecules per cm$^3$) multiplied by $r^2$ (in arcsecond$^2$). 
Right panels: wind velocity projected on the meridian plane 
(centre right) and the equatorial plane (extreme right). 
In the polar region the gas velocity $V_{out}$ is confined 
to meridian planes and tangent to parabolas of equation 
$z^2=\beta \xi$. Right: In the equatorial region the gas 
velocity is confined to planes parallel to the equatorial plane 
and tangent to hyperbolic spirals with a constant radial component 
$V_{rad}$ and a rotation velocity $V_{rot}$ having a power law dependence on $r$.}
  \label{fig11}
\end{center}
\end{figure*}

\subsection{Protostar L1527 IRS}
\label{sec3.5} 

\begin{figure*}
\begin{center}
  \includegraphics[width=.245\linewidth,trim={0cm 0cm 0.4cm 0cm},clip]{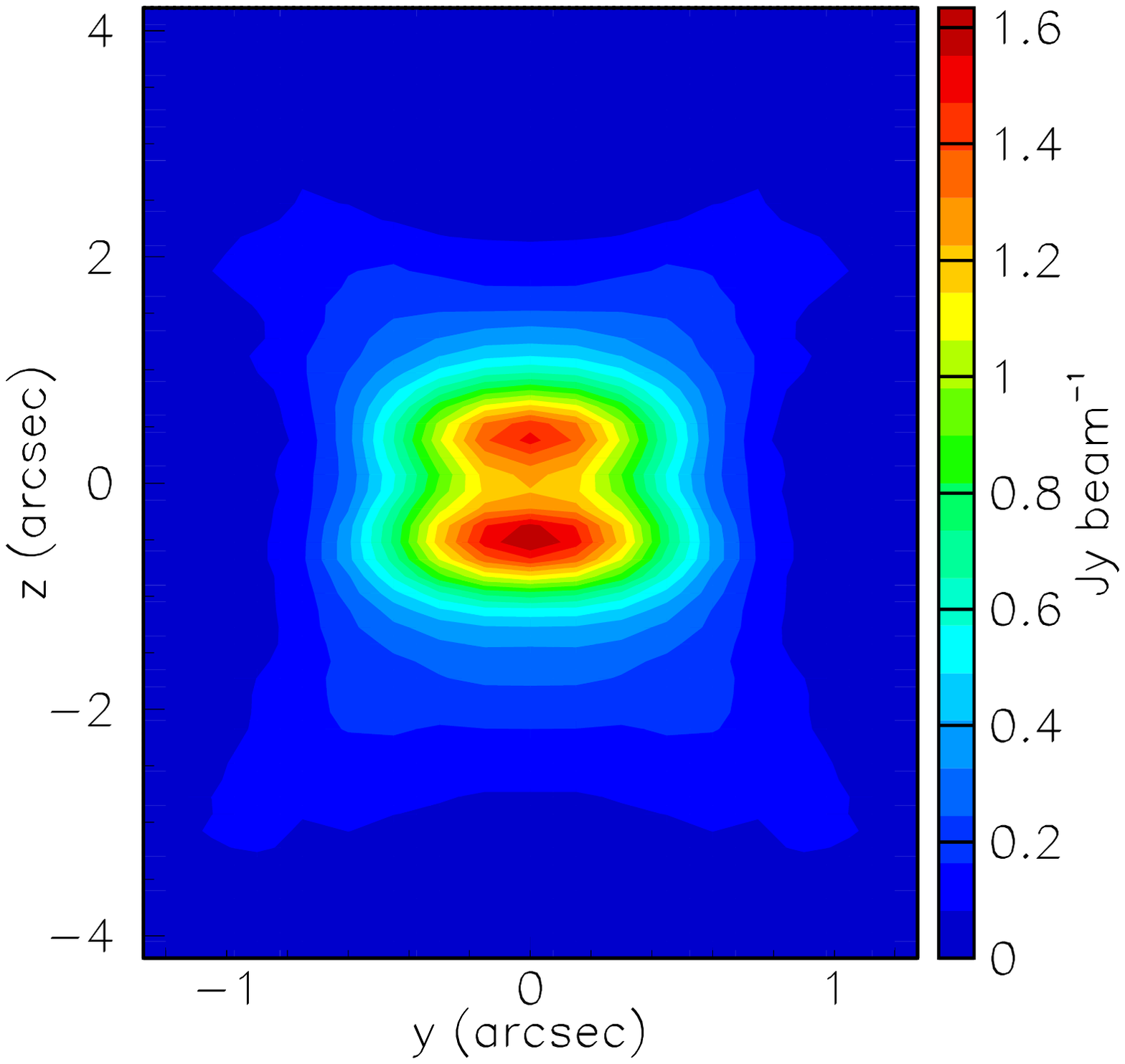}
  \includegraphics[width=.245\linewidth,trim={0.4cm 0cm 0cm 0cm},clip]{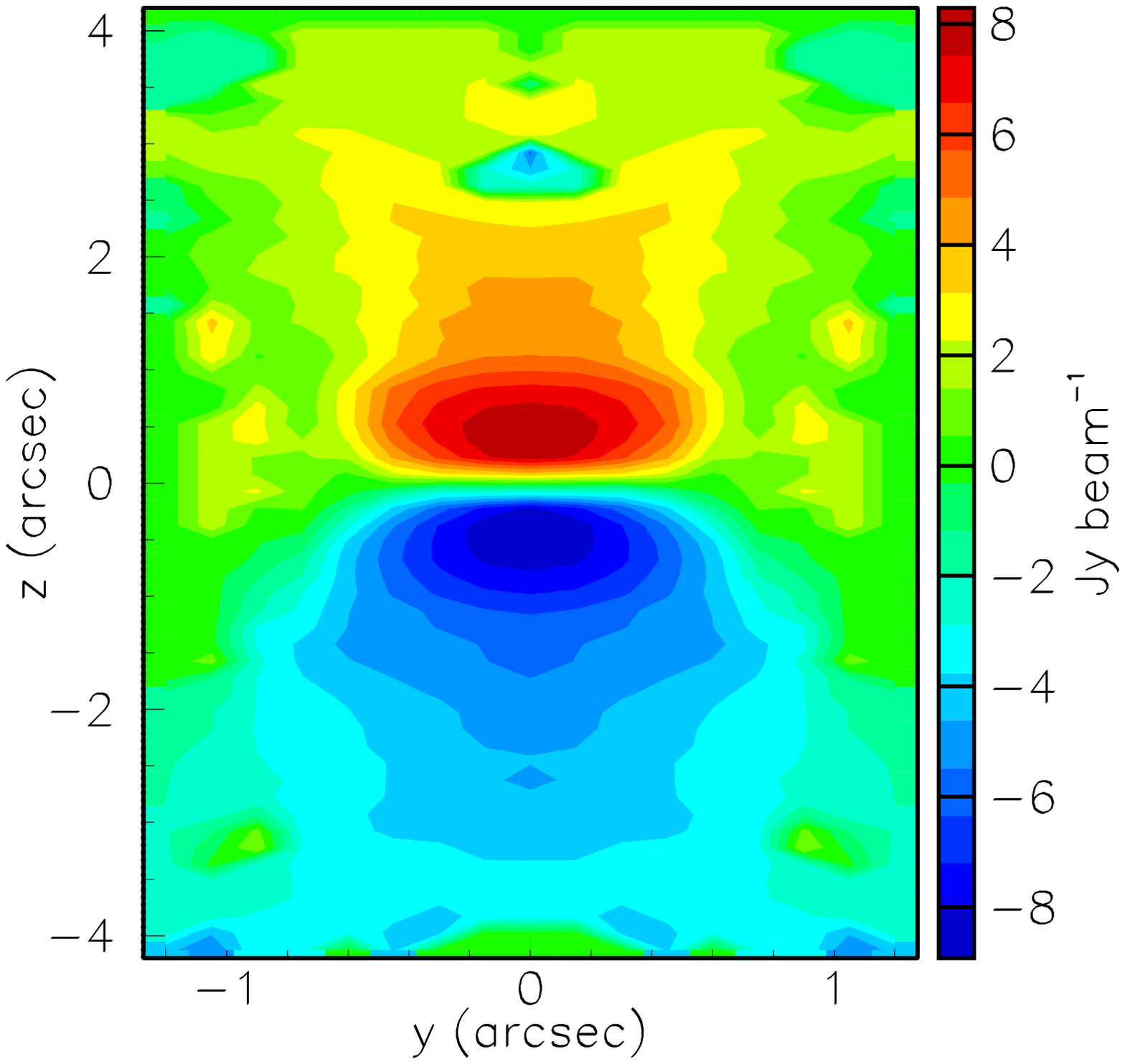}
  \includegraphics[width=.245\linewidth,trim={0.4cm 0cm 0cm 0cm},clip]{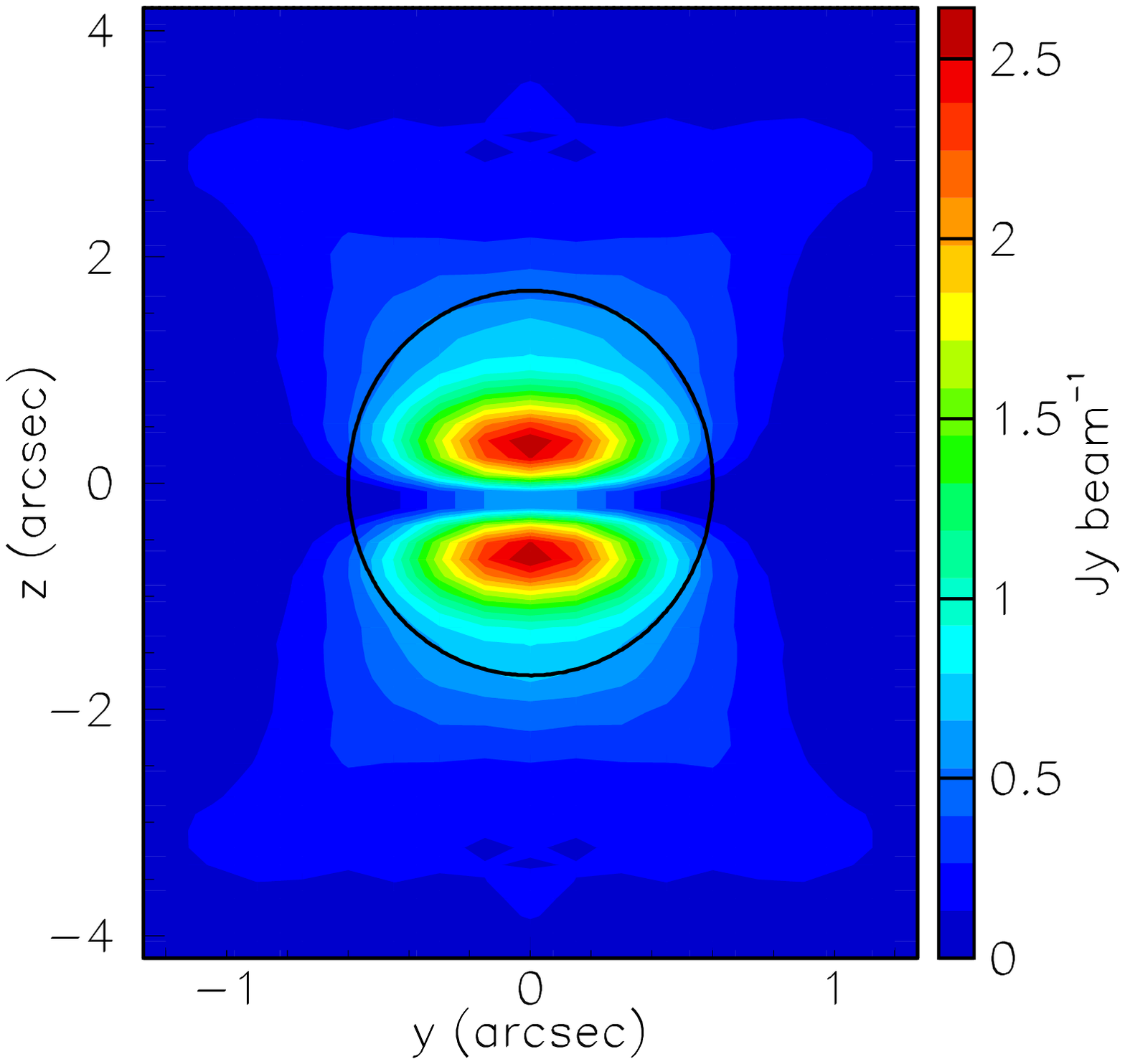}
  \includegraphics[width=.245\linewidth,trim={0.4cm 0cm 0cm 0cm},clip]{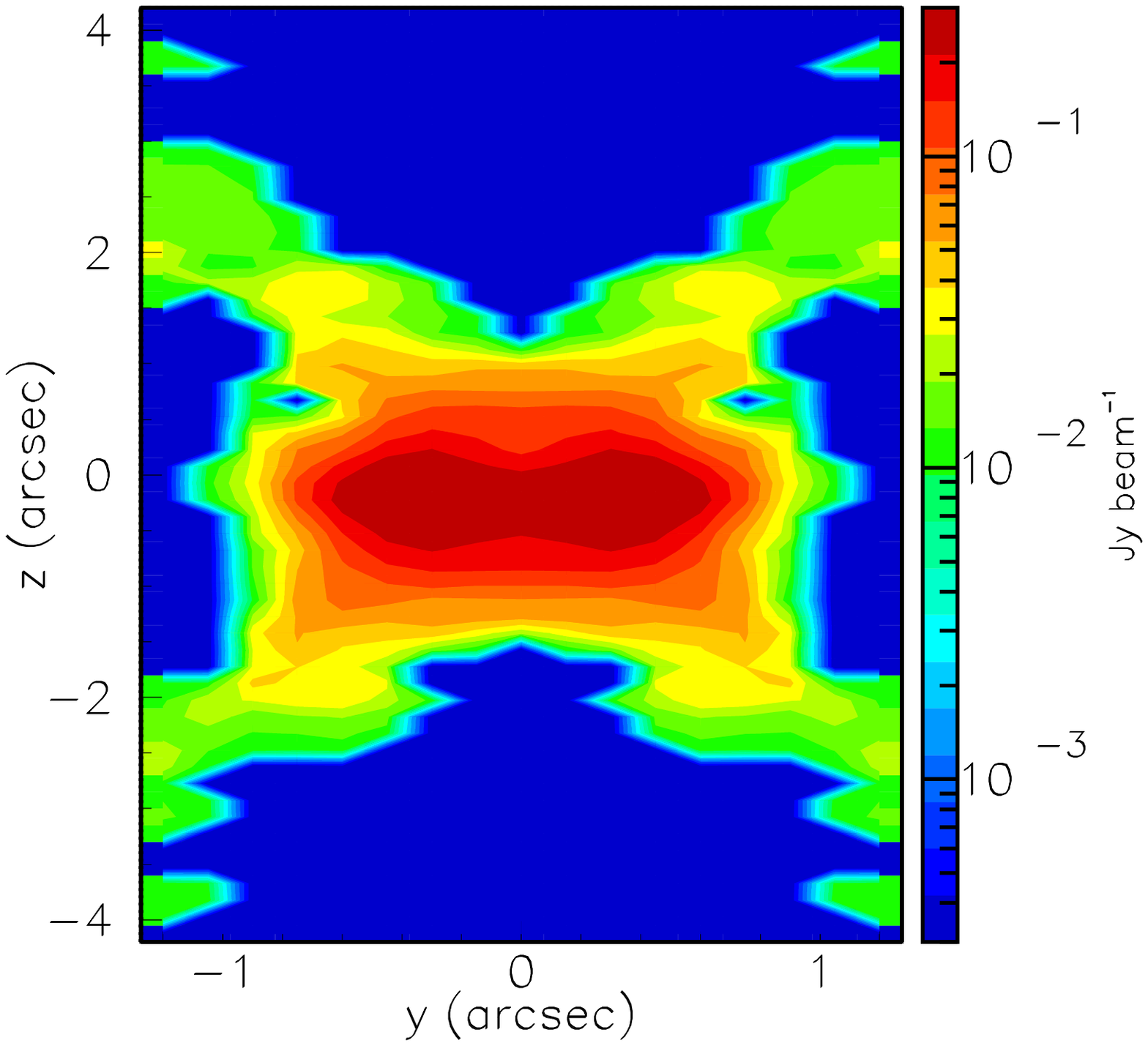}
  \caption{Protostar L1527: Sky maps of $F$ (extreme left panel), of $<V_x>$ 
(centre left panel), of $F_{rot}$ (centre right panel) and $F_{fall}$ (extreme right panel).}
  \label{fig12}
\end{center}
\end{figure*}

Many considerations developed in the present article may in fact have 
applications in a much broader context. We illustrate this point in 
the present sub-section with the example of a young protostar, L1527 IRS. 
It is in its earliest stages of star formation, with a CSE of about 
one solar mass, and has been recently observed by \cite{Tobin2012} 
in the continuum (1.3 and 3.4 mm wavelength) and on the $^{13}$\mbox{CO(2-1)} 
line at the Submillimeter Array (SMA) and the Combined Array for 
Research in Millimeter Astronomy (CARMA). Observations of the \mbox{C$^{18}$O(2-1)} 
line made at ALMA have been reported by \cite{Ohashi2014}. All above observations 
have established the presence of in-falling gas onto a rotating disk surrounding 
the protostar and having its axis close to the $y$ axis. Here, we use ALMA 
observations (2012.1.00647.S) of the \mbox{C$^{18}$O(2-1)} line having a 
slightly better spatial resolution than those used by \cite{Ohashi2014}. 
We refer the reader to a
detailed analysis by \cite{TuanAnh2016}. As the axis of the rotating 
disk is very close to the $y$ axis, we redefine $f(y,z,V_x)$ as  
east-west symmetrized flux densities, $\frac{1}{2}[f(y,z,V_x)+f(-y,z,V_x)]$. 
Figure~\ref{fig12} displays maps of the integrated flux $F$, redefined this 
way, and of the mean Doppler velocity $<V_x>$. They are dominated by rotation, 
red-shifted north and blue-shifted south. In the case of pure rotation about 
the $y$ axis, we expect \mbox{$f(y,z,V_x)=f(y,-z,-V_x)$} to cancel for $zV_x<0$ 
(one side of the $y$ axis to be blue-shifted and the other side red-shifted). 
As a result,
\begin{equation} \label{eq21}
f_{rot}(y,z,V_x)=|f(y,z,V_x)-f(y,-z,V_x)|
\end{equation}
is equal to $f(y,z,V_x)$ everywhere. 

On the contrary, in the case of pure in-fall, \mbox{$f_{rot}(y,z,V_x)=0$}. We may then define 
\begin{equation} \label{eq22}
f_{fall}(y,z,V_x)=f(y,z,V_x)-f_{rot}(y,z,V_x)
\end{equation}
as a measure of the amount of flux density associated with in-fall. Maps of $F_{rot}$ and $F_{fall}$ are displayed in Figure~\ref{fig12}. The map of $F_{rot}$ reveals the shape of the rotating CSE in the northern and southern hemispheres and $F_{fall}$, much smaller than $F_{rot}$, displays a clear enhancement of the in-falling flux along the diagonals of the map, in conformity with what is expected from accretion onto a flared disk. When swapping the $y$ and $z$ axes, the present situation is reminiscent of the Red Rectangle, the difference being that here rotation and in-fall combine everywhere while in the Red Rectangle case rotation and expansion are confined to different stellar latitudes. As in the Red Rectangle case, the effective emissivity can be calculated in space by solving the relevant integral equation and the kinematics can be estimated assuming power law dependences of the rotation and in-fall velocities on $r$. A significant difference between protostars and evolved stars is that the former are embedded from the beginning into dense clouds from which they build their CSE, while the latter build their CSE from their own atmosphere; as a result the observation of protostar CSEs is subject to absorption complications that are absent from the observations of the CSEs of evolved stars. This is the case here where the Doppler velocities nearing the star systemic velocity are completely obscured. A detailed analysis of the 
consequences \citep{TuanAnh2016} is beyond the scope of the present sub-section, the only purpose of which is to illustrate the very strong similarities between the studies of expanding CSEs of evolved stars and in-falling CSEs of protostars.  
\subsection{Multi-line observations}
\label{sec3.6}
Many observations are not limited to a single molecular line, but often include flux densities associated with two or more different lines of a same molecule, for example two rotation lines of carbon dioxide. In such cases, very precious information is carried by the ratio of the measured flux densities. In making such a ratio, the gas density drops out and the emission probabilities are well known: if absorption can be neglected, the ratio of the flux densities is a direct measure of the relative population of the emitting quantum states, namely a direct measure of temperature. In many cases, the distance from the star is large enough for absorption to be small, but in any case the study of the line ratio provides very important information: it measures the temperature when absorption can be neglected and when the gas is in thermal equilibrium and has been in such a state for a long time, but if such hypotheses are not valid, it carries information about the importance and nature of their violation. Making good use of multi-line observations should not be limited to mapping the flux ratio measured in each pixel: in principle, the comparison between the observed lines can -- and should -- be extended to any other quantity that has been used in the analysis.

\subsection{Asymmetries}
\label{sec3.7}
In practice, real stars are often very far from obeying the axial symmetry assumed in the present article. A much more lumpy landscape is often to be expected. As a result, two different issues need to be considered: one is to which extent does this impact on the value and interest of a description that assumes axisymmetry, or even central symmetry; the other is to which extent deviations from the model can be simply described. While the answer to such questions is bound to carry some subjectivity, an essential preliminary consideration is to evaluate quantitatively the importance of the observed deviations.

Independently from any model, deviations from central symmetry can directly be evaluated from the asymmetry parameter 
\begin{equation} \label{eq23}
A_{cntr}(y,z,V_x)=\frac{f(y,z,V_x)-f(-y,-z,-V_x)}{f(y,z,V_x)+f(-y,-z,-V_x)}
\end{equation}

Sky maps and P-V diagrams of this quantity tell both the importance and the nature of the observed deviations: they may be distributed more or less randomly in $y$, $z$ and $V_x$ or be concentrated in well-defined lumps. A systematic study of this asymmetry is an important complement to the studies described in the previous sections.
 
Revealing deviations from rotation invariance about the star axis cannot be done so directly. Partial information is contained in the symmetry -- or lack of -- about the $z$ axis (or $z'$ axis if $\psi$ differs from 0) but extracting the full information requires a detailed analysis of the dependence of $\chi^2$ on the star longitude (which is known in the model but unknown in the data). The outstanding quality of ALMA data allows for evaluating such a $\chi^2$ much more reliably than was previously possible. Ideally, if the quality of the data would allow for a reliable evaluation of measurement uncertainties one could work in the $uv$ plane and evaluate the degree of confidence attached to the fits and the uncertainties and correlation coefficients attached to each of the model parameters in units of standard deviations. In general, however, one is far from such a situation. Even under ideal conditions, a significant lumpiness of the observed fluxes usually prevents a very faithful description, the model being at best a crude approximation of reality. As a result, the best fit values of $\chi^2$ significantly exceed the number of degrees of freedom and one must rely on subjective arguments to obtain a sensible evaluation of the uncertainties attached to each of the model parameters. 

\section{Summary and conclusion}
\label{sec4}
The aim of the present work is to help constructing 3-D models of the CSE of 
axisymmetric radio sources, essentially evolved stars or protostars.
The accent is on the kinematics, central to the
astrophysics of AGB stars, more than on the morphology, for
which exists an abundant literature, in particular in the domain of
the deprojection of optical images of Planetary Nebulae \citep[][and references therein]{Steffen2011,Wenger2013}.
It does not have the pretension to give recipes, not even strict guidelines, 
but simply to illustrate possibly more quantitative approaches to data
analysis than often used in the current literature. Each star is a particular 
case: before starting modelling its CSE, one must first take into account the 
information available, optical, infrared or radio, and the physics considerations 
that are of relevance. In the present section, we summarize the results obtained 
in the preceding sections.

A number of preliminary points should be systematically examined before undertaking 
any modelling: they imply having a first close look at the data without requiring 
the help of a model. In particular the sky maps of $RF(y,z)$, of $<V_x>$ and of 
$<|V_x|>$ and their projections on the $R$ and $\varphi$ axes provide essential 
information. In addition to obtaining a global evaluation of the steepness of 
the decrease with $r$ of the effective emissivity, one learns from them to which 
extent central symmetry is obeyed; whether there is any evidence for the effective 
emissivity to depart from spherical symmetry; whether axial symmetry is revealed 
on the $<V_x>$ map; whether there is evidence for significant velocity gradients.  
 
In a spirit of minimizing qualitative arguments and quantifying as much as possible 
the results of the analysis, use should be made of simple quantities that carry 
important information. Examples have been given, such as $\chi^2_\psi$ and $A(\varphi)$, 
which may help the evaluations of the orientation of the star axis in space and of the 
relative contributions of expansion along the star axis and rotation about it. 
They are not unique, other indicators and discriminators may be found better 
adapted to such and such a particular case. 

Particular attention must be given to correlations between model parameters, 
such as between the inclination of the star axis on the line of sight, $\theta$, 
and the elongation of the effective emissivity and/or wind velocity distribution 
along the star axis. If a $\chi^2$ minimization method is used to adjust the model 
parameters, one should look at the dependence of $\chi^2$ on $\theta$, 
leaving all other parameters free to vary. 

As well known from the practice of such studies, P-V diagrams carry very 
precious information. However, we have argued that using polar rather than 
Cartesian coordinates in the sky plane is often better suited to making the 
best of them. In particular, the $\varphi$ vs $V_x$ diagrams have the 
advantage of being simply translated along the $\varphi$ axis when $\psi$ 
is varied, a useful feature when $\psi$ is not known with precision, and 
their symmetry properties tell about the relative purity -- or lack of -- 
of  expansion vs rotation. One must keep in mind that when expansion and 
rotation contribute significantly to the wind velocity in a same region 
of space, causing spiralling, the determination of the star axis is difficult. 
However, when expansion and rotation contribute significantly in different 
regions of space, this comment does not apply and their contributions to 
the P-V diagrams are simply superimposed. 

The above arguments, as important as they may be, are not sufficient to construct a model. Physics considerations must also be taken in due account. While a precise hydrodynamic description is well beyond the scope of the present considerations, some of its most obvious features must be kept in mind when deciding on the form given to the effective emissivity and velocities. In particular the validity of the stationarity hypothesis needs to be critically discussed.

The concrete examples that have been given as illustrations of the above arguments have contributed new results to earlier analyses of the CO emission of the CSE of AGB stars. The study of RS Cnc has given an opportunity to comment on the hypothesis of stationarity of the wind regime and on a possible rotation of the equatorial volume about the star axis. In both cases, limits have been given to their possible contributions. In the case of EP Aqr, a picture has been drawn that unifies the presentation of the two models that had been proposed earlier, one with isotropic but $r$-dependent winds, the other with a bipolar outflow. Evidence has been given in favour of the latter. Results obtained in the case of the post-AGB Red Rectangle have been briefly recalled. Using Mira Ceti as an example, we have commented about the value of constructing, at least locally, a 3-D representation under the hypothesis of radial expansion at constant velocity. Such a representation is particularly useful to visualise the topology of the effective emissivity and reveal the presence of possible detached arcs, cavities, or isolated lumps. Finally, the example of a protostar, L1527, has been used to underline very strong similarities between the expanding CSEs of evolved stars and the in-falling CSEs of protostars.

\section*{Acknowledgements}
We are deeply indebted to the anonymous referee for many
comments and suggestions that helped considerably improving the
quality of the manuscript. We are very grateful to the ALMA partnership, who are making their data available to the public after a one year period of exclusive property, an initiative that means invaluable support and encouragement for Vietnamese astrophysics. We particularly acknowledge friendly and patient support from the staff of the ALMA Helpdesk. This paper makes use of the following ALMA data: ADS/JAO.ALMA\# 2011.0.00223.S, 2012.1.00524.S, 2013.1.00047.S, 2012.1.00647.S. ALMA is a partnership of ESO (representing its member states), NSF (USA) and NINS (Japan), together with NRC (Canada), NSC and ASIAA (Taiwan), and KASI (Republic of Korea), in cooperation with the Republic of Chile. The Joint ALMA Observatory is operated by ESO, AUI/NRAO and NAOJ. The paper makes also use of observations carried out with the IRAM Plateau de Bure Interferometer and the IRAM 30-m telescope. IRAM is supported by INSU/CNRS (France), MPG (Germany) and IGN (Spain). We express our deepest gratitude to Professors Anne Dutrey, Stephane Guilloteau and Thibaut Le Bertre for the support, the interest in our work, and for very useful comments on the present manuscript. Financial support is acknowledged from VNSC/VAST, the NAFOSTED funding agency under grant number 103.99-2015.39, the World Laboratory, the Odon Vallet Foundation and the Rencontres du Viet Nam.


\begin{thebibliography}{99}
\bibitem[\protect\citeauthoryear{Decin et al.}{2015}]{Decin2015}
Decin, L., et al. 2015, A\&A, 574, A5
\bibitem[\protect\citeauthoryear{Hoai et al.}{2014}]{Hoai2014}
Hoai, D. T., et al. 2014, A\&A, 565, A54
\bibitem[\protect\citeauthoryear{Ihrke et al.}{2008}]{Ihrke2008}
Ihrke, I., Kutulakos, K., Lensch, H., Magnor, M. \& Heidrich, W.
2008, Proc. Eurographics EG '08 Annex, pp. 87-108
\bibitem[\protect\citeauthoryear{Leahy}{1991}]{Leahy1991}
Leahy, D.A. 1991, A\&A, 247, 584
\bibitem[\protect\citeauthoryear{Lintu et al.}{2007}]{Lintu2007}
Lintu, A., Lensch, H.P.A., Magnor, M., El-Abed, S. \& Seidel, H.-
P. 2007, Proc. IEEE/EG Int'l Symp. Volume Graphics, Hege \&
Machiraju eds., 9
\bibitem[\protect\citeauthoryear{Magnor et al.}{2004}]{Magnor2004}
Magnor, M., Kindlmann, G., Hansen, C. \& Duric, N. 2004, Proc.
IEEE Visualization Conf., 83
\bibitem[\protect\citeauthoryear{Magnor et al.}{2005}]{Magnor2005}
Magnor, M., Kindlmann, G., Hansen, C. \& N. Duric, N. 2005,
IEEE Trans. Visualization and Computer Graphics, 11/5, 485
\bibitem[\protect\citeauthoryear{Nhung et al.}{2015a}]{Nhung2015a}
Nhung, P. T., et al. 2015a, RAA, 15, 5
\bibitem[\protect\citeauthoryear{Nhung et al.}{2015b}]{Nhung2015b}
Nhung, P. T., et al. 2015b, A\&A, 583, A64 
\bibitem[\protect\citeauthoryear{Ohashi et al.}{2014}]{Ohashi2014}
Ohashi, N., et al., 2014, ApJ, 796, 131
\bibitem[\protect\citeauthoryear{Palmer}{1994}]{Palmer1994}
Palmer, P.L. 1994, MNRAS, 266, 697
\bibitem[\protect\citeauthoryear{Sabbadin}{1984}]{Sabbadin1984}
Sabbadin, F. 1984, MNRAS, 210, 341
\bibitem[\protect\citeauthoryear{Sabbadin et al.}{2000}]{Sabbadin2000}
Sabbadin, F., Cappellaro, E., Benetti, S. et al. 2000, A\&A, 355
\bibitem[\protect\citeauthoryear{Steffen et al.}{2011}]{Steffen2011}
Steffen, W., Koning, N., Wenger, S., Morisset, C. \& Magnor, M.
2011, IEEE Transactions on Visualization \& Computer Graphics,
17/4, 454
\bibitem[\protect\citeauthoryear{Tobin et al.}{2012}]{Tobin2012} 
Tobin, J.J., et al. 2012, Nature, 492/7427, 83
\bibitem[\protect\citeauthoryear{Tuan-Anh et al.}{2015}]{TuanAnh2015} 
Tuan-Anh, P., et al. 2015, RAA, Vol.15, No.12, 2213
\bibitem[\protect\citeauthoryear{Tuan-Anh et al.}{2016}]{TuanAnh2016} 
Tuan-Anh, P., et al. 2016, submitted to MNRAS, arXiv:1604.03801
\bibitem[\protect\citeauthoryear{Wenger et al.}{2013}]{Wenger2013} 
Wenger, S., Lorenz, D. \& Magnor, M. 2013,
Computer Graphics Forum (Proc. of Pacific Graphics PG), 32/7, 93
\bibitem[\protect\citeauthoryear{Winters et al. }{2007}]{Winters2007} 
Winters, J. M., et al. 2007, A\&A, 475, 559
\end{thebibliography}

\appendix
\section{}
\subsection{Transformation relations between space and star coordinates}
\label{seca1}
The star axis projects on $z'$, the transformed of $z$, on the sky $(y,z)$ plane. In $(x,y,z)$ coordinates, the star axis is a vector $(\cos\theta, \sin\theta \sin\psi, \sin\theta \cos\psi)$. For $\psi=0$, the star axis projects on $z$ on the sky plane, for $\theta=0$ it is along $x$. The $(x,y,z)$ to $(\xi,\eta,\zeta)$ transformation relations read:

\begin{equation}\label{aeq1}
\begin{split}
&\xi=x\cos\theta+z\sin\theta \cos\psi-y\sin\theta \sin\psi\\
&\eta=z\sin\psi+y\cos\psi\\
&\zeta=-x\sin\theta+z\cos\theta \cos\psi-y\cos\theta \sin\psi
\end{split} 
\end{equation}

We define $r=\sqrt{x^2+y^2+z^2}$ and $R=\sqrt{y^2+z^2}$, \mbox{$\partial r/\partial x=x/r$} and write $y=R\cos\varphi$  and $z=R\sin\varphi$. A point in the star frame having latitude $\alpha$ (positive along $\xi$) and longitude $\omega$ is defined as $(\xi,\eta,\zeta)=r(\sin\alpha,\cos\alpha \cos\omega,\cos\alpha \sin\omega)$ with $-90^\circ<\alpha<90^\circ$ and $-180^\circ<\omega<180^\circ$. On the equator the zero of longitude $(\omega=0)$ is along $\eta=y'$.

Hence the transformation relations between $(x,y,z)$ and $(r,\alpha,\omega)$ read:

\begin{equation}\label{aeq2}
\begin{split}
&x=r(\sin\alpha \cos\theta-\cos\alpha \sin\omega \sin\theta)\\
&R\cos(\varphi-\psi)=r\cos\alpha \cos\omega\\
&R\sin(\varphi-\psi)=r(\sin\alpha \sin\theta+\cos\alpha \sin\omega \cos\theta)\\
&\omega=\text{atan}[-\frac{x}{R}\sin \theta+\sin (\varphi-\psi)\cos \theta,\cos (\varphi-\psi)]\\
&\alpha=\text{asin}\{[x\cos\theta+R\sin(\varphi-\psi)\sin\theta]/\sqrt{x^2+R^2}\}\\
&r=\sqrt{R^2+x^2}
\end{split}
\end{equation}

Setting $\psi=0$, a point $(\xi,\eta,\zeta)=(b, a \times \cos\omega, a\times \sin\omega)$, with $a$ and $b$ constant, projects on $(y,z)$ in the sky plane. In the meridian plane of the star, it is a fixed point at \mbox{$(a=r\cos\alpha, b=r\sin\alpha)$}. In the sky plane
\begin{equation} \label{aeq3}
(\frac{y}{a})^2+(\frac{z-b\sin\theta}{a\times \cos \theta})^2=1
\,\,\textnormal{and}\,\,\,
a^2-(\frac{z-b\sin\theta}{\cos\theta})^2=y^2
\end{equation}

\begin{figure*}
\begin{center}
  \includegraphics[width=.7\linewidth,trim={0cm 0cm 0cm 0cm},clip]{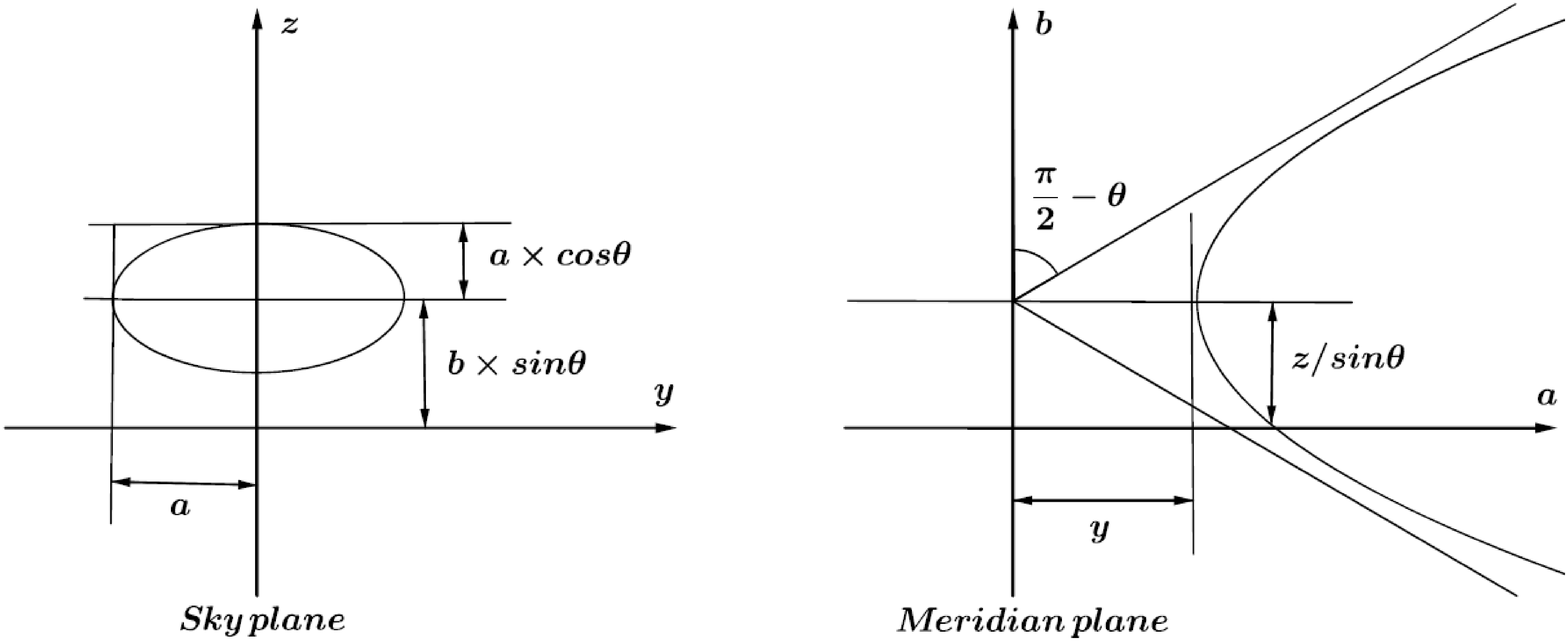}
  \caption{Left: Ellipse spanned in the sky plane by a fixed point $(a,b)$ in the meridian plane. Right: hyperbola spanned in the meridian plane by a fixed point $(y,z)$ in the sky plane.}
  \label{figa1}
\end{center}
\end{figure*}

Namely pixel $(y,z)$ spans an hyperbola in the meridian plane of the star, while the point $(a,b)$ spans an ellipse in the sky plane (Figure~\ref{figa1}).

\subsection{Effective emissivity, wind velocity and their symmetries}
\label{seca2}
Under the hypothesis of rotation invariance about the star axis, $V_{rad}$, $V_{ax}$ and $V_{rot}$ are functions of $r$ and $\sin\alpha$ exclusively, independent of $\omega$. Moreover, under the additional hypothesis of central symmetry, or equivalently of reflexion symmetry about the equatorial plane, $V_{rad}$ and $V_{rot}$ are even functions and $V_{ax}$ an odd function of $\sin\alpha$. It is useful to define two symmetry operations: $S_1$ changing $x$ in $-x$ and $z$ in $-z$ but leaving $y$ invariant and $S_2$ changing $y$ in $-y$ but leaving $x$ and $z$ invariant. Table~\ref{tablea1} summarizes symmetry properties of several quantities assuming rotational invariance about the star axis, central symmetry (or equivalently north-south symmetry with respect to the equator of the star) and $\psi=0$. One may then rewrite the Doppler velocity as the sum of an expansion and a rotation components: $V_x=V_{x1}+V_{x2}$ with 
\begin{equation} \label{aeq4}
\begin{split}
&V_{x1}=(x/r)V_{rad}+\cos\theta V_{ax} \\
&V_{x2}=-\frac{R\cos(\varphi-\psi)\sin\theta}{r\cos\alpha}V_{rot}
\end{split}
\end{equation}

One sees from Table~\ref{figa1} that, for $\psi=0$, $V_{x1}$ is odd under $S_1$ and even under $S_2$ while $V_{x2}$ is even under $S_1$ and odd under $S_2$. However, when the value of $\psi$ differs from zero, this is no longer true and maps of $<V_x>$, $<V_{x}^2>$ or, for that matter, any other moment, do not reveal the position of the star axis: their symmetry properties depend on both the value of $\psi$ and the relative importance of rotation and expansion.

\begin{table}
\begin{center}
\caption{Symmetry properties of several quantities under transformations $S_1$ and $S_2$ for $\psi=0$ under the hypothesis of rotational invariance and central symmetry in the star frame.}
\begin{tabular}{|c|c|c|c|}
  \hline
  \hline
  & $S_1$ & $S_2$ & $S_1 \times S_2$\\
  \hline
  $x$&$-x$&$x$&$-x$\\
  $y$&$y$&$-y$&$-y$\\
  $z$&$-z$&$z$&$-z$\\
  $r$&$r$&$r$&$r$\\
  $R$&$R$&$R$&$R$\\
  $\varphi$&$-\varphi$&$180^\circ-\varphi$&$180^\circ+\varphi$\\
  $\alpha$&$-\alpha$&$\alpha$&$-\alpha$\\
  $\rho$&$\rho$&$\rho$&$\rho$\\
  $V_{rad}$&$V_{rad}$&$V_{rad}$&$V_{rad}$\\
  $V_{ax}$&-$V_{ax}$&$V_{ax}$&-$V_{ax}$\\
  $V_{rot}$&$V_{rot}$&$V_{rot}$&$V_{rot}$\\
  $V_{x1}$&$-V_{x1}$&$V_{x1}$&$-V_{x1}$\\
  $V_{x2}$&$V_{x2}$&$-V_{x2}$&$-V_{x2}$\\
  \hline
  \hline
\end{tabular}
\label{tablea1}
\end{center}
\end{table}

\subsection{P-V diagrams}
\label{seca3}
\begin{figure*}
\begin{center}
  \includegraphics[width=.5\linewidth,trim={0cm 2cm 0cm 3.5cm},clip]{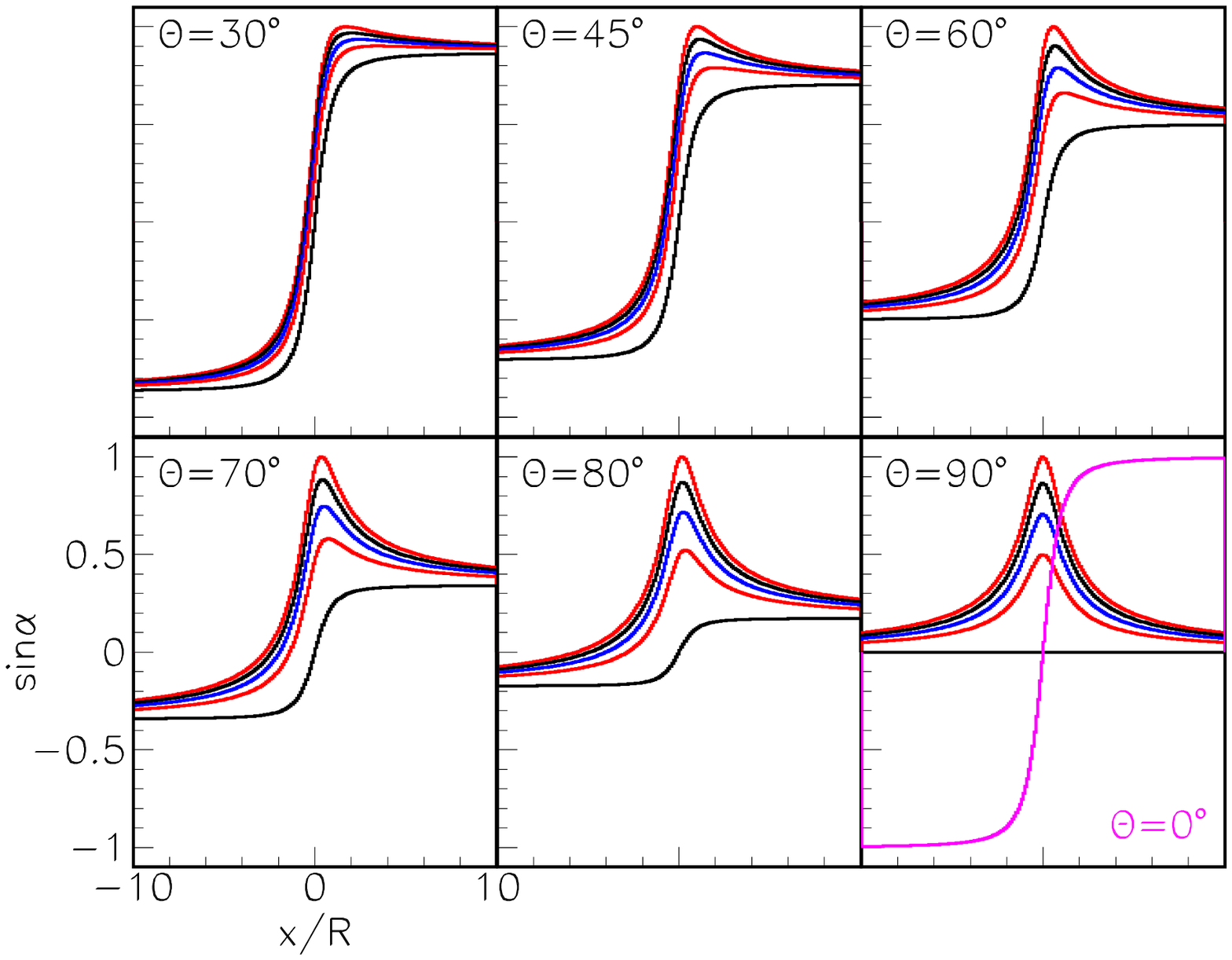}
  \caption{Distributions of $\sin\alpha=(u\cos\theta+\sin\varphi \sin\theta)/\sqrt{1+u^2}$ vs $u=x/R$. Each panel is for a fixed value of $\theta$ and in each panel, from down to up, $\varphi$ varies as $0^\circ$, $30^\circ$, $45^\circ$, $60^\circ$ and $90^\circ$. The last panel also includes the case $\theta=0^\circ$ for which $\sin\alpha$ is independent from $\varphi$. Values of $\sin\alpha$ for $\sin\varphi<0$ are obtained by symmetry about the origin.}
  \label{figa2}
\end{center}
\end{figure*}

Setting $\psi=0$, from the coordinate transformation relations between the star frame and the sky frame we obtain an important relation (Figure~\ref{figa2}) between the star latitude $\alpha$ and $x$,
\begin{equation}\label{eq4}
r\sin\alpha=x\cos\theta+z\sin\theta
\end{equation}

For a given pixel, once we know $x$, we also know \mbox{$r=\sqrt{x^2+R^2}$} and $\alpha$ 
from the above relation: we can calculate the effective density and the gas velocity, 
and therefore the flux density $f(y,z,V_x)=\rho (x,y,z)dx/dV_x$. 
The dependence of the Doppler velocity on $x$ is given by the relation 
$rV_x=xV_{rad}-y\sin\theta(\cos\alpha)^{-1}V_{rot}$ where $V_{ax}$ has been set to zero.

Figure~\ref{figa2} displays a number of interesting features that are worth a mention. When $x \rightarrow -\infty∞$, $\sin\alpha \rightarrow -\cos\theta$; then, for $\sin\varphi>0$, when $x$ increases, so does $\sin\alpha$ until it reaches 0 for $u=-\sin\varphi \tan\theta$. Then, it keeps increasing and crosses the value $\cos\theta$ when \mbox{$u=(\cos2\theta+\sin^2\theta \cos^2\varphi)/(\sin2\theta \sin\varphi)$}. For $u=\sin\varphi/\tan\theta$, it reaches its maximum, $\sin\varphi/\sqrt{1-\cos^2\theta \cos^2\varphi}$. In particular, when $\sin\varphi=\pm 1$ ($z$ axis) it reaches the poles \mbox{$(\sin\alpha=\pm 1)$} at $u=\pm 1/\tan\theta$. It then decreases slowly and for \mbox{$x \rightarrow +\infty$}, $\sin\alpha \rightarrow \cos\theta$. For $\theta=0$, $\sin\alpha =u/\sqrt{1+u^2}$ is independent of $\varphi$. For $\theta =90^\circ$, $\sin\alpha =\sin\varphi /\sqrt{1+u^2}$. For $\theta =90^\circ$, \mbox{$\sin\alpha =(u+\sin\varphi )/\sqrt{2(1+u^2)}$}. For $u=0$, $\sin\alpha =\sin\varphi \sin\theta$. For $\sin\varphi =0$, $\sin\alpha =u \cos\theta/\sqrt{1+u^2}$ and for $\sin\varphi =\pm 1$, $\sin\alpha =(u\cos\theta \pm \sin\theta )/\sqrt{1+u^2}$.

\subsection{Velocity spectra}
\label{seca4}
In the case of pure rotation, setting $V_{rot}=V_0\cos\alpha $, the spectral distribution 
takes a particularly simple form that illustrates well what happens: 
\begin{equation} \label{aeq5}
dN/dV_x= r^2R^{-1}\rho (r)[(\cos\varphi \sin\theta V_0)^2-V_x^2]^{-1/2}
\end{equation}

Namely the spectral distribution $dN/dV_x$ increases from its value at $V_x=0$ 
to infinity at the value $V_{max}=\cos\varphi \sin\theta V_0$, with 
\begin{equation} \label{aeq6}
dN/dV_x \sim(V_{max}^2-V_x^2)^{-1/2} \sim (2V_{max})^{-1/2}(V_{max}-V_x)^{-1/2}
\end{equation}

For case $A$, the same qualitative behaviour is obtained: the main difference is that $V_{rot}$ being now smaller in the equatorial region, the amplitude of the $\varphi$ oscillations is also smaller. Finally, in case $B$, one is either inside the cone, in which case $V_{rot}=0$, or outside, in which case $V_{rot}=V_0$ and 
\begin{equation} \label{aeq7}
V_x=-\frac{R\cos\varphi \sin\theta}{r\cos\alpha} V_0
\end{equation}

Outside the cone, $|V_x|$ reaches its maximal values $V_0$ for $|y|\sin\theta =r\cos\alpha $ and, as $\cos\alpha > 1/\sqrt{2}$, $R |\cos\varphi |\sin\theta/r>1/\sqrt{2}$, namely $r=R$ $(x=0)$ and $\cos^2\varphi \sin^2\theta >1/2$. Indeed, for pure rotation in case $B$, the $\varphi$ oscillation of the extreme $V_x$ values stays flat at maximum as long as $\cos\varphi >1/(\sqrt{2} \sin\theta)$. In particular, at $\theta =90^\circ$, this happens over $\pm 45^\circ$ around $\varphi=0$ ($mod\, 180^\circ$). A flat can only be seen if $\sin^2\theta >1/2$, meaning $\theta>45^\circ$.

\begin{figure*}
\begin{center}
  \includegraphics[width=.7\linewidth,trim={0cm 0cm 0cm 0cm},clip]{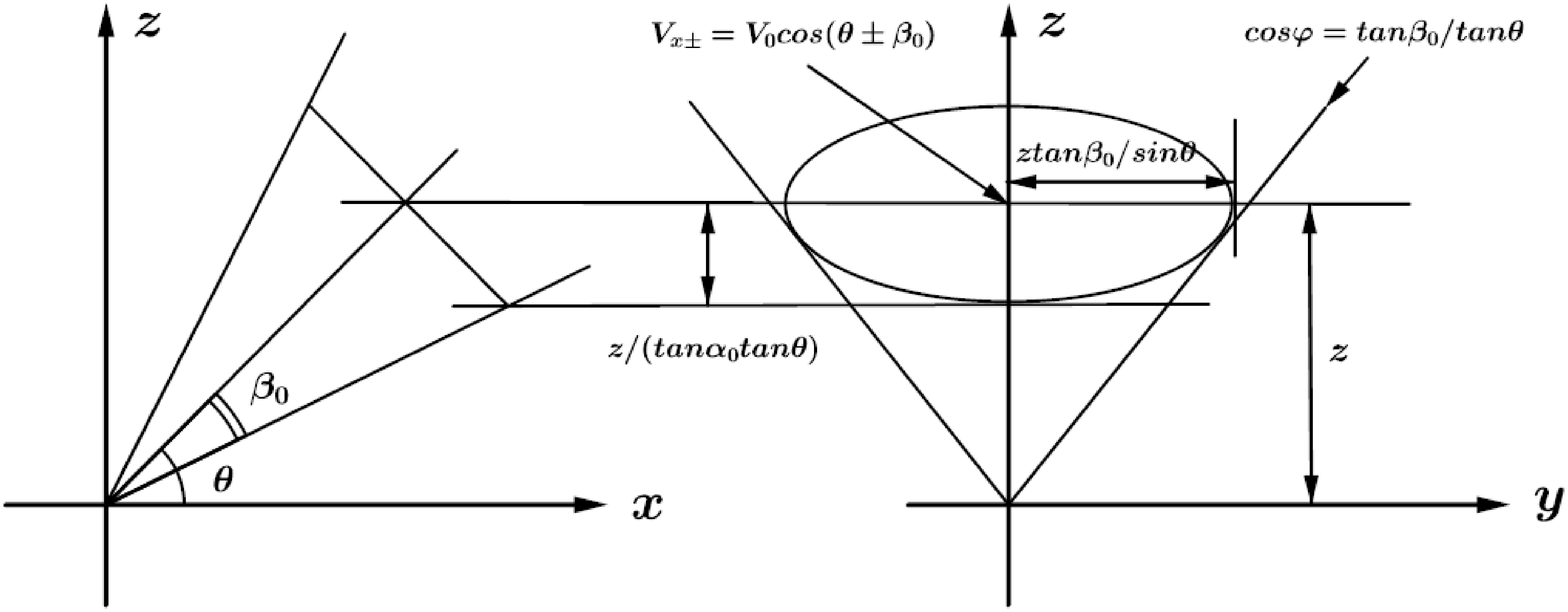}
  \caption{Conical bipolar outflow projected on the $(x,z)$ plane (left) and the sky plane (right).}
  \label{figa3}
\end{center}
\end{figure*}

In the case of pure expansion, when the bipolar outflow is confined inside a cone of latitude $\alpha_0$, nothing changes inside the cone but the gas velocity, and therefore $V_x$, cancel outside the cone. The cone splits space in two regions, its interior and its exterior. On the sky plane, the limit is given by the relation $\cos\varphi_{lim}=\tan\beta_0/ \tan\theta$ with $\beta_0=90^\circ-\alpha_0$ measuring the aperture of the cone. Note that any value of $\varphi$ can be associated with a point outside the cone while only values of $\varphi>\varphi_{lim}$ can be associated with regions inside the cone. For $\varphi>\varphi_{lim}$ one has therefore $V_x=0$. This happens only if $\cos\varphi_{lim}<1$, namely $\theta >\beta_0$. For smaller values of $\theta$ the interior of the cone covers the full sky plane. In general (Figure~\ref{figa3}) the spectrum retains therefore the general form $(V_0^2-V_x^2)^{-3/2}$ but is confined to an interval smaller than the maximum interval $[-V_0,V_0]$.

\subsection{Evaluating the value of $\psi$}
\label{seca5}
We define a quantity $\chi^2_{\psi}$ as $\chi^2_{\psi}=\chi^2_{\psi exp}+\chi^2_{\psi rot}$ where

\begin{equation} \label{aeq9}
\begin{split}
\chi^2_{\psi exp}=\sum_{exp}\{&[f(y,z,V_x)-f(y,0,V_x)]\\
&-[f(y,-z,-V_x)-f(y,0,-V_x)]\}^2/\Delta^2\\
\chi^2_{\psi rot}=\sum_{rot}\{&[f(y,z,V_x)-f(0,z,V_x)]\\
&-[f(-y,z,-V_x)-f(0,z,-V_x)]\}^2/\Delta^2.\\
\end{split}
\end{equation}

\begin{figure*}
\begin{center}
  \includegraphics[width=.33\linewidth,trim={0cm 1cm 1.5cm 1cm},clip]{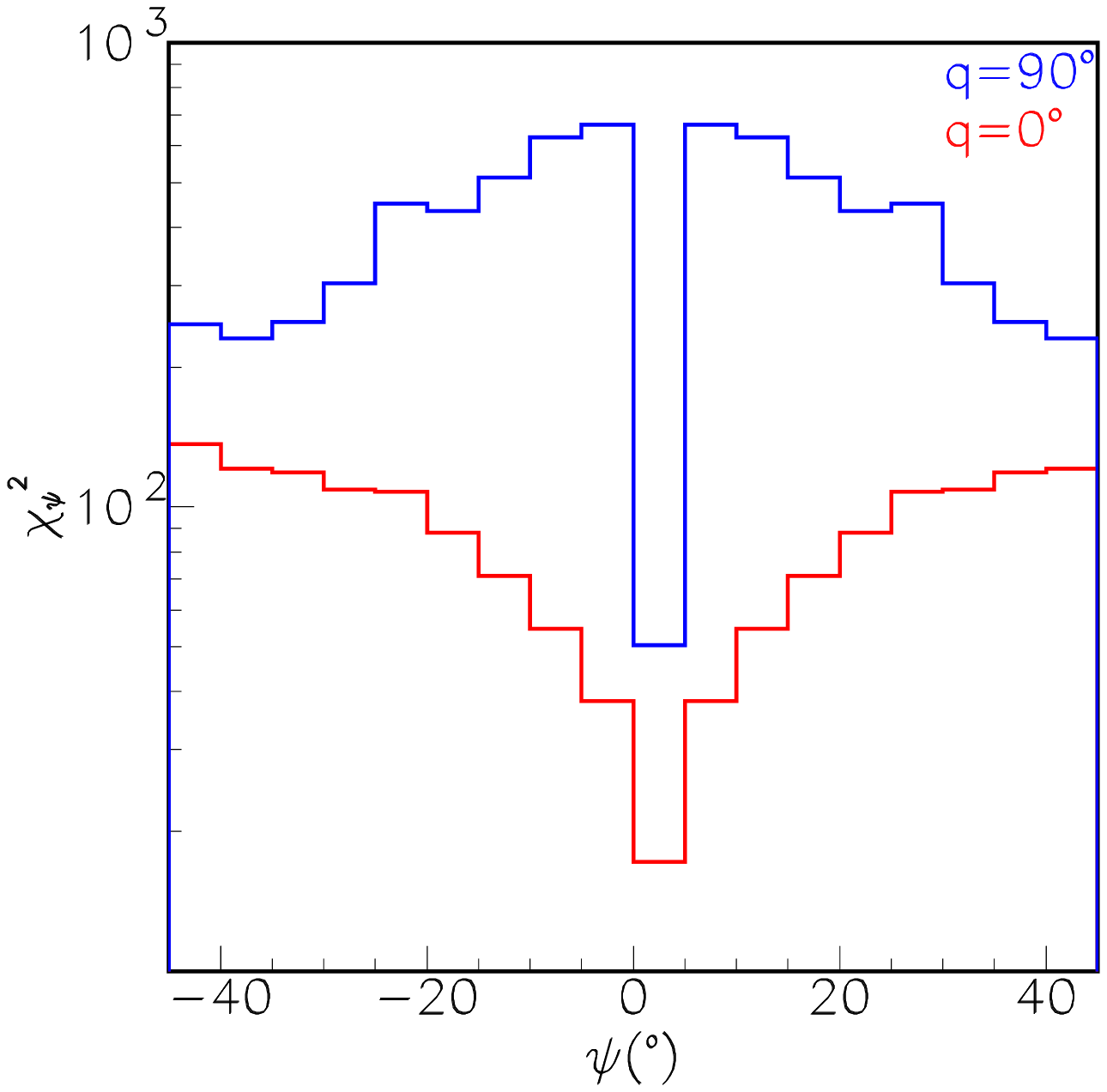}
  \includegraphics[width=.33\linewidth,trim={0cm 1cm 1.5cm 1cm},clip]{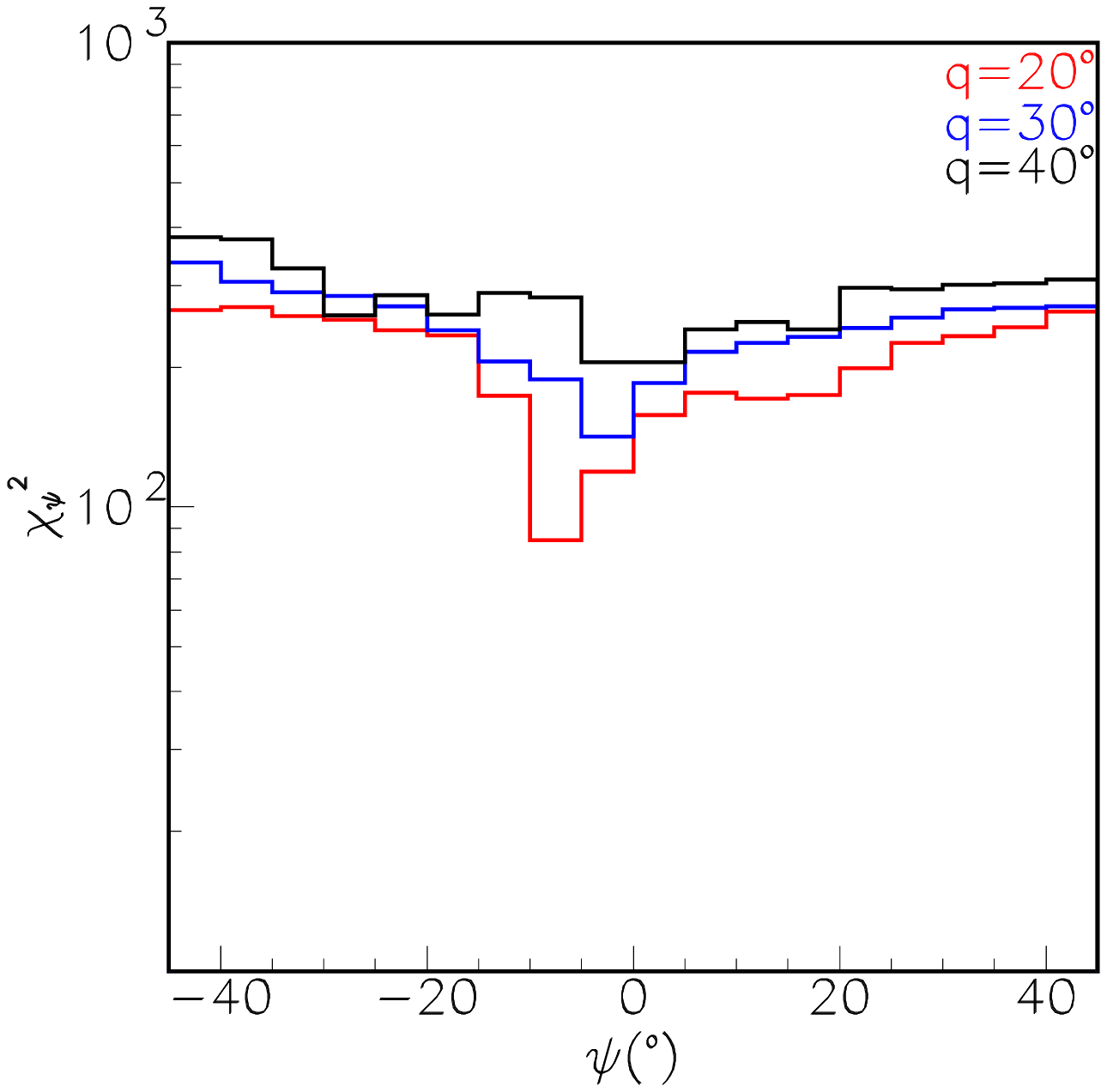}
  \includegraphics[width=.33\linewidth,trim={0cm 1cm 1.5cm 1cm},clip]{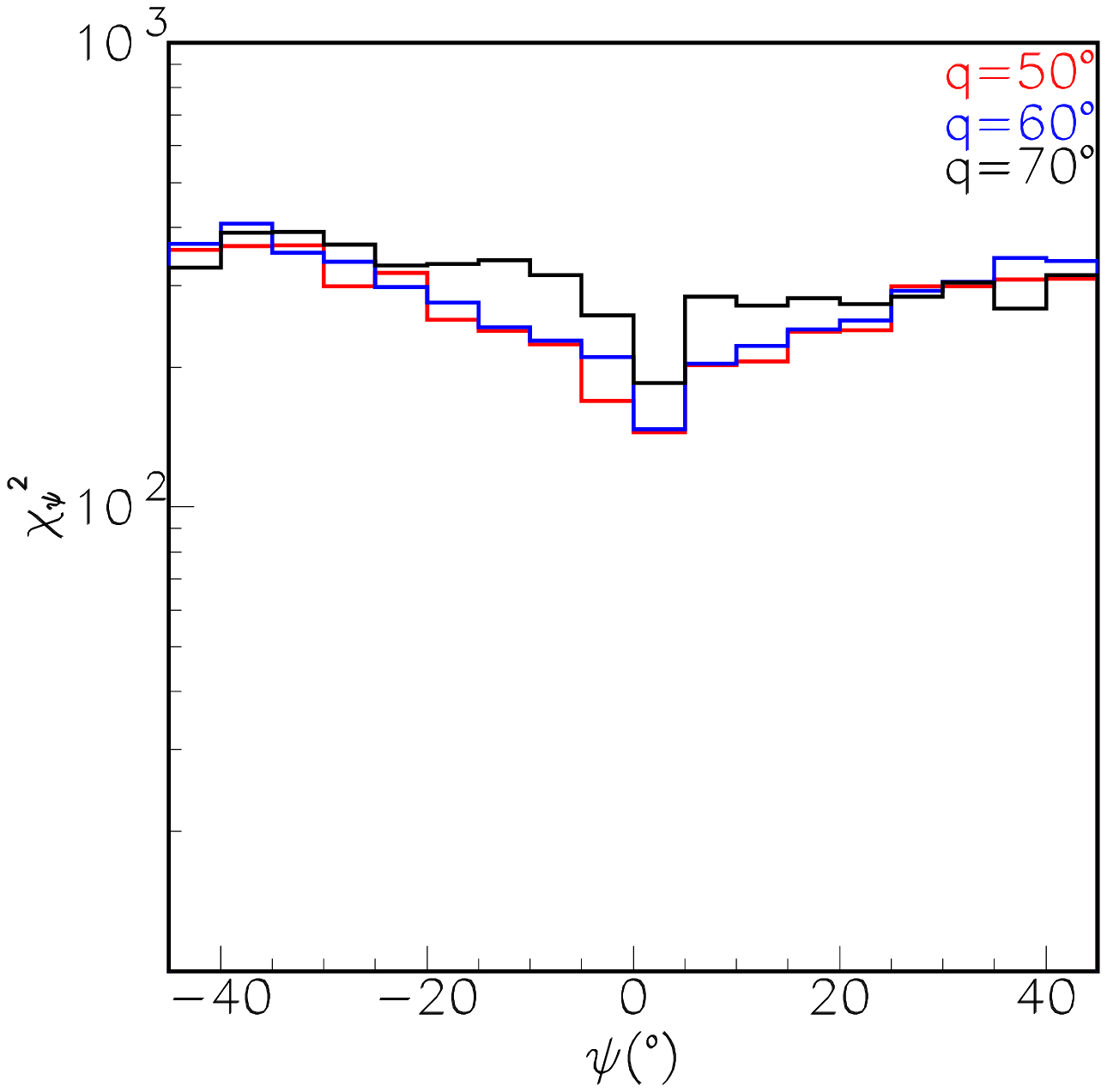}
  \caption{Distributions of $\chi^2_{\psi}$ as a function of $\psi$ for case $A$ and $\theta =45^\circ$. The left panel is for pure expansion ($q=0$, red) or pure rotation ($q=90^\circ$, blue). The middle and right panels are for mixed expansion and rotation, their relative importance being measured by the angle $q$ (see text) with values of 20$^\circ$ (red), 30$^\circ$ (blue) and 40$^\circ$ (black) in the middle panel, 50$^\circ$ (red), 60$^\circ$ (blue) and 70$^\circ$ (black) in the right panel. The projection of the star axis on the sky plane is at $\psi =0$.}
  \label{figa4}
\end{center}
\end{figure*}

The sum $\sum_{exp}$ runs over pixels $(y,z)$ located just a bit north of the $y$ axis ($z>0$ and small) and the sum $\sum_{rot}$ over  pixels $(y,z)$ located just a bit east of the $z$ axis ($y>0$ and small). In addition, both sums run over the whole velocity spectrum. In each square bracket, the subtraction of the flux density measured on the axis, rotation on the $y$ axis and expansion on the $z$ axis, leaves us with the contribution that changes sign under reflection about the axis, expansion about the $y$ axis and rotation about the $z$ axis. Therefore, if the $z$ axis is the projection on the sky plane of the star axis, $\chi^2_{\psi}$ nearly cancels when subtracting the two square brackets from each other. The value of $\psi$ associated with the star axis is therefore expected to be that for which the quantity $\chi^2_{\psi}$, calculated for flux densities rotated by $\psi$ about the origin on the sky map, reaches a minimum. The quantities $\Delta$ stand for the uncertainties attached to the measured flux densities. Here, we use for $\Delta$ the square root of the sum of the four flux densities appearing in the two square brackets of the relevant sum (which would be proper for purely statistical uncertainties). Figure~\ref{figa4} illustrates the difficulty of evaluating the proper value of $\psi$ in the mixed case. It is drawn for $\theta =45^\circ$ and for $V_{rad}=V_0\;\cos q \; \sin ^2\alpha$ and $V_{rot}=V_0\;\sin q \; \cos ^2\alpha$ for different values of $q$, namely of the relative importance of rotation vs expansion. The minimum is generally very shallow and in one case, $q=20^\circ$, it is off by $\sim 5^\circ$. Better indicators than $\chi^2_{\psi}$ may possibly be conceived, but they will always reveal the difficulty of measuring $\psi$ when rotation and expansion compete for inclinations around $\theta =45^\circ$.

\subsection{Telling rotation from expansion}
\label{seca6}
Assuming $-V_{rot}>0$ and writing $S_{\pm}(y,z)=\sum_{\pm}|V_x|f(y,z,V_x)$ where the sum extends over $\pm V_x>0$, we have for pure rotation
\begin{equation} \label{aeq10}
\begin{split}
&S_+(y,\pm z)=S_0;\qquad S_-(y,\pm z)=0;\\
&S_+(-y,\pm z)=0;\qquad S_-(-y,\pm z)=S_0
\end{split}
\end{equation}
and for pure radial expansion
\begin{equation} \label{aeq11}
\begin{split}
&S_+(\pm y,z)=S_1;\qquad S_-(\pm y,z)=S_2;\\
&S_+(\pm y,-z)=S_2;\qquad S_-(\pm y,-z)=S_1
\end{split}
\end{equation}

\begin{figure*}
\begin{center}
  \includegraphics[width=.3\linewidth,trim={0.5cm 0cm 0cm 0cm},clip]{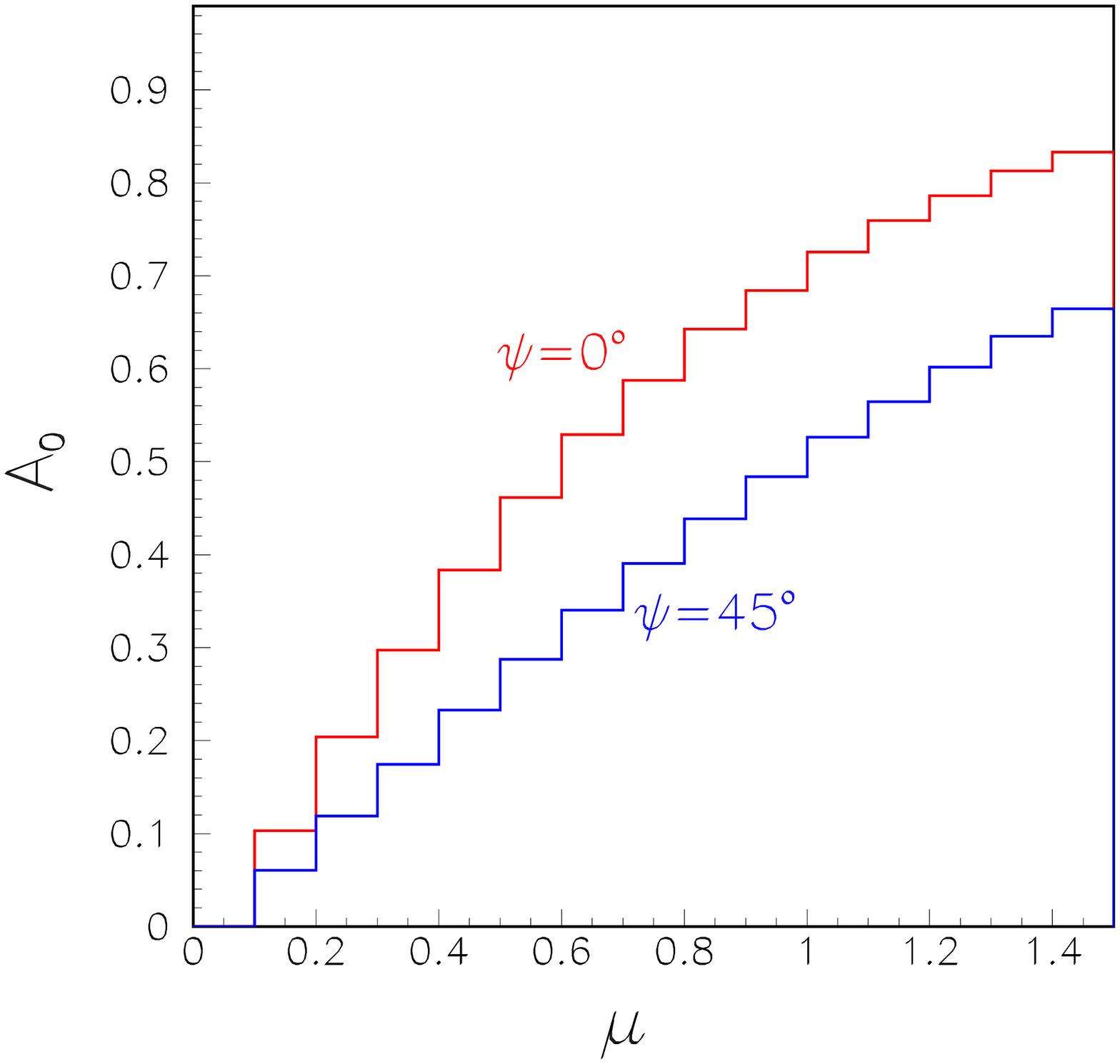}
  \hspace{1cm}
  \includegraphics[width=.3\linewidth,trim={0cm 0cm 0.5cm 0cm},clip]{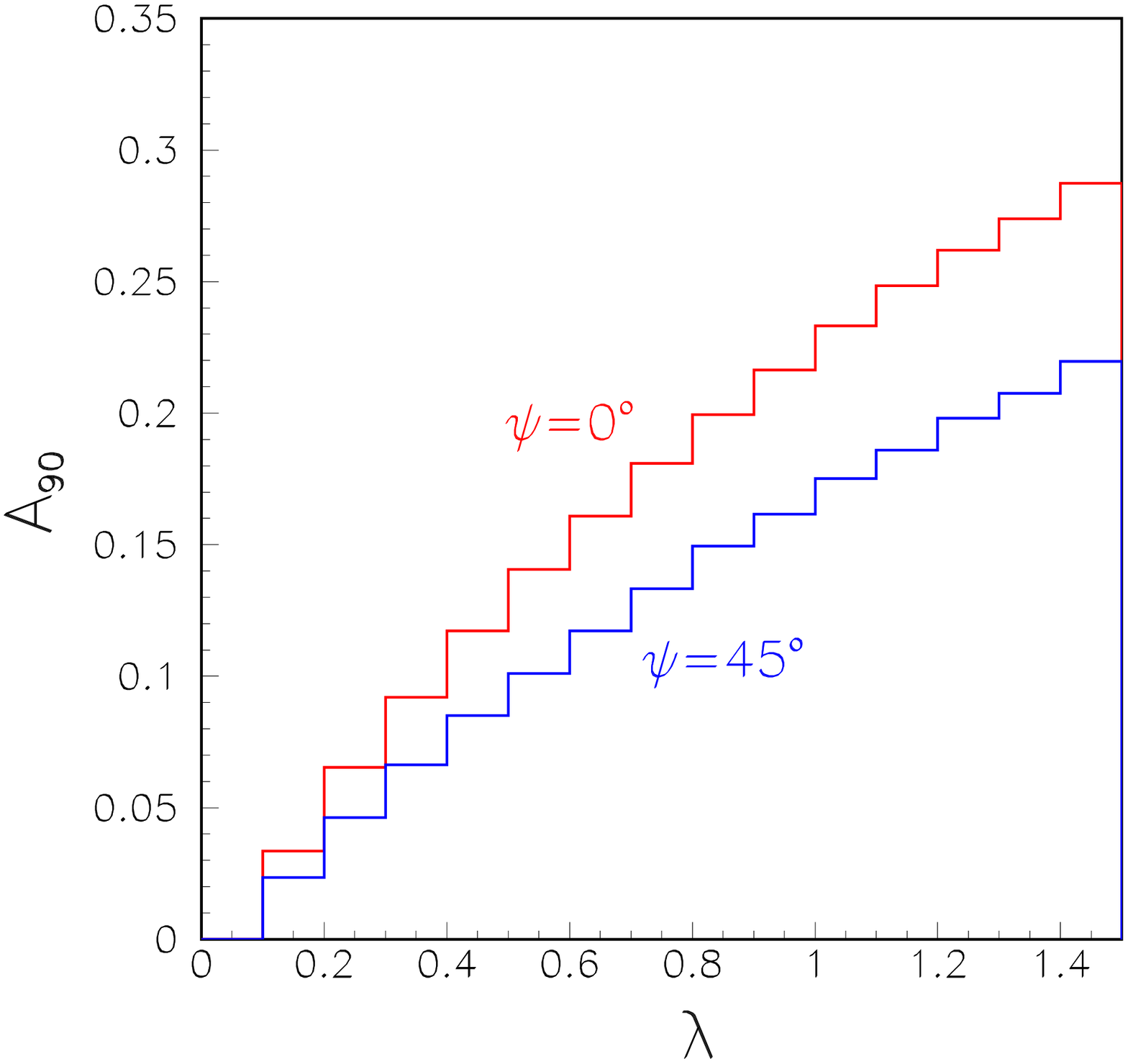}
  \caption{Dependence of $A_0$ on $\mu$ (left) and of $A_{90}$ on $\lambda$ (right) for $\psi=0$ and $\psi=45^\circ$ separately.}
  \label{figa5}
\end{center}
\end{figure*}

Hence, writing $A_{as}=(S_+-S_-)/(S_++S_-)$, \mbox{$|A_{as}(y,\pm z)|=1$} for pure rotation 
while for pure radial expansion $|A_{as}(\pm y,z)|=(|S_1-S_2|)/(S_1+S_2)$. 
Averaging over $R$, \mbox{$A(\varphi)=<|A_{as}(R,\varphi)|>=1$} for pure rotation and 
\mbox{$A(\varphi)=<(|S_1-S_2|)/(S_1+S_2)>$} for pure expansion, which cancels for both 
$\theta =90^\circ$ and $\theta =0^\circ$ and is always smaller than unity. In the 
case of pure rotation, $A(\varphi)=1$ independently from both $\varphi$ and $\theta$. 
The value of $A(\varphi)$ at $\varphi =0^\circ$, $\varphi =90^\circ$ or its integral over 
$\varphi$ should therefore be good indicators of the relative importance of rotation over 
expansion. However, as soon as rotation is mixed with some expansion, even when the latter 
is isotropic, $A(\varphi)$ starts departing from unity. Indeed, in the present article, 
for example in Figure~\ref{fig3}, we usually considered expansion cancelling on the star 
equator. In reality, even in the presence of a clear bipolar outflow, some expansion may 
take place in the equatorial region. Moreover, the CSE may start its evolution with a 
small bipolar outflow and no global expansion or, on the contrary, with a small isotropic 
expansion. In the latter case, when the bipolar outflow starts to emerge, it is but a 
small perturbation to the isotropic expansion. We take this in consideration by writing 
$V_{rad}=V_0(1+\lambda \sin^2\alpha )$ and $V_{rot}=\mu V_0\cos^2\alpha$ ($V_{rot}$ must 
cancel at the poles) and use an isotropic effective emissivity $\rho =\rho_0/r^2$. The 
parameters $\lambda$ and $\mu$ measure the amount of respectively polar expansion 
and equatorial rotation that add to the isotropic expansion. We find that, to a 
good approximation, $A(90^\circ)=A(270^\circ)\sim A_{90}\sin2\theta$ and 
$A(0^\circ)=A(180^\circ)\sim A_0\sin\theta$ with $A_{90}$ independent of $\mu$ 
(the $\cos\varphi$ factor in the contribution of rotation to $V_x$ is the cause) 
and $A_0$ independent of $\lambda$. Both $A_{90}$ and $A_0$ increase with 
respectively $\lambda$ and $\mu$ as shown in Figure~\ref{figa5}.

\begin{figure*}
\begin{center}
  \includegraphics[width=.32\linewidth,trim={0cm 0cm 0cm 0cm},clip]{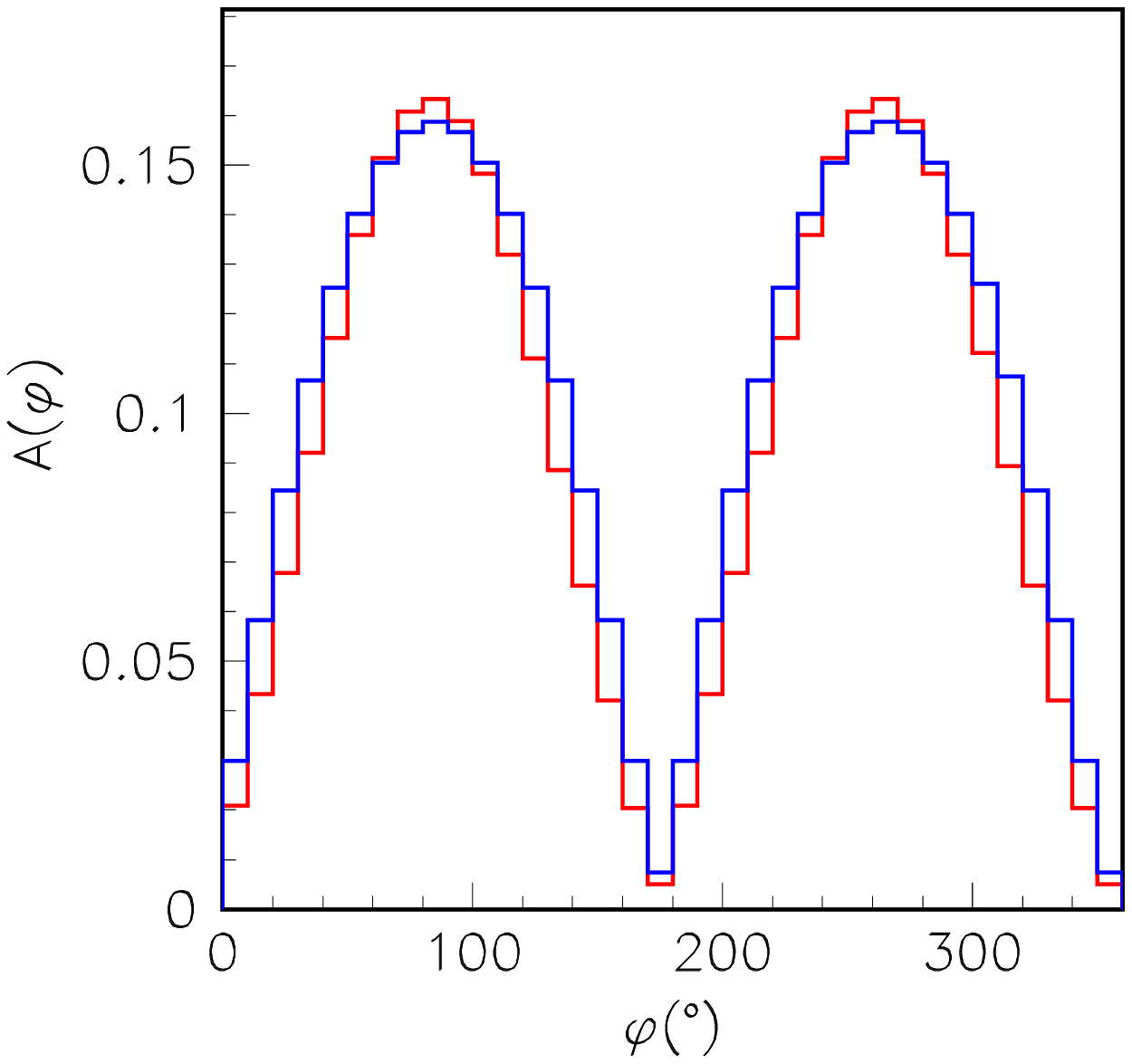}
  \includegraphics[width=.32\linewidth,trim={0cm 0cm 0cm 0cm},clip]{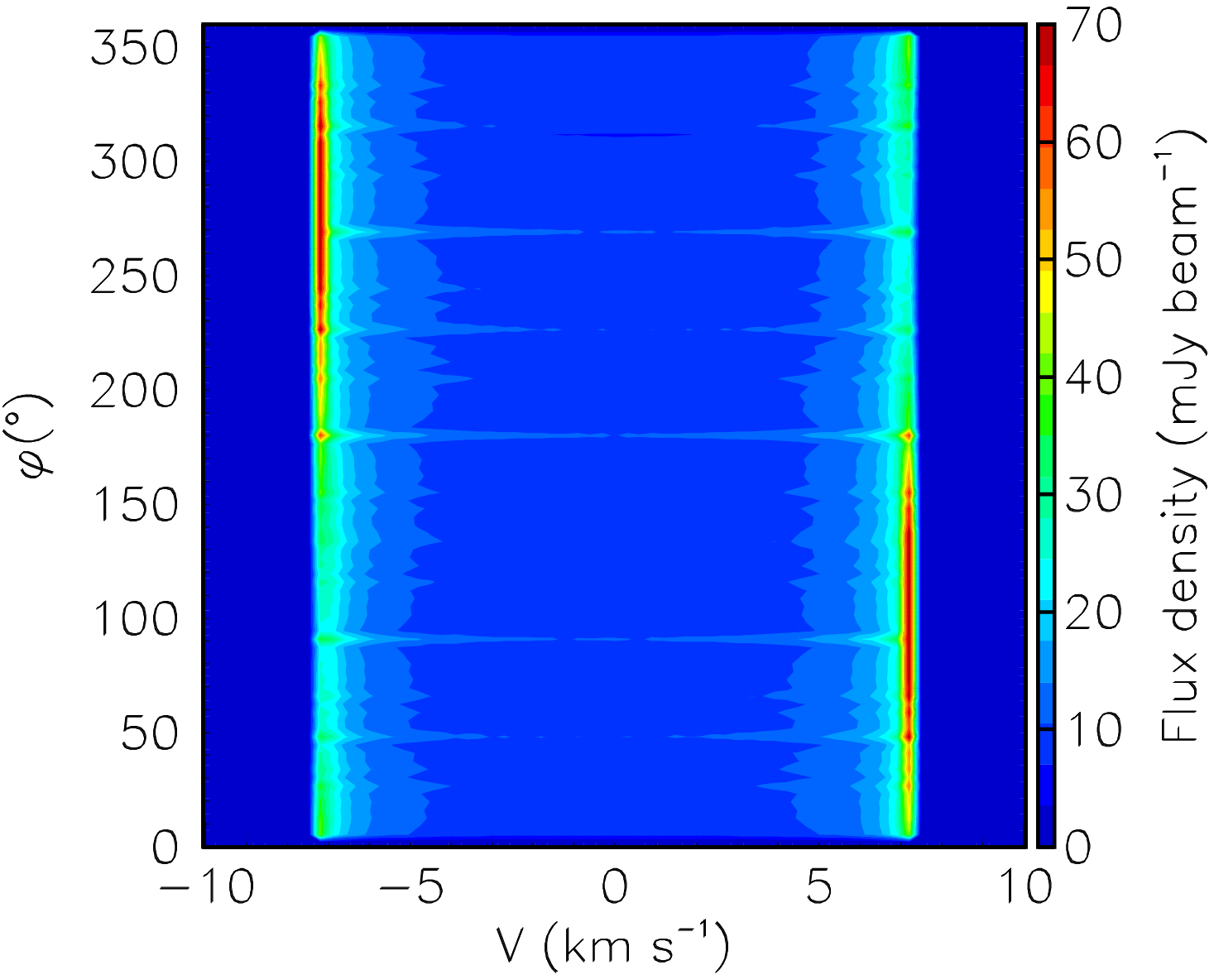}
  \hspace{.4cm}
  \includegraphics[width=.32\linewidth,trim={0cm 0cm 0cm 0cm},clip]{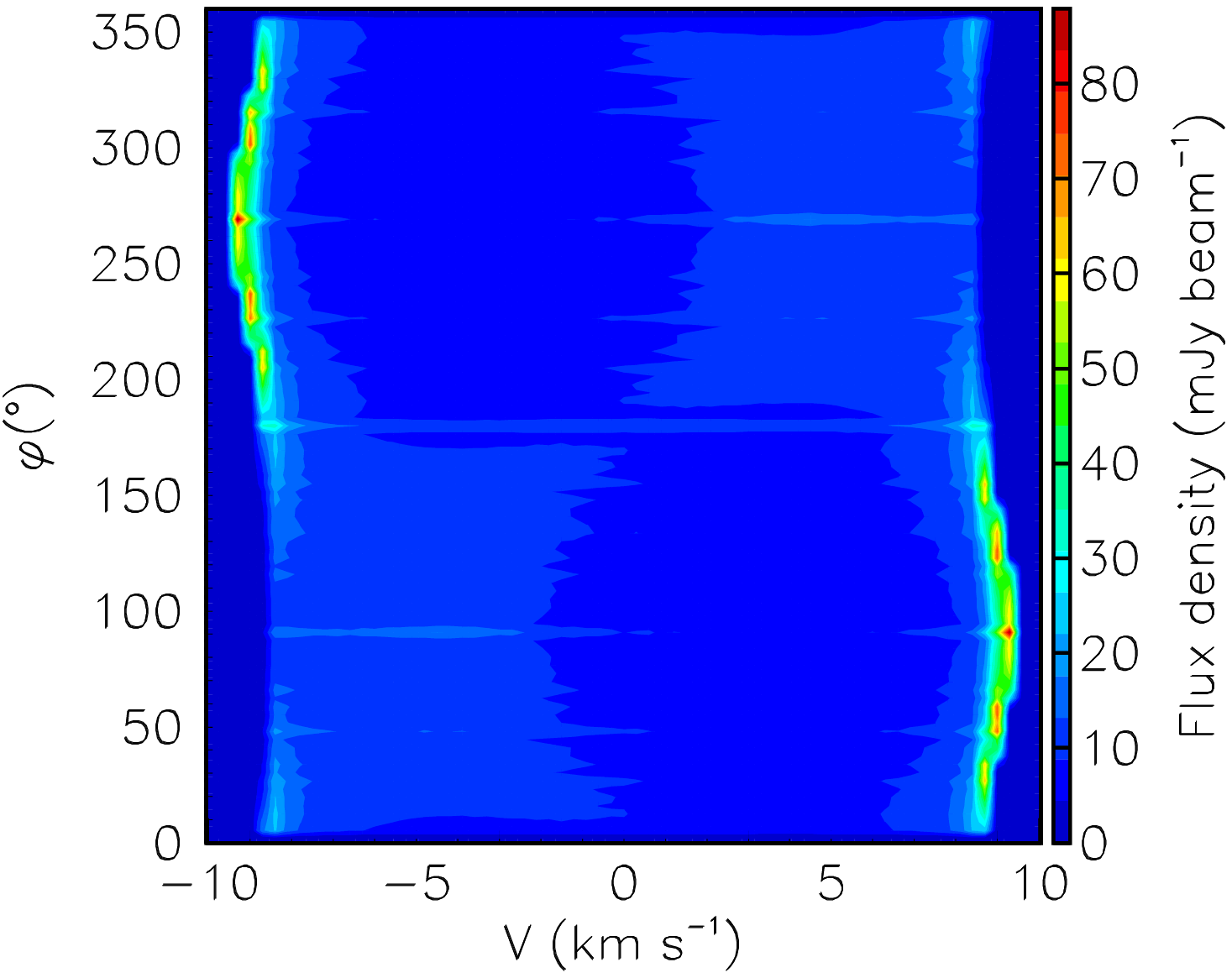}
  \caption{Comparison of the effect of a small rotation about the $y$ axis (case 1, $\mu =0.25$) with that of a bipolar expansion along the $z$ axis producing the same value of $A(\varphi)$ (case 2, $\lambda=0.6$). Left panel: distributions of $A(\varphi)$ for case 1 (blue) and case 2 (red). Middle and right panels: P-V diagrams for cases 1 and 2 respectively.}
  \label{figa6}
\end{center}
\end{figure*}

\begin{figure*}
\begin{center}
  \includegraphics[width=.33\linewidth,trim={0cm 1cm 0cm 1cm},clip]{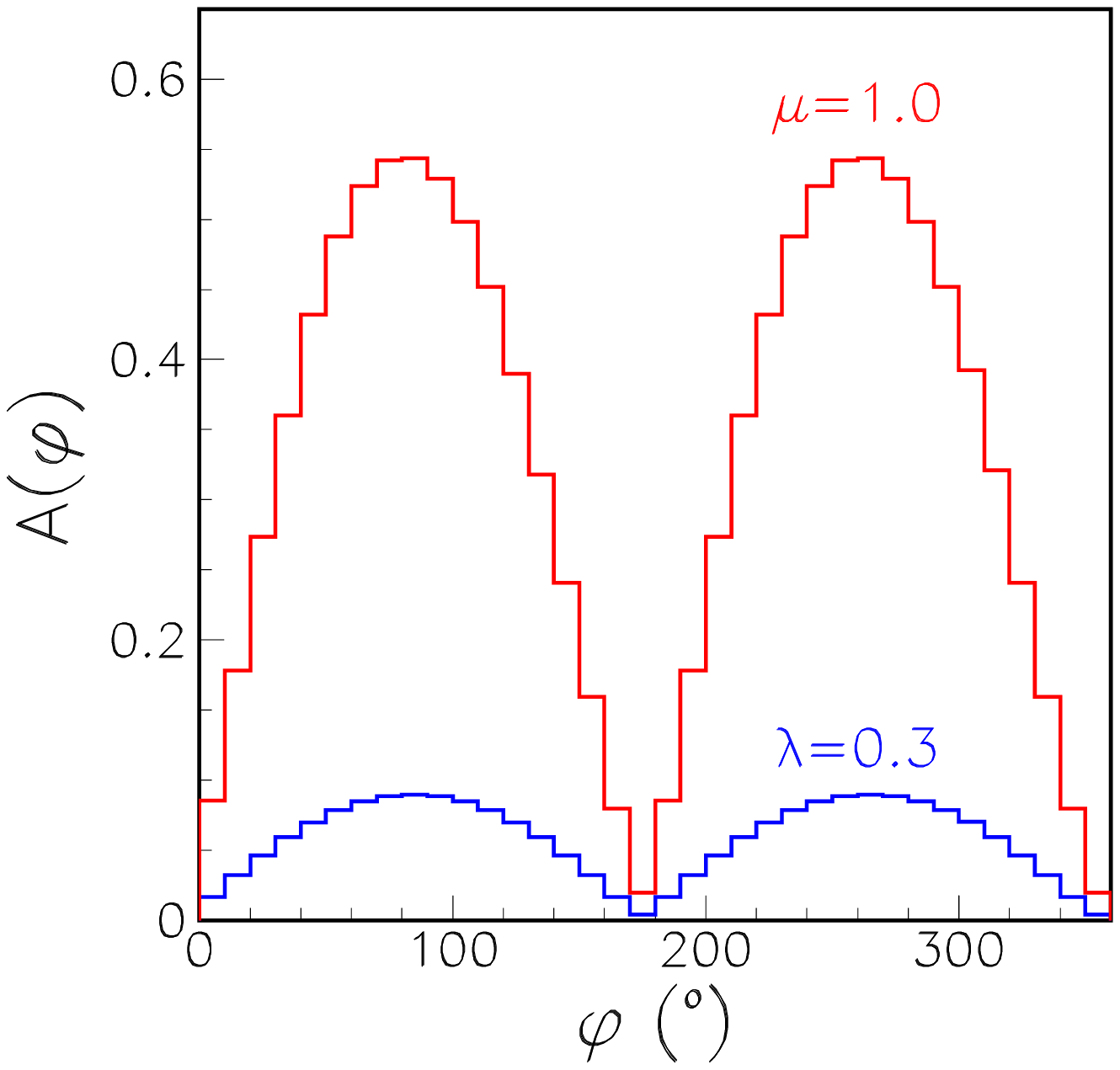}
  \includegraphics[width=.33\linewidth,trim={0cm 1cm 0cm 1cm},clip]{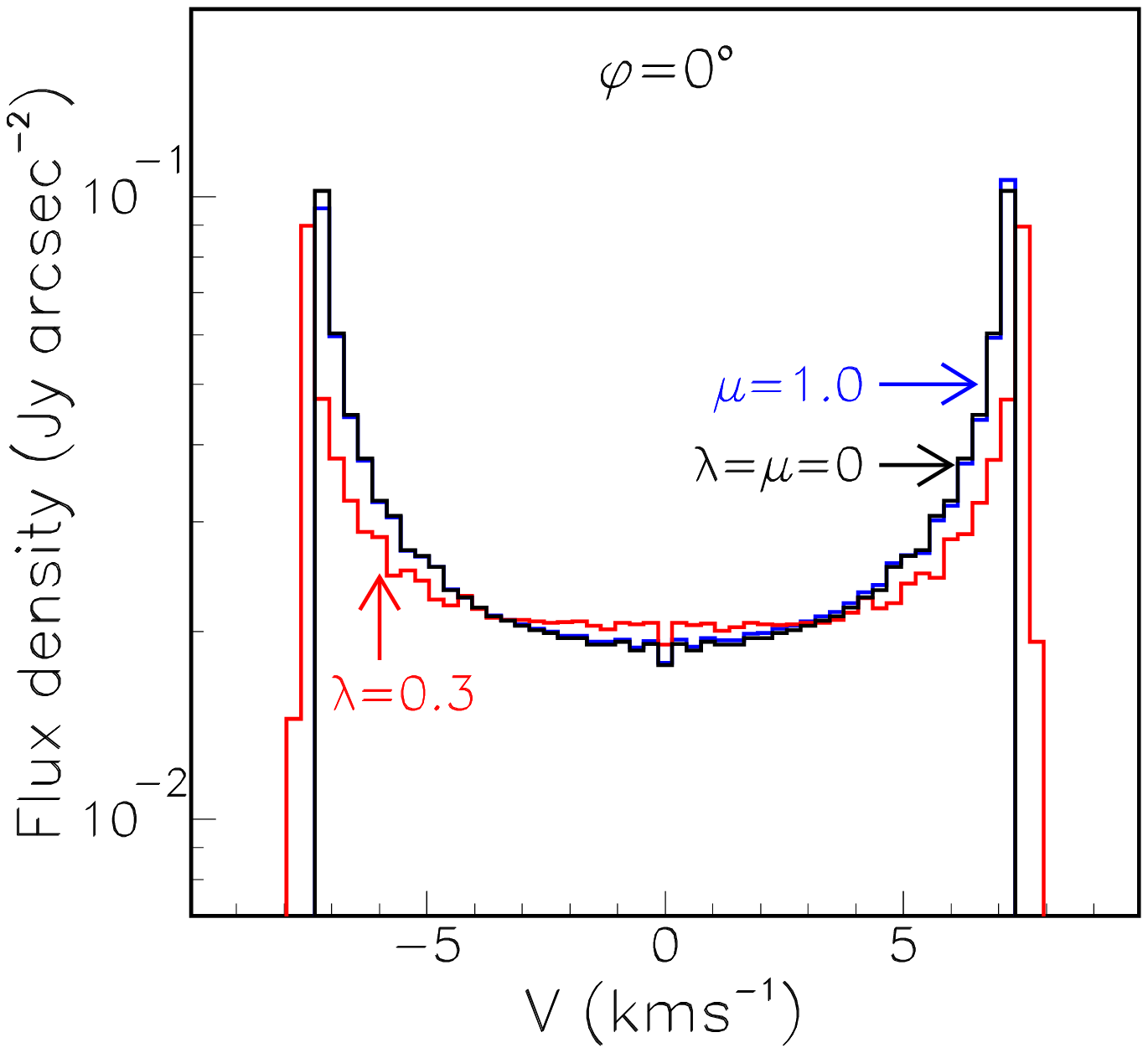}
  \includegraphics[width=.33\linewidth,trim={0cm 1cm 0cm 1cm},clip]{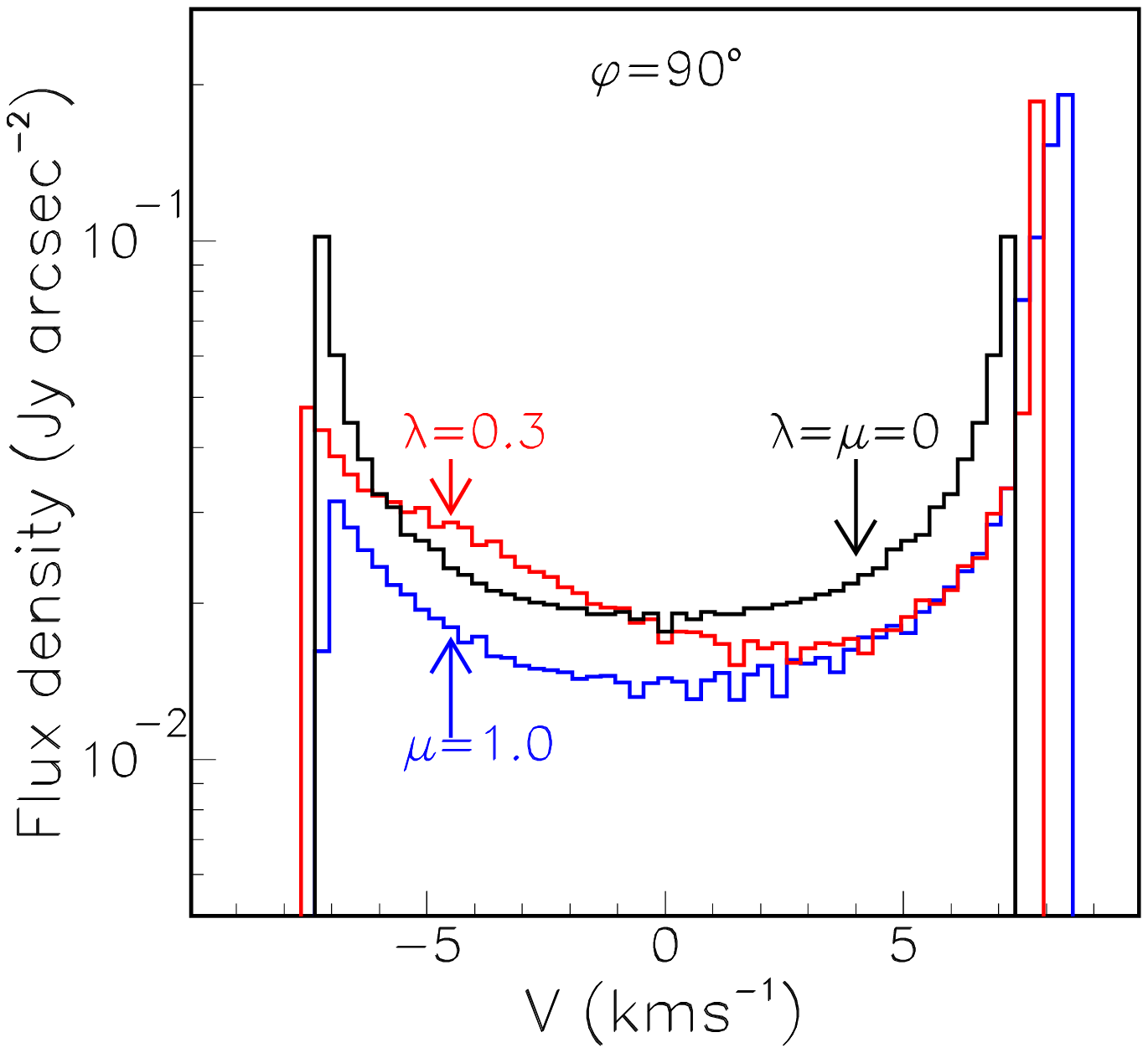}
  \caption{Comparison of the effect of a rotation about the $y$ axis (case 1, $\mu =1$) with that of a bipolar expansion along the $z$ axis (case 2, $\lambda =0.3$) producing similar effects on the P-V diagram. Left panel: distributions of $A(\varphi)$. Middle and right panels: Doppler velocity distributions obtained at $\varphi=0$ ($y$ axis) and $\varphi =90^\circ$ ($z$ axis) respectively. Black is for $\lambda =\mu =0$, blue for case 1 and red for case 2.}
  \label{figa7}
\end{center}
\end{figure*}

The above considerations suggest comparing the P-V diagrams of a $\lambda =0$ 
case with a $\mu =0$ case for a same value of $\theta$. We choose $\theta =45^\circ$ 
in order to maximize the effect of expansion and adjust $\mu$ and $\lambda$ to 
have similar values of $A(\varphi)$.  The associated P-V diagrams, shown in 
Figure~\ref{figa6}, display small but significant differences. If, on the contrary, 
we adjust $\mu$ and $\lambda$ to have similar effects on the P-V diagrams, 
we obtain much larger values of $A(\varphi)$ in the $\lambda =0$ case (rotation) 
than in the $\mu =0$ case (expansion). Figure~\ref{figa7} displays the 
Doppler velocity distributions obtained in this case on the positive $y$ and $z$ axes. 
The small differences illustrate the difficulty of telling expansion from rotation 
when they are small perturbations to a global isotropic expansion.

\subsection{Inclination $\theta$ of the star axis with respect to the line of sight}
\label{seca7}
We first consider the dependence on $\theta$ of the integrated flux $F$. Its partial derivative with respect to $\varphi$, $F'_{\varphi}(R,\varphi )$, is expected to be an indicator of $\theta$. Indeed, writing $\rho (r,\sin\alpha )$ the effective emissivity and using $\partial (\sin\alpha )/\partial \varphi=(R/r)\cos\varphi \sin\theta$, we obtain 
\begin{equation} \label{aeq12}
F'_{\varphi} (R,\varphi )=R\cos\varphi \sin\theta \int \rho ' _{\sin\alpha }(r,\sin\alpha )\frac{1}{r} dx,
\end{equation} 
namely
\begin{equation} \label{aeq13}
\sin\theta =F'_{\varphi}(R,\varphi)[R\cos\varphi \int \rho ' _{\sin\alpha } (r,\sin\alpha )\frac{1}{r} dx]^{-1}
\end{equation}

The dependence of $F$ on $\varphi$ is indeed an indicator of $\sin\theta $, 
however it requires the preliminary knowledge of the dependence of the 
effective emissivity on the star latitude, measured by $\rho ' _{\sin\alpha } (r,\sin\alpha )$. 
It is not possible to evaluate $\theta$ independently from the elongation of the 
effective emissivity along the star axis.  Indeed, modelling the effective emissivity 
as $\rho =\rho_0(1+\lambda_{\rho} \sin^2\alpha)r^{-2}$ ($r$ measures in arcseconds), gives:

\begin{equation} \label{aeq14}
\begin{split}
F'_{\varphi}(R,\varphi )&=R\cos\varphi \sin\theta \int 2\rho_0\lambda_{\rho}\sin\alpha \frac{1}{r} dx\\
&=2R\rho_0\lambda_{\rho}\cos\varphi \sin\theta \int (x\cos\theta +z\sin\theta )\frac{1}{r^4} dx
\end{split}
\end{equation}

The $x$ term in the parenthesis does not contribute to the integral and, writing $u=dx/R$,

\begin{equation} \label{aeq15}
\begin{split}
F'_{\varphi}(R,\varphi )&=2\,y\,z\,\rho_0\,\lambda_{\rho}\,\sin^2\theta \int \sqrt{u^2+1}du\\
&=\pi\,y\,z\,\rho_0\,\lambda_{\rho}\,\sin^2\theta
\end{split}
\end{equation}

Namely the dependence of the integrated flux on $\varphi$ measures the product $\lambda_{\rho}\sin^2\theta$: the inclination angle $\theta$ of the star axis on the line of sight cannot be measured from the map of integrated fluxes independently from the elongation $\lambda_{\rho}$ of the effective emissivity along the star axis. Qualitatively, this result is expected to be always true. In the case of a spherically symmetric effective emissivity but axially symmetric wind velocity, a similar result applies. It is best seen in the case $B$ of Section~\ref{sec2.3}, which is particularly simple in this context. As seen in Figure~\ref{figa3}, changing $\theta$ does not change the limits on the sky plane of the expansion cone as long as the ratio $\tan\beta_0 /\tan\theta$ is kept constant. Using this prescription, one may then readjust the values taken by the expansion and rotation velocities in order to minimize the changes experienced by the measured flux densities. On the axis of the cone, projected on the sky plane as the $z$ axis, $V_x=V_0\cos\theta$: keeping this product constant will limit the effect of changing $\theta$ to small variations of the limits of the velocity spectrum. Similarly, outside the cone, the Doppler velocity varies between 0 on the $z$ axis ($y=0$) to $V_0\sin\theta$: again, keeping this product constant will limit the effect of changing $\theta$ to small variations of the limits of the velocity spectrum. 

In summary, when modelling the star with case $B$ using five parameters $\psi$, $\theta$, $\beta$, $V_{exp}$ and $V_{rot}$ ($V_{exp}$ inside the cone and $V_{rot}$ outside), very strong correlations affect the last four parameters leaving much arbitrariness in the value of $\theta$.  What we effectively measure are the ratio $\tan\beta_0 /\tan\theta$ and the products $V_{exp}\cos\theta$ and $V_{rot}\sin\theta$. Qualitatively, as for the effective emissivity, we measure the product of the inclination of the star axis on the line of sight by the elongation of the expansion cone. Another evidence for this ambiguity between inclination and elongation was obtained in Appendix~\ref{seca6} where $A(\varphi)$ was found to include a factor $\lambda \sin\theta$.

\subsection{Radial dependence of the CSE properties}
\label{seca8}
For $\psi=0$, in the case of a spherical effective emissivity, we may write 
\begin{equation} \label{aeq16}
\rho(r,\sin\alpha )=\sum \rho_nr^{-n}
\end{equation}
implying $F(R,\varphi)=\sum F_n(R,\varphi)$
\begin{equation} \label{aeq17}
\begin{split}
&\textnormal{with}\\
&F_n(R,\varphi)=\rho_n\int (x^2+R^2)^{-\frac{n}{2}}dx=\rho_nR^{1-n}I_n\\
&\textnormal{and}\\
&I_n=2\int (1+u^2)^{-\frac{n}{2}}du\\
\end{split}
\end{equation}
here, the integral runs from 0 to $+\infty$ and for $n=2$, $I_2=\pi$.

Hence 
\begin{equation} \label{aeq18}
F'_R(R,\varphi)=\sum (1-n)R^{-n}\rho_nI_n
\end{equation}
the $R$-dependence of the integrated flux is simply related to the 
$r$-dependence of the effective emissivity. In particular, in case of a pure 
power law, 
\begin{equation} \label{aeq19}
\rho(r,\sin\alpha)=\rho_nr^{-n}\\
\end{equation}
$R^{n-1}\;F(R,\varphi)$ is a constant over the sky plane.

If central symmetry is obeyed, the mean value of $V_x$ cancels. However, velocity gradients can be expected to show up on the expression of the square or absolute value of $V_x$. In the case of radial expansion and assuming a pure power law dependence of the effective emissivity on $r$, 
 
\begin{equation} \label{aeq20}
\begin{split}
I_n<|V_x|>&=R^{n-1}\int |x|V_{rad}r^{-n-1}dx\\
&=2\int uV_{rad}(1+u^2)^{-\frac{1}{2}(n+1)}du\\
&=\int V_{rad}v^{-\frac{1}{2}(n+1)}dv.
\end{split}
\end{equation}

In the absence of gradient,  $I_n<|V_x|>=\frac{1}{2}(n-1)V_{rad}$ is independent of $R$. 
In general, developing $V_{rad}$ in powers of $r$, the $R$-dependence of 
$<|V_x|>$ depends simply on the $r$-dependence of the expansion velocity.

In the case of pure rotation, $V_x=y\sin\theta(r\cos\alpha)^{-1}V_{rot}$ 
and writing $V_{rot}=V_0\cos\alpha r^{-k}$, a form making the enhancement 
of rotation in the equatorial region explicit, we obtain

\begin{equation} \label{aeq21}
\begin{split}
I_n<|V_x|>&=y\sin\theta R^{n-1}\int \frac{1}{\cos\alpha}V_{rot}r^{-n-1}dx\\ 
&=y\sin\theta V_0R^{n-1}\int r^{-n-k-1}dx\\
&=\cos\varphi \sin\theta V_0R^{-k}.
\end{split}
\end{equation}

The $R$-dependence of $<|V_x|>$ is the same as the $r$-dependence of $V_{rot}$.

\bsp	
\label{lastpage}
\end{document}